%% file: main-rev1.tex
\documentclass[journal]{IEEEtran}
\usepackage[english]{babel}
\usepackage{blindtext}
\usepackage{tikz,pgfplots}
\usetikzlibrary{calc}
\usepackage{tkz-euclide}
\pgfplotsset{compat=1.5}
\usepgfplotslibrary{polar}
\usepackage{epstopdf} 
\usepackage{amsmath}
\usepackage{wrapfig}
\usepackage{verbatim}
\usepackage{mathrsfs}
\usepackage{amssymb}
\usepackage{lipsum}  
\usepackage{accents}
\usepackage{acronym}
\usepackage{graphicx}
\usepackage{textcomp}
\usepackage{xcolor}
\usepackage{mathtools}
\usepackage{color,soul}
\usepackage{graphicx}
\usepackage{booktabs}
\usepackage{multirow}
\usepackage{tikz}
\usepackage[font={small}]{caption}
\usepackage[font={small}]{subcaption}
\usepackage{algorithm}
\usepackage{algorithmic}
\usepackage{fancyhdr}
\usepackage[symbol*]{footmisc}

\DeclareMathOperator*{\argmax}{arg\,max}
\DeclareMathOperator*{\argmin}{arg\,min}

\usepackage{caption}
\usepackage{subcaption}
\usepackage{cleveref}
\usepackage{url}

\newcommand{\eq}[1]{Eq.~\eqref{#1}}
\newcommand{\fig}[1]{Fig.~\ref{#1}}

\newcommand{\secref}[1]{Section~\ref{#1}}

\input{ACRONYMS.tex}

\newcommand{\rev}[1]{{\color{blue}#1}} 

\renewcommand{\rev}{}

\DefineFNsymbolsTM{otherfnsymbols}{%
  \textdagger    \dagger
  \textsection   \mathsection
  \textdaggerdbl \ddagger
  \textasteriskcentered *
  \textbardbl    \|%
  \textparagraph \mathparagraph
}%

\IEEEaftertitletext{\vspace{-2\baselineskip}}

\begin{document}

\title{JUMP: Joint communication and sensing with Unsynchronized transceivers Made Practical}

\author{Jacopo~Pegoraro$^{\dag  *}$, Jesus~O.~Lacruz$^{\ddag}$, Tommy~Azzino$^{\mathsection}$, Marco~Mezzavilla$^{\mathsection}$, \\Michele~Rossi$^{\dag \star}$, Joerg~Widmer$^{\ddag}$, and Sundeep~Rangan$^{\mathsection}$
\thanks{$^{*}$Corresponding author email \texttt{jacopo.pegoraro@unipd.it}.
$^{\dag }$These authors are with the University of Padova, Department of Information Engineering.
$^{\star }$This author is with the University of Padova, Department of Mathematics ``Tullio Levi-Civita''.
$^{\ddag }$These authors are with the IMDEA Networks Institute.
$^{\mathsection}$These authors are with the New York University.

This work was partially supported by the European Union under the Italian National Recovery and Resilience Plan (NRRP) of NextGenerationEU, partnership on “Telecommunications of the Future” (PE0000001 - program “RESTART”).}}

\markboth{Journal of \LaTeX\ Class Files,~Vol.~14, No.~8, August~2021}%
{Shell \MakeLowercase{\textit{et al.}}: A Sample Article Using IEEEtran.cls for IEEE Journals}


\maketitle

\begin{abstract}
Wideband \mbox{millimeter-wave} communication systems can be extended to provide radar-like sensing capabilities on top of data communication, in a cost-effective manner.
However, the development of  \textit{joint communication and sensing} technology is hindered by practical challenges, such as occlusions to the line-of-sight path and clock asynchrony between devices. The latter introduces \textit{time-varying} timing and frequency offsets that prevent the estimation of sensing parameters and, in turn, the use of standard signal processing solutions. Existing approaches cannot be applied to commonly used phased-array receivers, as they build on stringent assumptions about the multipath environment, and are computationally complex.
We present JUMP, the first system enabling \textit{practical} bistatic and asynchronous joint communication and sensing, while achieving accurate target tracking and micro-Doppler extraction in realistic conditions.
Our system compensates for the timing offset by exploiting the channel correlation across subsequent packets. Further, it tracks multipath reflections and eliminates frequency offsets by observing the phase of a dynamically-selected static reference path.
JUMP has been implemented on a 60~GHz experimental platform, performing extensive evaluations of human motion sensing, including non-line-of-sight scenarios. In our results, JUMP attains comparable tracking performance to a full-duplex monostatic system and similar micro-Doppler quality with respect to a phase-locked bistatic receiver.
\end{abstract}


\section{Introduction} \label{sec:intro}

\rev{\ac{jcs} has emerged as a potential game changer for next-generation wireless networks, endowing communication systems with radar-like capabilities to perceive their surroundings~\cite{zhang2021enabling, liu2023seventy}. 
\ac{jcs} system designs are categorized in radar-centric and communication-centric, depending on which of the two functionalities is regarded as primary~\cite{zhang2020perceptive, cui2021integrating}. Communication-centric \ac{jcs} is the most promising approach for a cost effective solution, as it leverages the ubiquitous communication hardware and waveforms for sensing, adding minimal overhead and modifications to existing devices and protocols~\cite{zhang2022integration, pegoraro2022sparcs}. To this end, communication-centric \ac{jcs} should maintain the typical wireless network configuration of \textit{separated} transmitter and receiver, with half-duplex capabilities (bistatic configuration)~\cite{lu2023integrated}, as opposed to impractical full-duplex designs~\cite{barneto2021full}. In this setting, sensing is performed by repurposing the \ac{cir} or \ac{csi} estimation processes to extract information about the environment~\cite{cui2023integrated}. This allows estimating physical parameters such as the distance and velocity of nearby targets of interest, as these are related to signal reflection delays and Doppler shifts~\cite{Mazahir_JCRsurvey, zhang2021overview}. 
However, a large-scale adoption of bistatic \ac{jcs} is impaired by the fact that network nodes are \textit{asynchronous}, i.e., they have different clock sources and oscillators for the \ac{rf} front-end~\cite{zhang2022integration}.}
This asynchrony causes a time-varying \ac{to} and \ac{cfo}. The former appears as a common delay shift for all propagation paths in the \ac{cir}, preventing the correct estimation of actual path delays. \rev{The latter is a random frequency shift that destroys phase coherence across subsequent packets, hindering the estimation of the Doppler shift and of the \ac{md} effect. The \ac{md} is a frequency modulation of the reflected signal around the main Doppler frequency induced by movements of a target or target components. It is the main signal feature used in fine-grained wireless movement sensing, and has a vast number of applications in target classification, human activity recognition, pervasive healthcare, and person identification, among others~\cite{pegoraro2023rapid, li2022integrated, meneghello2022sharp}.}
While existing communication systems use algorithms to compensate for \ac{to} and \ac{cfo}, they treat the sensing parameters (delay and Doppler shift) as part of the undesired offsets and {\it remove} them~\cite{chen2019residual}. This makes such existing techniques unfit to our sensing purpose, as delay and Doppler are the channel parameters that we use to describe the physical environment.
For these reasons, clock asynchrony is the main obstacle to large-scale practical implementations of \ac{jcs} systems~\cite{zhang2022integration,zhang2021overview}, where \ac{to} and \ac{cfo} due to clock asynchrony are to be removed while \textit{retaining} the delay and the Doppler shift.

In this work, we design and implement JUMP, a \ac{jcs} system that solves the problem of asynchrony and enables bistatic sensing in wideband communication systems, e.g., \ac{mmwave} networks, which are particularly problematic due to the high phase noise~\cite{rasekh2019phase}. 
The system builds on two main insights. 
\rev{
\textit{First}, JUMP exploits the slow change of the \ac{cir} delay profile compared to the packet rate, and uses a fast correlation-based method to estimate the relative \ac{to} across subsequent packets. By compensating for the \ac{to}, it allows for target localization and tracking without ambiguity, even under frequent occlusions of the \ac{los}.

\textit{Second}, it leverages the fact that the \ac{cfo} is \textit{constant} across the different signal propagation paths for the same packet. This property, paired with the accurate multipath resolution of wideband signals, allows identifying \textit{static} reference multipath components, whose frequency shift is only due to \ac{cfo} (being static, they do not contain the \ac{md} component). Hence, the signal collected from these static paths is used to remove the \ac{cfo} from the sensing paths, which contain the \ac{md} from the targets of interest. This allows aggregating phase-coherent \ac{cir} estimates from subsequent packets, enabling \ac{md} estimation despite the clock asynchrony. 

Existing bistatic sensing approaches use cross-antenna-based \ac{to} and \ac{cfo} compensation, using the signal at one antenna of a \ac{mimo} receiver as a reference~\cite{zeng2019farsense, zeng2020multisense, li2022csi,ni2021uplink}. JUMP solves several practical problems associated with these methods: \textit{(i)}~it works with commonly used phased-array receivers, which do not allow access to the signal at each antenna; \textit{(ii)}~it preserves the linearity of the signal and does not introduce cross-terms that require complex estimation algorithms; \textit{(iii)}~it correctly operates in weak \ac{los} and \ac{nlos} conditions. Moreover, JUMP estimates the full \ac{md} spectrum of the targets, including the contributions of their different moving parts, not just their main Doppler frequency.}

Unlike other \ac{jcs} approaches, it is a \textit{practical} solution that neither requires full-duplex hardware capabilities, as in~\cite{pegoraro2022sparcs,barneto2021full,pegoraro2023rapid, heino2021design}, nor any synchronization between the nodes, nor any modification to the underlying communication standard. We stress that these qualities make JUMP appealing in more general settings than just \ac{jcs}, e.g., in bistatic radar systems~\cite{ zhang2022integration,malanowski2012two}, in case synchronization through fiber links or \ac{gps} signals is not viable.  

\rev{We provide analytical insights into JUMP's \ac{to} and \ac{cfo} removal performance, complemented by numerical simulations. Moreover, we prototype it on a 60~GHz IEEE~802.11ay-based \ac{sdr} platform. This implements independent transmitter and receiver pairs, thus serving as a realistic testbed for asynchronous \ac{jcs}. In the implementation, we solve additional practical issues, such as symbol-level synchronization, which is usually neglected in \ac{jcs} research~\cite{zhang2022integration}.}
To benchmark the system's target tracking and \ac{md} extraction capabilities, we conduct a vast experimental campaign in two indoor environments, performing people tracking and \ac{md} signature extraction.
In the dataset collection, we augment the testbed with additional receivers in monostatic and bistatic phase-locked configurations, enabling a comparison with alternative \ac{jcs} approaches from the literature. 

The contributions of this work can be summarized as:

$1.$ To the best of our knowledge, JUMP is the first system that enables practical \ac{jcs} for asynchronous transceivers in realistic bistatic configurations, performing accurate target tracking and \ac{md} extraction. 

\rev{$2.$ We design a correlation-based algorithm to compensate for the relative \ac{to} across subsequent packets, attaining consistent bistatic target tracking even in challenging multitarget and \ac{nlos} scenarios.}

\rev{$3.$ We leverage the high multipath resolution of wideband transceivers to remove the \ac{cfo} using a reference propagation path instead of a reference antenna, as done in the \ac{jcs} literature. This largely simplifies the estimation of the Doppler spectrum, removing limiting assumptions on the multipath environment, or the necessity of a \ac{mimo} antenna array.}

$4.$ We build two JUMP prototypes and collect a large dataset of IEEE~802.11ay \ac{cir} measurements for human motion tracking and \ac{md} extraction.

Our work shows promising results for bistatic asynchronous \ac{jcs} systems compared to monostatic full duplex and bistatic phase-locked ones.

\rev{The manuscript is organized as follows. In \secref{sec:rel-work} we discuss the related work, while \secref{sec:model} introduces the system model. JUMP is presented in \secref{sec:method}, along with a detailed explanation of each signal processing block. \secref{sec:analysis} contains an analysis of the \ac{to} and \ac{cfo} compensation errors, complemented by numerical simulations. In \secref{sec:implementation} we describe the implementation of JUMP on an experimental prototype. \secref{sec:results} contains experimental results demonstrating JUMP's promising performance. Finally, in \secref{sec:discussion} we underline key aspects of our design and future research directions, while concluding remarks are given in \secref{sec:conclusion}.}

\section{Related work}\label{sec:rel-work}

\noindent \textit{Full-duplex \ac{jcs}.}
The use of full-duplex technology for \ac{jcs} has been advocated in several works. In~\cite{barneto2021full,heino2021design}
the authors have investigated antenna arrays, beamforming, and waveform design to enable full-duplex \ac{jcs} in \ac{mmwave} 5G systems. 
In~\cite{pegoraro2023rapid, pegoraro2022sparcs}, full-duplex has been used as a practical solution to \ac{cfo} for IEEE~802.11ay sensing, presenting algorithms to detect multiple targets and their \ac{md} signatures in indoor scenarios.
However, full-duplex entails strong self-interference, which remains a challenging and unsolved problem in real communication systems, as it requires self-interference cancellation techniques~\cite{barneto2021full}, which are not yet a mature technology. As an alternative, the transmitted power can be reduced to avoid saturating the receiver, but this is only suitable for short-range indoor use~\cite{pegoraro2023rapid}.

\noindent \rev{ \textit{Bistatic, asynchronous \ac{jcs}.} }
Some methods are available to provide accurate synchronization between communication network nodes~\cite{zhang2022integration}, but require the use of \ac{gps} signals or the cooperation of devices. In systems including a single transmitter-receiver pair or non-cooperating nodes, two main methods have been proposed: \ac{cacc}~\cite{ni2021uplink} and
\ac{casr}~\cite{zeng2019farsense, zeng2020multisense, li2022csi}. Both use the signal collected at one of the receiver array's antennas as a reference to remove the \ac{to} and the \ac{cfo}. These approaches have three main drawbacks that are solved by JUMP. First, they require a \ac{mimo} antenna array at the receiver, which is not available in commercial \ac{mmwave} systems, as \textit{phased-arrays} are preferred for their lower implementation cost and complexity. Second, they entail a higher complexity in the estimation of the sensing parameters, by doubling the number of parameters or introducing non-linearity~\cite{li2022csi,ni2021uplink}. Third, for their correct operation, they require strong assumptions about the multipath environment, such as the continuous presence of a dominant \ac{los} link between the transmitter and the receiver, or the presence of a single sensing target~\cite{li2022csi} in the monitored space.
Contrarily, JUMP has no requirements for the receiver array and handles environments with multiple moving targets and \ac{los} occlusions. \rev{Other techniques have been proposed that rely on partially overlapping subbands to remove \ac{to} and \ac{cfo}, e.g.,  \cite{xie2019precise, zhu2018pi}. These require the availability of such subbands, which is rarely the case, and they can not work with \ac{sc} systems. 

Recently, a Kalman filter-based technique for removing the \ac{to} in bistatic asynchronous \ac{jcs} has been proposed~\cite{chen2023kalman}. Differently from JUMP, however, \cite{chen2023kalman} estimates the \textit{joint Doppler plus \ac{cfo}} frequency shift, which does not contain useful information about the target's movement. JUMP can instead remove the \ac{cfo} and retain the Doppler shifts induced by the multiple moving parts of the target, enabling fine-grained sensing applications.}



\noindent \textit{Bistatic sub-6~GHz Wi-Fi sensing.}
Several studies have proposed ways to perform activity recognition~\cite{meneghello2022sharp,li2019wi, chen2018wifi} using Wi-Fi \ac{ofdm} \ac{csi}. Indeed, most of the \ac{cfo} removal techniques described in the previous section have been originally proposed for Wi-Fi.
The main drawback of sub-$7$~GHz sensing lies in its inherent low-ranging resolution, which is a direct consequence of the relatively low transmission bandwidth ($40$-$80$~MHz are typical values). 
There, reflected paths can only be resolved with an accuracy of a few meters, which causes major performance degradation with multiple concurrently moving subjects~\cite{korany2020multiple}. 

\noindent \textit{\ac{mmwave} bistatic \ac{jcs}.} 
A few works have addressed bistatic \ac{jcs} in \ac{mmwave} systems, assuming synchronized transmitter and receiver 
\cite{pucci2022performance,kanhere2021target,gao2022integrated}. Others have focused on tracking targets based on distance and \ac{aoa}, without addressing sensing tasks that require phase analysis or aggregation of signal samples over coherent processing intervals, such as Doppler speed estimation~\cite{garcia2020polar}. 
JUMP instead performs \ac{md} estimation in addition to tracking, and removes the unrealistic assumption of having time-synchronized nodes.

\noindent \textit{Radar systems.} Bistatic radar systems have been widely studied and adopted~\cite{malanowski2012two, kuschel2019tutorial, richards2010principles}. In the so-called \textit{active} bistatic radar setup, the \ac{to} and \ac{cfo} are eliminated by connecting the transmitter and the receiver through high-speed fiber links, achieving phase locking, or exploiting \ac{gps} as the common clock source. In the \textit{passive} radar case, the receiver exploits signals of opportunity from non-cooperative transmitters. Two receiving channels are used, one receiving the direct \ac{los} signal and the other pointing towards the target of interest. Then, offsets are removed by cross-correlating the \ac{los} signal with the reflected one~\cite{samczynski20215g}. Besides requiring complex implementation and being heavily dependent on the type of transmitted signal, these systems cannot handle occlusions of the \ac{los} link~\cite{kuschel2019tutorial}.

\section{Bistatic system model}\label{sec:model}

In this section, we introduce the bistatic system model.

\subsection{CIR for asynchronous transceivers}\label{sec:cir}

The \ac{cir} estimation process consists of transmitting a pilot signal from a transmitter~(TX) device that, after being reflected by nearby objects or humans, is collected at the receiver~(RX). The latter correlates the known pilot waveform with the received one. The resulting \ac{cir} contains \ac{to} and \ac{cfo} that prevent accurate estimation of the targets' parameters~\cite{zhang2022integration,ni2021uplink}.
In our model, we highlight \textit{(i)}~the role of \ac{to} and \ac{cfo}, and \textit{(ii)}~the impact of discretizing the \ac{cir} for wideband systems. We use a \ac{sc} \ac{cir} model to simplify the understanding of the implementation and experimental results, which are based on \ac{sc} IEEE~802.11ay. However, JUMP is equally applicable to \ac{ofdm}-based JCS systems. \rev{The model is based on, e.g.,~\cite{pegoraro2023rapid,kumari2017ieee}, but contains significant modifications to take into account \ac{to}, \ac{cfo}, and the bistatic geometry.} 
\subsubsection{Timing offset}\label{sec:del-offset}

In wireless communications the TX and the RX clocks are not synchronized and thus exhibit an unstable relative clock drift over time~\cite{ zhang2022integration,zhang2021overview}. In the absence of an absolute timing reference, the RX performs packet detection and \textit{coarse} synchronization. For example, in IEEE~802.11ad/ay packet detection is done with an autocorrelation method which returns a rough estimation of the start of the packet~\cite{liu2014all}. This operation does not yield a synchronization at the symbol-level, thus a residual \ac{to} is included into the channel estimate and removed via equalization. Note that these operations are performed separately on each packet, and the \ac{to} changes across subsequent packets depending on the selected synchronization point. In \ac{jcs} we are interested in computing the exact delays of the signal propagation paths, without the residual synchronization error, as these are linked to the distance of the sensing targets. Hence, including the \ac{to} in the channel estimates results in a wrong estimation of the delays. Due to the time-varying nature of the \ac{to}, the cumulative estimation error across time can add up to tens of nanoseconds after just a few milliseconds~\cite{zhang2022integration}. This uncertainty in the delay easily leads to large errors in estimating the target's position (several meters), even within short time intervals~\cite{zhang2021overview}.

\subsubsection{Frequency offset}\label{sec:freq-offset}


Clock drifts can be typically ignored over the course of the preamble of a single communication packet, as their variation is negligible over short time periods (micro-second level). Therefore, in communication systems such as the IEEE~802.11ad/ay the \ac{cfo} is estimated and compensated for each packet separately. This is done by computing the total \textit{phase} error induced by the \ac{cfo} using an autocorrelation technique~\cite{liu2014all} and removing it from the received samples. 
In \ac{jcs} instead, we are interested in retrieving the Doppler shift caused by the target's movement speed.
Therefore, compensating for the \ac{cfo} for each packet separately is not a viable option, as this method removes the \ac{cfo} \textit{and} the Doppler shift, as they both appear in the phase of the received signal and they can not be easily separated. This is further discussed in \secref{sec:md-extr}. In addition, estimating the phase differences and Doppler shifts caused by moving targets requires processing the received samples \textit{coherently across subsequent packets}. The \ac{cfo} can not be considered constant within such extended time intervals (several milliseconds), and it makes the phase of signal samples across different packets \textit{incoherent}, preventing the estimation of Doppler shifts. 

\ac{cfo} is especially severe in \ac{mmwave} systems. As an example, it can reach hundreds of kHz with 60~GHz devices, while for a target moving at 10~m/s the Doppler shift is 4~kHz. This is shown in \fig{fig:md-off}, where we plot the \ac{cfo}-affected normalized Doppler spectrum and that obtained by JUMP from a walking person. Without removing the \ac{cfo}, the \ac{md} is completely corrupted and does not carry any useful information about the movement.

\rev{\subsubsection{Transmitted signal model}\label{sec:txsig-model}
We denote by $s[i]$ the discrete TX symbol sequence, by $L$ the number of TX symbols, and by $p_{\rm tx}(t)$ the pulse-shaping filter.
Assuming unit energy per TX symbol and a sampling period equal to \mbox{$\Delta \tau = 1/B$}, with $B$ the TX bandwidth, the \ac{sc} TX signal is
\begin{equation}\label{eq:tx-signal-model}
x(t) = \sum_{i=0}^{L-1} s[i] p_{\rm tx}\left(t - i\Delta\tau\right).
\end{equation}
In a directional \ac{jcs} system employing phased antenna arrays, different \acp{bp} are used to steer the signal energy towards the intended direction. 
The TX signal is thus multiplied by the analog beamforming vector, $\mathbf{u}$, obtaining $\mathbf{x}(t) = \mathbf{u} x(t)$.
In our simulations and implementation, $s[i]$ is a set of complementary Golay sequences used for channel estimation in IEEE~802.11ay, as detailed in \secref{sec:implementation}. However, here we consider it to be a single sequence of pilot symbols for the sake of generality.}

\subsubsection{CIR model} \label{sec:cir-model}
\begin{figure}[t!]
	\begin{center}   
		\centering
		\subcaptionbox{\ac{md} with \ac{cfo}.\label{fig:md-off}}[2.3cm]{\includegraphics[width=2.3cm]{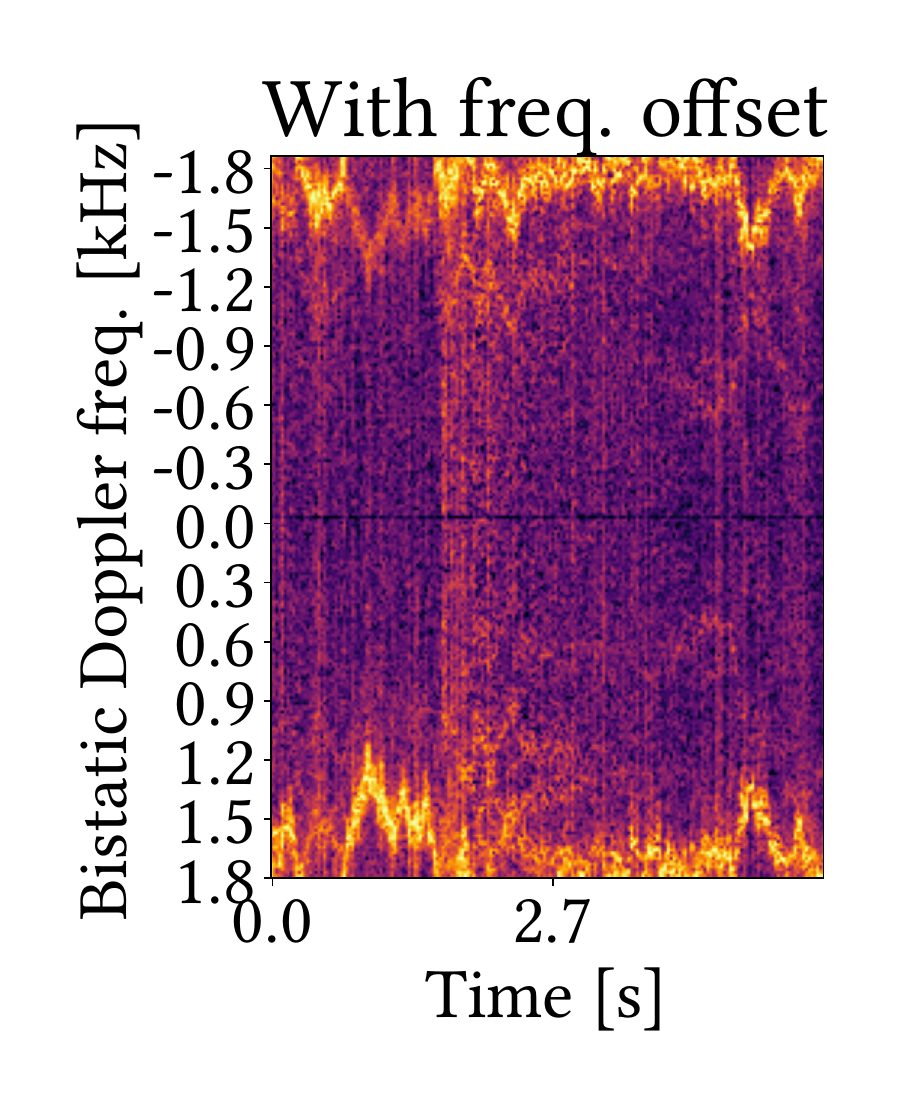}}
        \subcaptionbox{JUMP \ac{md}.\label{fig:md-jump}}[2.cm]{\includegraphics[width=2.cm]{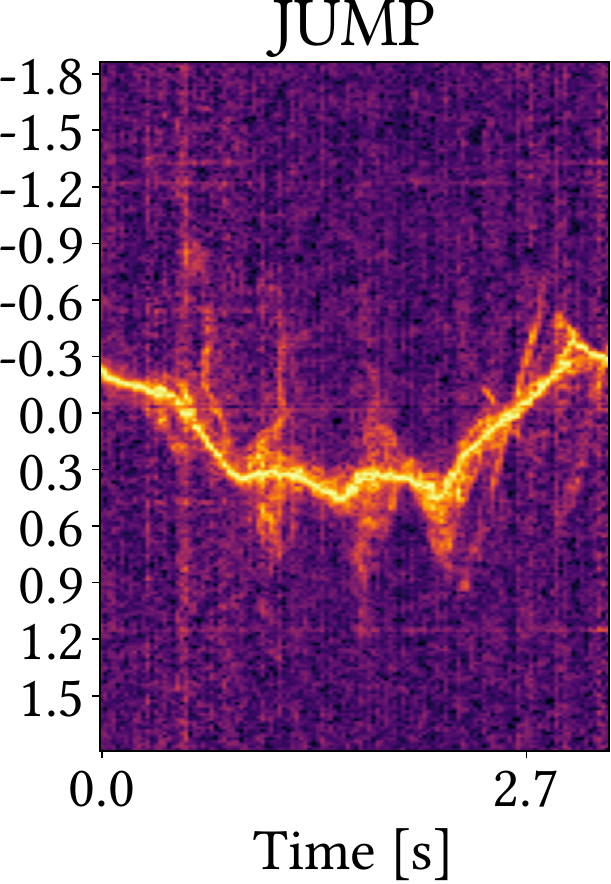}}
        \subcaptionbox{CIR magnitude.\label{fig:cir-mag}}[3.6cm]{\includegraphics[width=3.6cm]{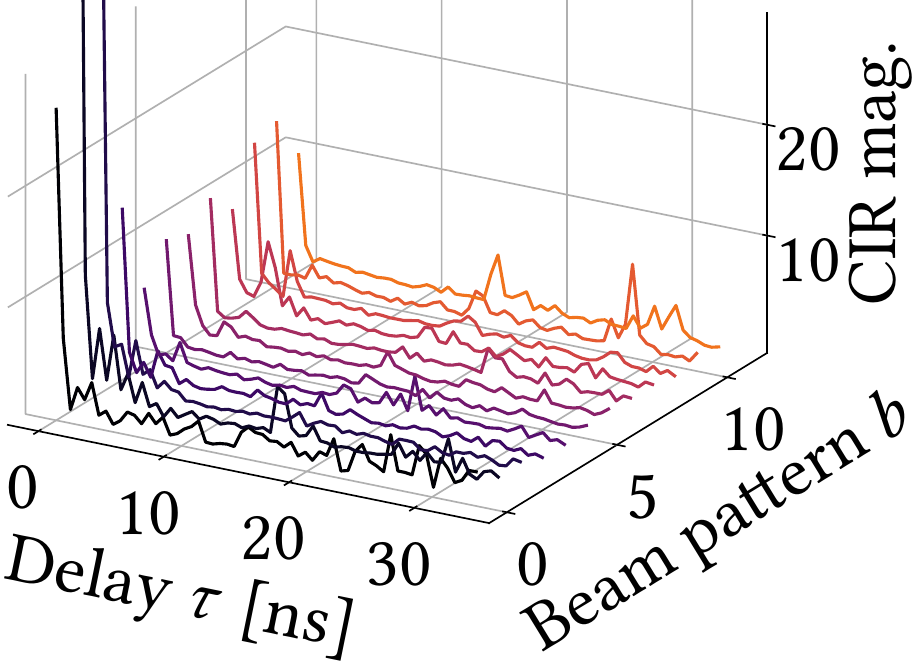}}
		\caption{\ac{md} spectrogram affected by the \ac{cfo} (a) compared to JUMP's \ac{md} (b). (c) shows the CIR magnitude for different BPs.}
		\label{fig:cir-offset}
	\end{center}
\end{figure}

\rev{
We denote by $N(t)$ the number of propagation paths at time $t$, by $\alpha_{m}(t)$ the complex signal attenuation coefficient of the $m$-th reflection, while  and $\tau_m (t)$ is its delay. $f_{{\rm D}, m}(t)$ is the Doppler frequency due to the movement of the $m$-th reflector. Moreover, we call: $\mathbf{v}$ the vector of phase shifts used at the RX phased array (RX beamforming vector), $\theta_m, \varphi_m$ the \ac{aod} and \ac{aoa} of the $m$-th reflection, respectively, and \mbox{$\mathbf{a}_{\rm tx}(\theta_m), \mathbf{a}_{\rm rx}(\varphi_m)$} the corresponding TX and RX array steering vectors. Considering uniform linear arrays at the TX and RX we have $\mathbf{a}_{\rm tx}(\theta) = [1, \dots, e^{-j (M_{\rm tx}-1)\pi \sin \theta}]^{\mathsf{T}}$, $\mathbf{a}_{\rm rx}(\theta) = [1, \dots, e^{-j (M_{\rm rx}-1)\pi\sin \theta }]^{\mathsf{T}}$, where $M_{\rm tx}$ and $M_{\rm rx}$ are the number of antennas at the TX and RX, respectively. 
Given these definitions, and calling $\delta(\cdot)$ the Dirac delta function, the continuous-time \ac{cir} model is 
\begin{equation}\label{eq:cir-continuous}
    h(t, \tau) = \sum_{m=1}^{N(t)} A_{m}(t) e^{j2\pi f_{{\rm D}, m}(t)t} \delta(\tau - \tau_m (t)),
\end{equation}
with \mbox{$A_{m}(t) = \alpha_m(t) \mathbf{v}^{\mathsf{H}}\mathbf{a}_{\rm rx}(\varphi_m) \left(\mathbf{a}_{\rm tx}(\theta_m)\right)^{\mathsf{H}} \mathbf{u}$},
where superscript $(\cdot)^{\mathsf{H}}$ indicates the complex conjugate transpose. Note that we include in the \ac{cir} model also the effect of TX and RX beamforming, that are combined in the complex-valued coefficients $A_m(t)$. $\alpha_m(t)$ instead accounts for the combined effect of the propagation loss and the target's \ac{rcs}~\cite{richards2010principles}.}

\rev{\subsubsection{Received signal model}\label{sec:rxsig-model}
The RX signal is the result of a convolution between the TX signal and the \ac{cir}. In addition, it is also affected by \ac{to} and \ac{cfo}, which we denote by $\tau_{\rm o}(t)$ and $f_{\rm o}(t)$, respectively. The RX applies a receive filter, $p_{\rm rx}(t)$, and we define \mbox{$p(t) = p_{\rm tx}(t) * p_{\rm rx}(t)$} to be the convolution between the TX and RX filters. 
The noise-free RX signal is
\begin{equation}\label{eq:rx-signal}
    y(t) = \sum_{m=1}^{N(t)}  A_{m}(t)  e^{j2\pi (f_{{\rm D}, m}(t) + f_{\rm o}(t))t} x'(t-\tau_{m}(t) - \tau_{\rm o}(t)),
\end{equation}
with \mbox{$x'(t)=\sum_{i=0}^{L-1}s[i]p(t - i/B )$}.
Due to the presence of \ac{to}, the TX signal is further delayed with respect to the sole propagation time. The \ac{cfo} instead causes an additional frequency shift that is added to the Doppler effect. We remark that \ac{to} and \ac{cfo} are \textit{constant} across the signal paths, and hence they are independent of index $m$.

The RX signal is sampled with period $\Delta \tau$, obtaining a discrete sequence of samples for each packet. We denote by $T$ the time between two subsequent packets. In deriving the expression of the sampled RX signal we make the following common  assumptions: (i)~the combined filter $p(t)$ satisfies the Nyquist condition, i.e., \mbox{$p(i\Delta \tau) = \delta[i]$} (Kronecker delta function), and (ii)~the number of propagation paths and their parameters (amplitudes, Doppler shift, and delay) can be considered constant within a short processing interval of $K$ packets, provided that $T$ is sufficiently small that the propagation environment changes slowly compared to it~\cite{zhang2021overview}. We stress that the \ac{to} and \ac{cfo} instead vary in each packet, hence they retain their dependence on time.
The delay resolution of the system after discretization is determined by the sampling period $\Delta \tau$, thus being inversely proportional to the TX bandwidth. Due to the finite resolution, reflections with delay difference below $\Delta \tau$ are not \textit{resolvable} at the RX. Hence, out of the $N$ true signal propagation paths, only \mbox{$N_{\rm r} \leq N$} resolvable paths appear in the RX signal. We index such paths by variable $n$ to distinguish them from the true propagation paths.
Due to sampling, the $n$-th path delay and the \ac{to} are approximated as multiples of the sampling period, \mbox{$\tau_{n} = \rho_n  \Delta\tau$} and \mbox{$\tau_{\rm o}(kT) = \rho_{\rm o} (kT) \Delta \tau$}.

Although they cannot be distinguished in the delay domain, unresolvable paths could be discerned based on the Doppler shift. The $n$-th resolvable path is the superposition of $N_n$ non-resolvable paths, with \mbox{$N = \sum_{n=1}^{N_{\rm r}}N_n$}.
The $i$-th RX symbol in packet $k\in [0, \dots, K-1]$ is
\begin{eqnarray}\label{eq:discr-rx-signal}
        y[k,i] &= &\sum_{n=1}^{N}\tilde{h}_{n}(kT)   s[i -\rho_n -\rho_{\rm o} (kT)], \\
    \tilde{h}_{n}(kT)&=&\sum_{\nu=1}^{N_n}A_{n, \nu} e^{j2\pi (f_{{\rm D}, n, \nu} + f_{\rm o}(kT))kT},
\end{eqnarray}
where $A_{n, \nu}$ and $f_{{\rm D}, n, \nu}$ are the complex gain and the Doppler frequency of the $\nu$-th superimposed path within resolvable path $n$, respectively.
In \eq{eq:discr-rx-signal} we neglect fractional components of the $n$-th path delay and the \ac{to}, since these cause sensing errors which are below the resolution of the system. 
\eq{eq:discr-rx-signal} represents the \textit{noiseless} RX symbols. In our model, we consider $y[k, i]$ to be affected by additive complex white Gaussian noise distributed as $\mathcal{CN}(0, \sigma_w^2)$, where $\sigma_w^2$ is the noise variance. Recalling we assumed unit TX symbol energy, the \ac{snr} is denoted by $\Gamma = |\tilde{h}_1(kT)|^2 / \sigma_w^2$ where $\tilde{h}_1(kT)$ is the gain of the \ac{los} propagation path.}

\rev{\subsubsection{Estimated CIR model}\label{sec:est-cir}
\ac{cir} estimation is carried out by cross-correlating the received samples with the known transmitted pilot sequence~$s[i]$. This yields
\begin{equation}\label{eq:cir-discrete}
        h[k,\ell] = \sum_{n=1}^{N_{\rm r}} \tilde{h}_{n}(kT) \psi_n[\ell -\rho_n -\rho_{\rm o} (kT)],
\end{equation}
for $k\in [0, \dots, K-1]$, where $\psi_n[\ell]$ is the cross-correlation of $s[i]$ with its delayed and frequency-shifted version (due to Doppler and \ac{cfo}) for resolvable path $n$~\cite{richards2010principles, kumari2017ieee}.
In our implementation, we use complementary Golay sequences which exhibit perfect autocorrelation property when the frequency shift is zero~\cite{802.11ay}. This condition does not hold in \eq{eq:cir-discrete}, but such non-ideality can be neglected if the product of the Doppler plus \ac{cfo} frequency and $L\Delta \tau$ (the duration of a pilot sequence) is much smaller than $1$, as shown in~\cite{kumari2017ieee}. As an example, with a \ac{cfo} of $100$~kHz, a Doppler shift of $4$~kHz, $\Delta \tau = 0.568$~ns, and $L=128$ we get $0.008 \ll 1$.
It follows that the estimated \ac{cir} at a specific time is represented as a vector of complex channel gains, or \textit{taps}, indexed by $\ell=0, \dots, L-1$. The $\ell$-th tap is related to a corresponding distance, $d_{\ell}=c\ell\Delta \tau$, with $c$ being the speed of light.}

\rev{Note that, as a result of \eq{eq:cir-discrete}, the \ac{snr} on the peaks of the estimated \ac{cir} is increased with respect to $\Gamma$ as a result of the coherent integration of the signal. We denote the coherent integration gain by $G$. The exact gain depends on the frequency shift of the received signal (due to Doppler and \ac{cfo}), and is upper bounded by $G \leq L$, which is the coherent gain in case of no frequency shift~\cite{richards2010principles}.
}

\rev{
In the following, the TX is considered to be directional, whereas the RX is assumed to use a quasi-omnidirectional \ac{bp}. Note that this is coherent with how existing \ac{mmwave} commercial devices equipped with phased-arrays operate. This is not restrictive, as our method also works in the symmetric case where the RX listens to the signal using directional \acp{bp}, and the TX is omnidirectional.}

\rev{
To account for the effect of analog TX beamforming, we denote by $h_b[k, \ell]$ the $\ell$-th tap of the \ac{cir} at time $kT$ obtained using \ac{bp} $b$, i.e., when the TX uses beamforming vector~$\mathbf{u}_b$. The complex gain of a path is denoted by $A_{\nu, n, b}$, which contains the antenna gain given by \ac{bp} $b$ along the direction that points to reflector $n$, together with the propagation loss and the \ac{rcs}~\cite{richards2010principles}. A visual representation of the magnitude of the \ac{cir} for different \acp{bp} is shown in \fig{fig:cir-mag}.}

\subsection{Bistatic sensing configurations}\label{sec:bist-sens}
Next, we detail how the bistatic geometry affects the multipath reflection delays and Doppler frequencies. 

\subsubsection{Bistatic reflection delay} 
\begin{figure}
     \centering
     \includegraphics[width=8cm]{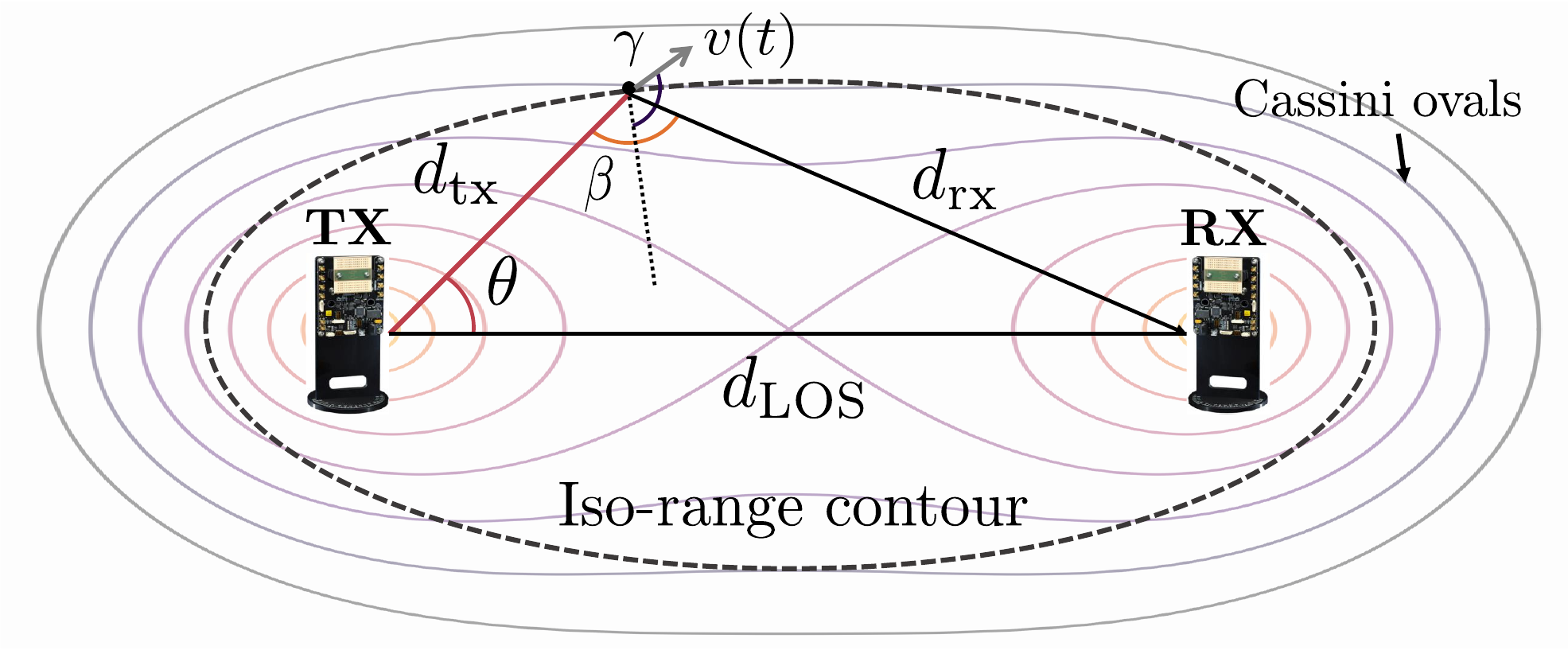}
     \caption{Schematic representation of the bistatic geometry.}
     \label{fig:bistatic-geom}
\end{figure}

We denote by $\tau$ the \textit{relative} delay of a generic multipath reflection, measured relative to the \ac{los} propagation path along which the first signal copy arrives at the RX. Also, we denote by $d_{\rm LOS}$ the \ac{los} distance between the TX and the RX. $d_{\rm tx}$ and $d_{\rm rx}$ are the distances from the TX to the target and from the target to the RX, respectively, as shown in \fig{fig:bistatic-geom}. The following relation holds $c\tau = d_{\rm tx} + d_{\rm rx} - d_{\rm LOS}$, which states that the relative delay is the time needed for the signal to propagate along the excess length of the reflected path ($d_{\rm tx} + d_{\rm rx}$) with respect to the \ac{los} path ($d_{\rm LOS}$). 
Note that all reflectors located on the ellipse with focii coinciding with the TX and the RX will yield the same measured relative delay due to having the same $d_{\rm tx} + d_{\rm rx}$. This elliptical region is usually termed \textit{iso-range} contour.
In \fig{fig:bistatic-geom}, we also represented the \ac{aod} of the reflection, called $\theta$, as the angle formed by $d_{\rm tx}$ and $d_{\rm LOS}$.
\rev{Exploiting the geometric relations shown in \fig{fig:bistatic-geom}, we can relate $d_{\rm LOS}$, $d_{\rm tx}$, and $d_{\rm rx}$ through the law of cosines
\begin{equation}\label{eq:law-cosines}
    d_{\rm rx}^2 = d_{\rm tx}^2 + d_{\rm LOS}^2 - 2d_{\rm tx} d_{\rm LOS}\cos \theta.
\end{equation}
The distance of the reflector with respect to the TX, for a given delay $\tau$, is obtained as
\begin{equation}\label{eq:drx-eq}
   d_{\rm tx}(\tau) = \frac{(c\tau + d_{\rm LOS})^2 - d_{\rm LOS}^2}{2(c\tau + d_{\rm LOS} - d_{\rm LOS} \cos \theta)}.
\end{equation}
}
The capability of a bistatic system to distinguish two reflectors located at different distances (hence causing different reflection delays) is termed \textit{range resolution}. As described next, this value depends on the \ac{cir} delay resolution $\Delta\tau$.
We define $\beta$ as the \textit{bistatic angle}, i.e., the angle formed by the segments connecting the TX to the target and the target to the RX. Due to the elliptical shape of the iso-range contours, the range resolution depends on the bistatic angle as $\Delta d \approx c \Delta\tau / (2\cos(\beta/2))$~\cite{kuschel2019tutorial}. This is a crucial aspect of bistatic systems. In practice, it prevents accurate localization of the reflectors when they approach the \ac{los} segment, as $\Delta d \rightarrow +\infty$ when $\beta \rightarrow \pi$. The elliptical area around the \ac{los} segment in which the resolution degrades is typically termed \textit{forward scattering} region. 


\subsubsection{Bistatic SNR} In bistatic systems, \rev{for scattering-type reflections}, the expected \ac{snr} at the RX is inversely proportional to the square of the product $d_{\rm tx}d_{\rm rx}$~\cite{kuschel2019tutorial}. For this reason, the locii of points along which the \ac{snr} is constant follow the so-called \textit{Cassini ovals}, as shown in \fig{fig:bistatic-geom}.
As such, compared to a monostatic system, reflectors far from the TX device can still yield high \ac{snr} if they are sufficiently close to the RX.

\subsubsection{Bistatic \ac{md} spectrum}\label{sec:bist-doppl-spec}

\rev{Consider the bistatic \ac{md} spectrum of the $n$-th resolvable path of the \ac{cir}. According to \eq{eq:cir-discrete}, it contains the contribution of $N_{n}$ reflectors, each having a possibly different speed, together with the \ac{cfo}. 
Moreover, the reflectors' movement speed maps to a different Doppler shift depending on the motion's direction and the bistatic angle. 
Denoting by $\gamma_{n, \nu}$ the angle between the bisector line of the bistatic angle $\beta_{n, \nu}$ and the velocity vector of the $\nu$-th reflector superimposed resolvable path $n$, with magnitude $v_{n, \nu}$, we have~\cite{kuschel2019tutorial}
\begin{equation}\label{eq:md-bistatic}
   f_{{\rm D}, n, \nu} = \frac{2 v_{n, \nu}}{\lambda} \cos \gamma_{n, \nu} \cos \frac{\beta_{n,\nu}}{2},
\end{equation}
where $\lambda = c/f_c$ is the wavelength of the carrier $f_c$ of the TX signal. As a consequence, in a bistatic scenario, the Doppler shift depends on the bistatic angle in addition to the direction of motion.
The standard method for extracting the \ac{md} spectrum applies a \ac{stft} to windows of subsequent \ac{cir} estimates~\cite{pegoraro2023rapid}. 
Taking a processing window of $K$ packets, the spectrum will present peaks at frequencies $f_{{\rm D}, n, \nu}$. However, in our asynchronous \ac{jcs} setting, this approach is not directly applicable as the Doppler frequency is added to the time-varying \ac{cfo}.}
Therefore, we cannot directly estimate the Doppler shift without removing the \ac{cfo} first. 

\section{Methodology} \label{sec:method}

\begin{figure}
     \centering
     \includegraphics[width=7cm]{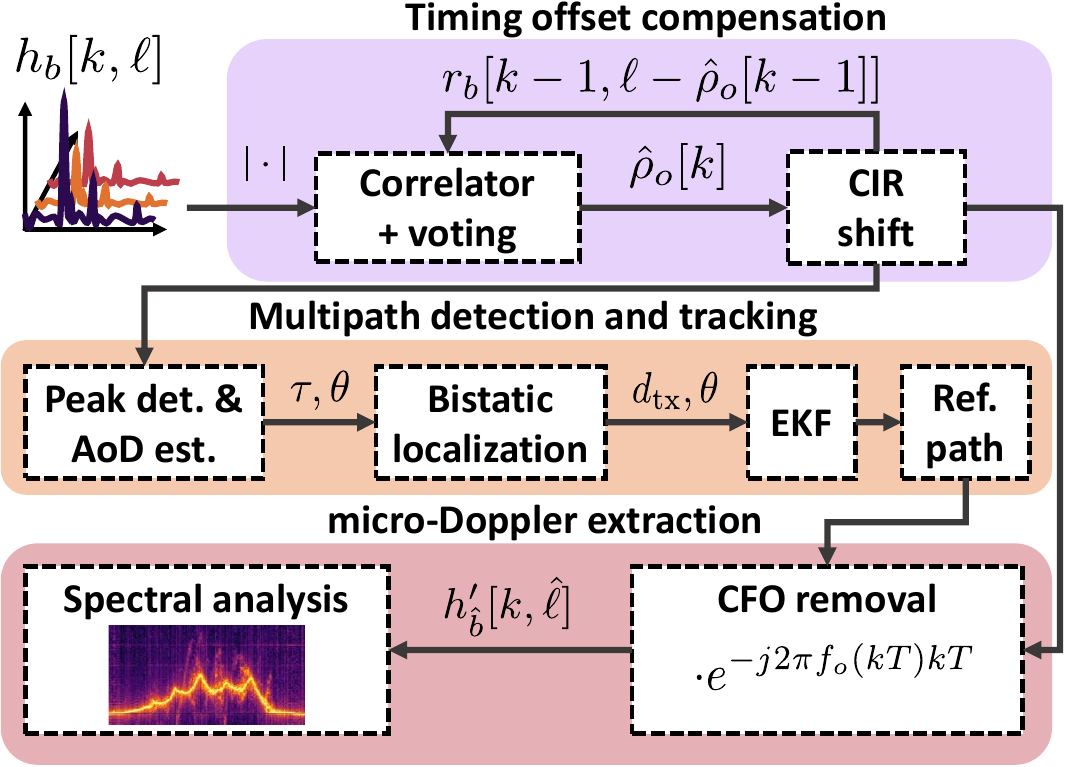}
     \caption{Block diagram of the proposed method.}
     \label{fig:block-diag}
\end{figure}
This section presents JUMP, whose block diagram is shown in \fig{fig:block-diag}. It includes three main steps, as described next.

\textit{$1.$ Timing offset compensation:} After obtaining the \ac{cir} in the current timestep, we leverage the previous \ac{cir} to estimate the relative random \ac{to} between the two, as detailed in \secref{sec:timing-offest-comp}. The random \ac{to} is estimated by computing the correlation between the \ac{cir} magnitude profiles with different \acp{bp}, resulting in the candidate \acp{to}, and then combining them through a majority voting scheme (\secref{sec:del-off-est}). Subsequently, the current \ac{cir} is shifted in the delay domain, compensating for the random offset (\secref{sec:cir-shift}). 

\textit{$2.$ Multipath detection and tracking:} Once the \ac{to} is compensated for, we perform detection and tracking of the multipath reflections in the environment. This is achieved by extracting the peaks in the \ac{cir} magnitude, computing their \ac{aod}, and smoothing the measurements with an \ac{ekf} tracking algorithm. \secref{sec:multipath-det-track} provides a detailed description of this step. 
Finally, we select the strongest \textit{static} reflection from the set of tracked multipath components by measuring the variance of its 2D location estimated by the \ac{ekf} (\secref{sec:static-selection}). We use this path to remove the \ac{cfo} from the other reflections. 

\textit{$3.$ $\mu$D extraction:} To extract the \ac{md} signature of the sensing targets, we compensate for the \ac{cfo} by multiplying the \ac{cir} by a complex exponential with the phase of the previously selected static path (see \secref{sec:md-extr}). The key insight behind this operation is that the \ac{cfo} is the same on all the multipath reflections. 
Subsequently, we apply \ac{stft} to the \ac{cir} to extract the \ac{md}.

\subsection{Timing offset compensation}\label{sec:timing-offest-comp}

Here, we propose a method to compensate for the \ac{to} between two subsequent estimates of the \ac{cir}. 
We (reasonably) assume that, besides the reflections on the sensing targets, the background contains multipath reflections on \textit{static} objects, which are \textit{slowly} time-varying compared to the packet transmission rate. In addition, we consider that the packet transmission rate is high with respect to the movement speed of the sensing targets, which is a common assumption for radar and \ac{jcs}. 
As a consequence, occlusions to the \ac{los} lead to a progressive and ``slow'' disappearance of the corresponding path from the \ac{cir}, which happens over the course of several \ac{cir} estimates.

Consider two subsequent \ac{cir} estimates obtained from packets $k$ and $k-1$ using the same \ac{bp} $b$, and denote the magnitude of the \ac{cir} by $r_b[k, \ell] = |h_b[k, \ell]|$. If the two estimates are obtained at a sufficiently high sampling rate $T$, the propagation paths due to reflections on static objects do not change significantly, producing similar \ac{cir} magnitude profiles (envelopes).
\rev{\ac{to} is expressed as \mbox{$ \tau_{\rm o}(kT)= \rho_{\rm o}(kT) \Delta\tau$} with \mbox{$\rho_{\rm o}[k] \in \{0, 1, \dots, L-1\}$}, as in \secref{sec:rxsig-model}, defining $\rho_{\rm o}[k] \triangleq \rho_{\rm o}(kT)$.} Therefore, our assumption about the slowly varying multipath environment can be written as
$r_b[k, \ell + \rho_{\rm o}[k]]\approx r_b[k-1, \ell]$. This equality means that the \ac{cir} envelope is preserved between subsequent time steps. It is just shifted by $\rho_{\rm o}[k]$ in the path delay dimension, and this shift $\rho_{\rm o}[k]$ is constant across all \acp{bp} but varies on a per packet basis (with index $k$).

In the following, we propose our method to estimate $\rho_{\rm o}[k]$ and then use it to compensate for the \ac{to}, and shift the \ac{cir} back to a standard timing reference.

\subsubsection{Estimation of the timing offset}\label{sec:del-off-est}
Due to the similarity of the \ac{cir} envelopes in subsequent packets, a good estimate of the shift $\rho_{\rm o}[k]$ can be obtained by maximizing the cross-correlation between the two \acp{cir} magnitudes as follows
\begin{equation}
\hat{\rho}_b[k]= \argmax_{\rho} \sum_{\ell=0}^{L-1} r_b[k, \ell + \rho] r_b[k-1, \ell].
\label{eq:rho_estimate}
\end{equation}
In \eq{eq:rho_estimate}, the estimate $\hat{\rho}_b[k]$ depends on the \ac{bp} used in the estimation of the \ac{cir}, although $\rho_{\rm o}[k]$ is the same across all \ac{bp}.
As such, a more robust estimate considers the $P$ \ac{cir} estimates obtained for each packet, one per \ac{bp}. We turn $\hat{\rho}_b[k]$ onto a one-hot vector representation
$\hat{\mathbf{\Lambda}}_b[k] \in \{0,1\}^L$, which equals $1$ at index $\hat{\rho}_b[k]$ and $0$ elsewhere. Then, we use majority voting over the \acp{bp} $\hat{\rho}_b[k], b=1,\dots,P$, i.e., 
\begin{equation}\label{eq:maj-vote}
\hat{\rho}_{\rm o}[k] = \argmax_{i} \sum_{b=1}^P \hat{\Lambda}_{b, i}[k],
\end{equation}
\rev{where $\hat{\Lambda}_{b, i}[k]$ is the $i$-th component of $\hat{\mathbf{\Lambda}}_b[k]$.}
\eq{eq:maj-vote} combines the estimated offsets from all the directions illuminated by the TX \acp{bp}. This procedure improves the robustness of the final decision, ignoring \textit{outlying} $\hat{\rho}_b[k]$ that can be produced by \acp{bp} pointing in directions without static reflectors.

When transitioning from a \ac{los} situation to a \ac{nlos} one, due to occlusion from a moving obstacle, $\rho_{\rm o}[k]$ can be large. This is caused by the fact that the first packet copy to be detected at the RX will not travel along the (blocked) \ac{los}. Still, the remaining static multipath reflections in the \ac{cir} are expected to remain constant, especially when considering a combination of all the \acp{bp} like in \eq{eq:maj-vote}. Therefore, our method reliably estimates the relative \ac{to} even in such cases, as proven by our \ac{nlos} results in \secref{sec:analysis} and in \secref{sec:results}.

\subsubsection{\ac{cir} shift}\label{sec:cir-shift}

Upon obtaining an estimate of the \ac{to}, its effect is compensated for by shifting the \ac{cir} in the delay domain by $\hat{\tau}_{\rm o}(kT)= \hat{\rho}_{\rm o}[k] \Delta\tau$. In practice, this is implemented as a shift of the \ac{cir} along indices $\ell = 0, \dots, L-1$ by $\hat{\rho}_{\rm o}[k]$ positions. The corrected \ac{cir} is obtained as
\begin{equation}\label{eq:circshift}
    h^{\rm c}_b[k, \ell] = h_b[k, \ell - \hat{\rho}_{\rm o}[k]], \quad \forall \ell = 0, 1, \dots, L'-1,
\end{equation}
where $L'< L$ is the length of the \ac{cir} portion of interest, which depends on the size of the monitored environment. The \ac{cir} taps having negative index after \eq{eq:circshift} are removed.
The \ac{cir} estimated from the previous packet and $h^{\rm c}_b[k, \ell]$ are aligned, thus allowing the computation of delays relative to the same reference point. 

\subsubsection{Initialization}\label{sec:to-init} Suppose that the sensing operation starts in a \ac{los} condition. The direct path between the TX and RX is the \textit{first} path to be detected at the RX. In this case, we initialize the \ac{cir} alignment process with respect to the \ac{los}. Subsequent \ac{cir} estimates are shifted to compensate for the \ac{to} as previously described, allowing consistent tracking of the locations of the reflectors.
If, instead, the \ac{los} is unavailable when the sensing process starts, the \ac{cir} alignment is initialized with the first reflection that arrives at the RX. This introduces an offset in the obtained delays, which is corrected \textit{a-posteriori} when the first packet traveling along the \ac{los} path is detected. Therefore, having visible \ac{los} during at least one timestep enables consistent tracking of the reflections.

\subsection{Multipath detection and tracking}\label{sec:multipath-det-track}

All the \ac{cir} estimates get the same timing reference using the \ac{to} compensation method from \secref{sec:timing-offest-comp}, which is represented by the first \ac{cir} tap in $h^{\rm c}_b[k, \ell]$. 
As commonly done in \ac{jcs} systems, we assume that the \ac{los} distance, $d_{\rm LOS}$, and the relative orientation between the TX and the RX, $\alpha$, are known a priori. $d_{\rm LOS}$ can be estimated by applying \textit{localization} methods to the RX node, for example,~\cite{shastri2022review}, while $\alpha$ is typically obtained from a beam alignment protocol~\cite{802.11ay}.
Unlike existing approaches, we are interested in tracking both the dynamic \textit{and} the static multipath components in the \ac{cir}, as the latter are used to remove the \ac{cfo}. This makes the detection and tracking much more challenging as we want to preserve the information about the static objects in the environment, so clutter mitigation cannot be applied.

\subsubsection{Delay measurement by peak detection}  \label{sec:del-meas}

The computation of the position of the peaks in the \ac{cir} magnitude after the alignment step of \secref{sec:timing-offest-comp} yields the relative delay of the reflection with respect to the \ac{los}, $\tau$. To reliably detect \ac{cir} peaks in the presence of clutter and background noise, we apply the \ac{cacfar} algorithm to the \ac{cir} magnitude after the \ac{to} compensation, denoted by $r^{\rm c}_b[k, \ell]$~\cite{richards2010principles}.
This operation consists of computing a dynamic threshold using a moving window. 
We call $p$ the index of a generic peak returned by \ac{cacfar} at time $kT$. The delay of the corresponding reflection is then obtained as $\tau = p\Delta \tau$. 

\subsubsection{Angle-of-Departure estimation}\label{sec:aod-est}
Upon receiving the signal reflected from an object in the environment, the RX computes the corresponding \ac{aod} relative to the TX, $\theta$, 
using a modified version of the algorithm proposed in~\cite{Lacruz_MOBISYS2020}. The latter computes the correlation between each \ac{bp} shape and the \ac{cir} profile for a specific channel tap and requires that the TX \ac{bp} shapes be known or estimated in advance.
Unlike~\cite{Lacruz_MOBISYS2020}, however, \rev{in our system model more than one reflection can overlap in the same peak in the \ac{cir} magnitude}, allowing multiple \ac{aod} to be detected for a single peak $p$ in the delay domain, and collecting them into set $\Theta_p$. In the following, the time index $k$ is omitted to simplify the notation. We call $g_b(\theta)$ the gain of \ac{bp} $b$ along direction $\theta$, and $\mathbf{r}[\ell] = [r^{\rm c}_1[\ell], \dots, r^{\rm c}_P[\ell]]^{\mathsf{T}}$ the vector of \ac{cir} magnitudes for a specific $\ell$. 

Our \ac{aod} estimation algorithm computes  
\begin{equation}\label{eq:aoa-corr}
    \Theta_p = \left\{\theta \in [0, \pi]\mbox{ s. t. } \sum_{b =1}^{P} g_b(\theta) \frac{r^{\rm c}_b[p]^2}{||\mathbf{r}[p]||_2^2} > \vartheta\right\},
\end{equation}
where $\vartheta$ is a threshold value used to adjust the sensitivity to multiple peaks in the correlation profile of the \ac{aod}.

\subsubsection{Distance estimation} \label{sec:dist-est}
\rev{
To track the position of the targets with respect to the TX, we compute $d_{\rm tx}(\tau)$ in \eq{eq:drx-eq} for $\tau = p\Delta\tau$ for each peak.
Note that $d_{\rm tx}$ depends on $p$ (already estimated) and $d_{\rm LOS}$ (assumed to be known) and $\theta$, which belongs to set $\Theta_p$. Hence, a set of candidate target locations is obtained for each \ac{cir} peak $p$, by collecting all pairs $(d_{\rm tx}, \theta)$, with \mbox{$\theta \in \Theta_p$}. 
A set of measurement vectors $\mathbf{z} = [d_{\rm tx}, \theta]^{\mathsf{T}}$ collects all such pairs, {\it for all} the identified peaks $p$, and represents the input of an \ac{ekf}~\cite{ribeiro2004kalman} that tracks the Cartesian position of the targets relative to the TX.
}
\subsubsection{Target tracking (EKF)}\label{sec:ekf}
The \ac{ekf} takes the measurement vectors $\mathbf{z}$ as input and, based on their evolution, obtains estimates of the states of the targets. The state is defined as the $x,y$ position of a target and its velocity components along the two axes, $\mathbf{x} = [x, y, \dot{x}, \dot{y}]^{\mathsf{T}}$. A constant-velocity model is adopted to approximate the targets' movement, where we denote by $\mathbf{F}$ the state transition matrix and by $\mathbf{w}, \mathbf{v}$ the state and measurement noises, respectively. The tracking process operates following the \ac{ekf} model equations $\mathbf{x}_{k+1} = \mathbf{F} \mathbf{x}_{k} + \mathbf{w}_k$ and $\mathbf{z}_{k} = g(\mathbf{x}_{k}) + \mathbf{v}_k$,
where $g(\mathbf{x}) = [\sqrt{x^2 + y^2}, \arctan{(y/x)}]^{\mathsf{T}}$ is the non-linear function relating the current state to the current measurements. For mea\-sure\-ments-to-tracks association, we adopt the cheap-nearest-neighbors joint probabilistic data association filter~\cite{shalom2009probabilistic,fitzgerald1986development}, while to handle the initialization and termination of tracks, we use the method described in~\cite{wagner2017radar}. \rev{Such track management strategy filters out most false targets by taking into account the tracks' evolution across time, and allows quick initialization of new tracks whenever new targets enter the monitored area.}

\subsubsection{Static path selection}\label{sec:static-selection}

As described in \secref{sec:ekf}, the state of each tracked multipath component contains an estimate of the current location of the reflector, i.e., the state components $x$ and $y$. 
To identify a reliable static path, we first compute the mean location across the last $K$ tracking steps for all the tracked reflectors. Then, we obtain the variance of the locations in the $K$ frames with respect to this mean value. We select the reflectors with a variance under a certain threshold $\sigma^2_{\rm thr}$ for all the $K$ frames. These are candidate reference paths, as their location is stable around a fixed position.
The \textit{strongest} path among them is the one providing the most reliable \ac{cir} phase value, having higher \ac{snr}. This path serves as a reference in the subsequent \ac{cfo} removal.

\subsection{\ac{md} spectrogram extraction}\label{sec:md-extr}

Here, we detail our \ac{cfo} removal strategy and subsequent \ac{md} spectrum extraction.
To compute the \ac{md} spectrograms of the tracked reflectors, we identify the \ac{cir} taps associated with the corresponding multipath components. This requires mapping the \ac{ekf} states of the tracked components to the corresponding element of the \ac{cir} in the delay and angle domains.
First, an estimate of the distance between the target and TX is $\hat{d}_{\rm tx} = \sqrt{x^2 + y^2}$. Then, denoting by $\alpha$ the angle between the TX and the RX, in the Cartesian reference system of the TX, an estimate of the \ac{aod} is given by $\hat{\theta} = | \arctan\left(y / x\right) - \alpha|$.
Using $\hat{d}_{\rm tx}$ and $\hat{\theta}$, we obtain an estimate of the distance between the target and the RX
\begin{equation}\label{eq:drx-hat}
    \hat{d}_{\rm rx} = \sqrt{\hat{d}_{\rm tx}  + d_{\rm LOS}^2 - 2d_{\rm LOS}\hat{d}_{\rm tx}\cos \hat{\theta}}.
\end{equation}
Once $\hat{d}_{\rm tx}$ and $\hat{d}_{\rm rx}$ are known, we estimate the \ac{cir} tap containing the reflection of the target through the estimated delay associated with a path of length $\hat{d}_{\rm tx} + \hat{d}_{\rm rx}$, as $\hat{\tau} = (\hat{d}_{\rm tx} + \hat{d}_{\rm rx} - d_{\rm LOS})/c$. From this estimate, the \ac{cir} tap corresponding to each target is obtained as the one minimizing the delay difference with $\hat{\tau}$, i.e., \mbox{$\hat{\ell} = \argmin_{\ell} |\hat{\tau}- \ell\Delta \tau |$}. Similarly, we select the \ac{bp} pointing in the direction of the target, $\hat{b}$, as the one having the strongest gain along the direction $\hat{\theta}$.

The key idea behind our \ac{cfo} correction method is that the term $f_{\rm o}(kT)$ is constant in all multipath components of the \ac{cir}. Therefore, we can isolate the \ac{cfo} component from the reference \textit{static} path and remove it from the reflections on the sensing targets. 
The static reference path, which could either be the \ac{los} path or a reflection on a stationary object (e.g., a wall), is identified as detailed in \secref{sec:static-selection}. Its corresponding \ac{cir} tap, $\hat{\ell}_{\rm s}$, is obtained as detailed above.
\rev{
Neglecting the \ac{bp} index, the expression of the $\hat{\ell}_{\rm s}$-th \ac{cir} tap, corresponding to the static path, is
\begin{equation}\label{eq:cir-static}
    h^{\rm c}[k, \hat{\ell}_{\rm s}] = \tilde{h}_{n
_{\rm s}}(kT) = A_{n_{\rm s}}  e^{j2\pi f_{\rm o}(kT)kT},
\end{equation}
where $n_{\rm s}$ is the index of the static path in $1, \dots, N_{\rm r}$. Note that since the path is static, even if it is a result of the superposition of multiple unresolvable reflections, these cannot be distinguished in the Doppler domain since their Doppler shift is equal to zero. 
As a result, \eq{eq:cir-static} shows that the $\hat{\ell}_{\rm s}$-th tap contains a single complex exponential component, whose phase only contains the \ac{cfo}, without additional Doppler frequencies. Therefore, the offset can be removed from any other \ac{cir} tap corresponding to a target, with indices $\hat{\ell},\hat{b}$, by computing
\begin{equation}\label{eq:cir-clean}
  h'_{\hat{b}}[k, \hat{\ell}] = \frac{h^{\rm c}_{\hat{b}}[k, \hat{\ell}]}{ e^{j\angle{h^{\rm c}[k, \hat{\ell}_{\rm s}]}} }
     =\sum_{\nu=1}^{N_{n_{\rm t}}} A_{n_{\rm t}, \nu, \hat{b}}e^{j2\pi f_{{\rm D}, n_{\rm t}, \nu }kT}, 
\end{equation}
where $\angle{ h^{\rm c}[k, \hat{\ell}_{\rm s}]} = \mathrm{arctan}\left(h_Q^{\rm c}[k, \hat{\ell}_{\rm s}] /  h_I^{\rm c}[k, \hat{\ell}_{\rm s}]\right)$ represents the phase of the reference path, letters I and Q denote the in-phase and quadrature components of $h^{\rm c}[k, \hat{\ell}_{\rm s}]$, and $n_{\rm t}$ is the index of the resolvable path corresponding to the target.

The \ac{md} spectrum of $h'_{\hat{b}}[k, \hat{\ell}]$ contains the time-varying contribution of the different reflections superimposing in \ac{cir} tap $\hat{\ell}$, as specified by \eq{eq:md-bistatic}. To compute it, we take the squared magnitude of the \ac{stft} of the cleaned \ac{cir} $h'$, obtaining $S_{\hat{b}}[w,q, \hat{\ell}] = |\mathrm{STFT}\{h'_{\hat{b}}[k, \hat{\ell}]\}|^2$, where $w$ and $q$ are the discrete frequency and \ac{stft} time-frame indices, respectively. 
Finally, for a tracked target, the \ac{md} spectrum is obtained by combining $Q$ \ac{cir} taps preceding and following $\hat{\ell}$, as $\mu\mathrm{D}[w,q]=\sum_{\ell = \hat{\ell}-Q}^{\hat{\ell}+Q}S_{\hat{b}}[w,q, \ell]$. This accounts for the extension of the target, which may exceed the system's ranging resolution.}

We stress that, through \eq{eq:cir-clean}, JUMP effectively removes the \ac{cfo} \textit{without affecting} $f_{{\rm D}, n_{\rm t}, \nu}(kT)$ (the Doppler term). Conversely, standard \ac{cfo} removal methods used in communication systems compensate for the cumulative phase error caused by the \ac{cfo}. This is done by following, e.g., the technique in~\cite{liu2014all}. Considering a single packet, we drop index $k$ and denote by $f_{\rm o}$ the \ac{cfo}, and by $f_{{\rm D}}$ the Doppler shift from a reflector. This causes a phase error of \mbox{$\phi_{\rm o} = 2 \pi (f_{\rm o} + f_{{\rm D}})m\Delta\tau$}, which increases \textit{linearly} with time across subsequent samples in the packet preamble (here indexed by $m$). Then, the phase of the autocorrelation of the received signal at lag $M$ amounts to $M \phi_{\rm o}$, where $M$ is the length of the pilot sequence used in the preamble (e.g., a Golay sequence in IEEE~802.11ay). This means that the cumulative phase error can be estimated by computing the phase of such autocorrelation, and dividing it by $M$. However, the estimated $\phi_{\rm o}$ contains $f_{{\rm D}}$, and it is therefore useless from a \ac{jcs} perspective. For \ac{jcs}, we instead aim at removing $f_{\rm o}$, while retaining $f_{{\rm D}}$.

\subsection{Impact of RX mobility}\label{sec:rx-mob}
\rev{One of the key assumptions we made so far is that TX and RX are static, as only in this case static reference paths can be identified and used to remove the \ac{cfo}. 
In this section, we discuss the impact of movements of the RX on JUMP. We show that if: (i)~the RX is capable of estimating the \ac{aoa} of the received multipath components, and (ii)~an onboard sensor (e.g., an accelerometer on a mobile device) is available that can estimate the RX velocity, then the effect of RX motion can be compensated for. A full analysis of TX and RX motion is however beyond the scope of this paper, and constitutes a primary future research direction.}

\rev{
Consider the scenario in \fig{fig:rx-mob}, where the RX device is moving with velocity $v_{\rm rx}$ and angles: $\eta$, with respect to the extension of the segment connecting TX and RX, and $\xi$, with respect to the extension of the segment connecting the sensing target and the RX. We assume the RX and the target velocities to be constant within a short processing time interval.}

\rev{
As a result of the RX motion, the phase of \ac{cir} along the propagation path caused by the target (\textit{sensing path}) is given by \mbox{$\phi_{\rm t}(kT) = 2\pi \left[f_{{\rm D}, n_{\rm t}, \nu} + f_{{\rm D}, {\rm t}}^{{\rm rx}} + f_{\rm o}(kT) \right]kT$},
where $f_{{\rm D}, {\rm t}}^{{\rm rx}} = \frac{v_{\rm rx}}{\lambda}\cos \xi$ is the Doppler shift caused by the movement of the RX on the sensing path, indexed by $n_{\rm t}$.
The phase of the \ac{cir} along the direct TX-RX path is $  \phi_{{\rm s}}(kT) = 2\pi [ f_{{\rm D}, {\rm s}}^{{\rm rx}}+ f_{\rm o}(kT) ]kT$,
where  $f_{{\rm D}, {\rm s}}^{{\rm rx}} = \frac{v_{\rm rx}}{\lambda}\cos \eta$ is the Doppler frequency caused by the RX along the TX-RX path. The expressions of $\phi_{\rm t}(kT)$ and $\phi_{{\rm s}}(kT)$ highlight that the Doppler shift caused by the RX movement is \textit{different} along the TX-RX path and the target-RX path, due to the different angles $\eta$ and $\xi$. This prevents aggregating the RX movement Doppler to the \ac{cfo} and removing it by using the phase of the static path.  
\begin{figure}[t!]
     \centering
     \includegraphics[width=0.7\columnwidth]{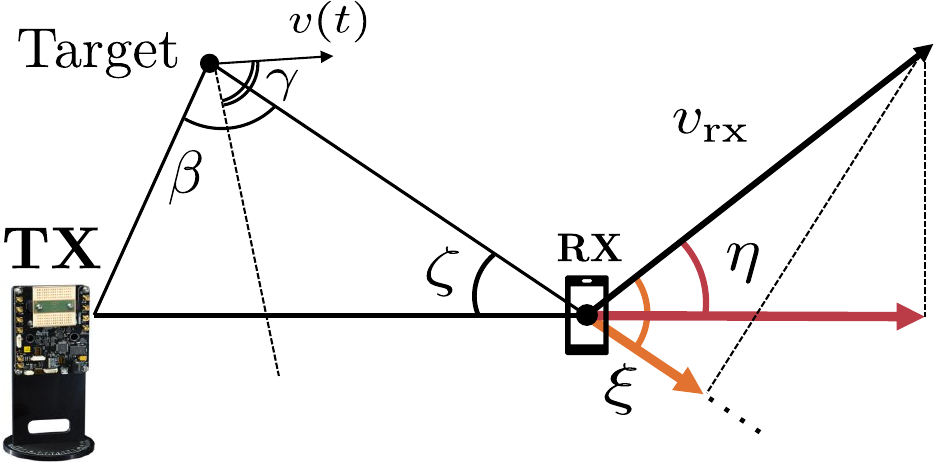}
     \caption{\rev{Schematic representation of the bistatic geometry with a moving RX}.}
     \label{fig:rx-mob}
     \vspace{-3mm}
\end{figure}
Indeed, if one uses \eq{eq:cir-clean} directly, the resulting phase of the sensing path, $\phi'_{\rm t}$,  contains a residual  frequency offset that depends on $v_{\rm rx}$, $\eta$, and $\xi$, i.e., 
\begin{equation}\label{eq:res-off}
    \phi'_{\rm t}(kT) = 2\pi \left[f_{{\rm D}, n_{\rm t}, \nu} +\frac{v_{\rm rx}}{\lambda} (\cos \xi - \cos \eta)\right]kT.
\end{equation}
Note that from the geometry in \fig{fig:rx-mob}, by calling the \ac{aoa} of the sensing path $\zeta$, we have \mbox{$\xi=\zeta +\eta$}.
It follows that if the RX can estimate the \ac{aoa}, $\hat{\zeta}$, and its own movement speed vector (e.g., from an onboard accelerometer), $\hat{v}_{\rm rx}, \hat{\eta}$, the additional offset in \eq{eq:res-off} can be compensated for by multiplying the \ac{cir} by a complex exponential with phase $-2\pi \frac{\hat{v}_{\rm rx}}{\lambda} [\cos ( \hat{\zeta} + \hat{\eta}) - \cos \hat{\eta}]kT$. An accurate estimate of $v_{\rm rx}$, $\zeta$, and $\eta$ allows JUMP to recover the correct Doppler frequency of the target.
Errors on the estimation of $v_{\rm rx}$, $\zeta$, and $\eta$ can lead to a residual phase error, which is amplified by the presence of the wavelength $\lambda$ in the denominator of~\eq{eq:res-off}. At \ac{mmwave} frequencies, due to the short wavelength, JUMP's robustness to RX movements heavily depends on the quality of the external velocity estimate from the RX accelerometer. 
}

\section{Analysis and numerical simulation
}\label{sec:analysis}
\rev{In this section, we analyze the \ac{to} and \ac{cfo} compensation capabilities of JUMP by providing insights on the residual timing and phase errors, as well as numerical simulations to validate our claims. We take IEEE~802.11ay as a reference, as it is also used in our testbed implementation. In this context, channel estimation fields appended to the packets are called beam-training (TRN) fields~\cite{802.11ay}. A TRN field includes $6$ TRN \textit{units}, each made of complementary Golay sequences of $L=128$ samples modulated with \ac{bpsk}, which can be transmitted using different antenna \acp{bp}. 

In the simulations, we set \mbox{$B=1.76$~GHz} (as in a standard IEEE~802.11ay channel), hence $\Delta \tau = 0.57$~ns and $\Delta d = 17$~cm. The inter-packet transmission time is $T = 0.27$~ms. We generate $10^4$ \ac{cir} realizations with a random number of scatterers between $2$ and $10$. Each scatterer is located at a random distance from the TX, chosen uniformly at random in $[1.5, 10]$~m. The \ac{cir} taps' amplitudes are computed using the bistatic radar equation~\cite{richards2010principles}, with a random \ac{rcs} for the scatterers in the interval~$[-20, 10]$~dBsm (dB per square meter). The transmitted signal is a single IEEE~802.11ay TRN field, for a total length of $768$~symbols. To evaluate JUMP's performance in intermittent LOS/NLOS conditions, in some of the simulations we modulate the \ac{cir} amplitude for the \ac{los} path with an exponential profile which decreases to zero. This simulates the blockage due to scatterers passing between the TX and the RX.

The \ac{to} and the phase shift due to \ac{cfo} are obtained as uniform random variables in the intervals $[0, 20 \Delta \tau]$~s and $[0, 2\pi]$, respectively.}
\begin{figure}[t!]
     \centering
    \includegraphics[width=\columnwidth]{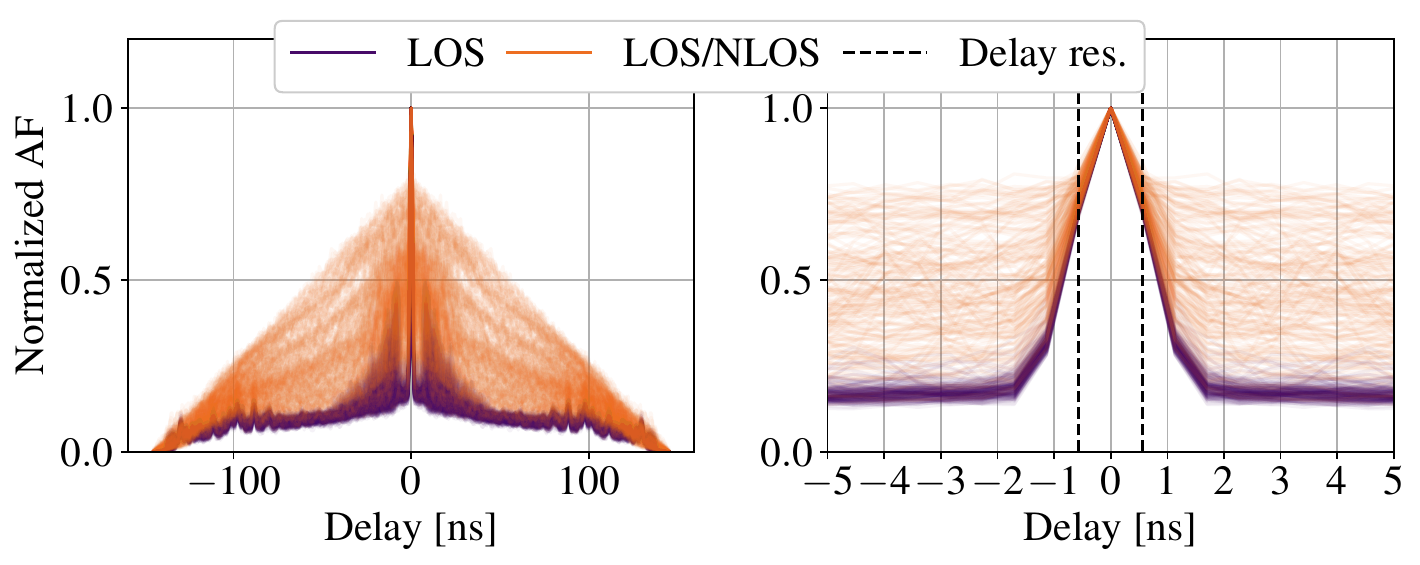}
    \includegraphics[width=\columnwidth]{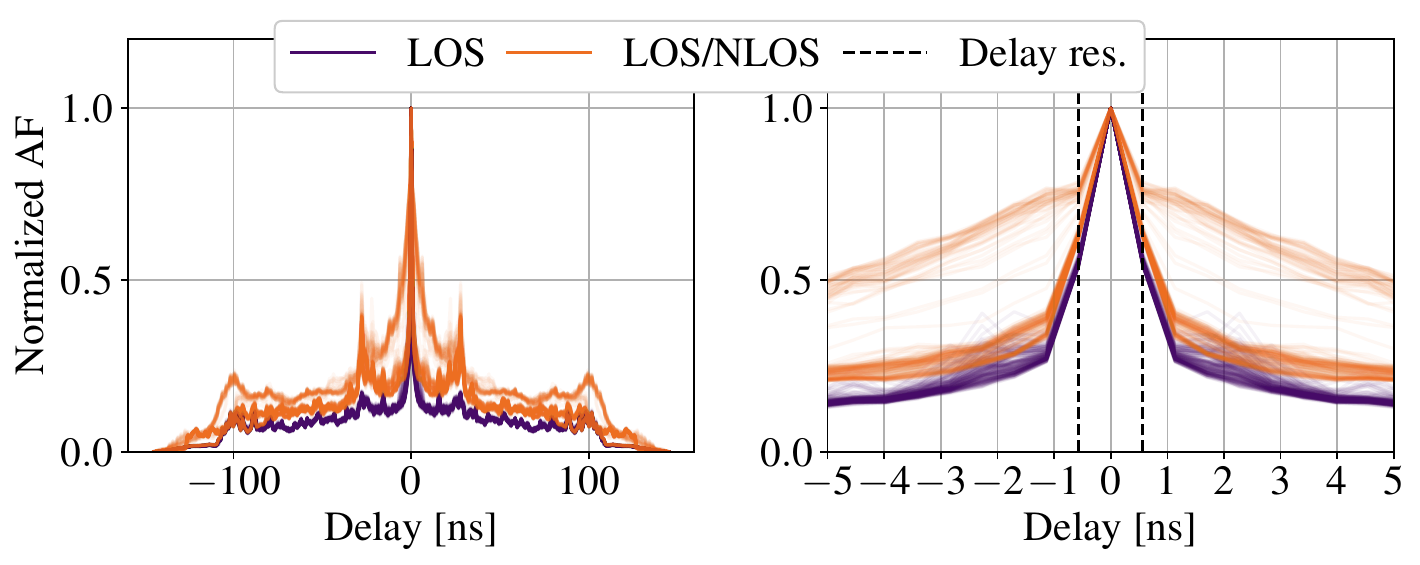}
    \caption{\rev{Ambiguity function of the \ac{cir} amplitude profiles obtained from real data for \ac{los} and intermittent \ac{los}/\ac{nlos} (a zoom around $\pm 5$~ns is shown on the right). Different curves represent different channel realizations, for a total of $1000$.}}
     \label{fig:af-analysis}
     \vspace{-3mm}
\end{figure}
\begin{figure*}[t!]
	\begin{center}   
		\centering
		\subcaptionbox{$B=1.76$~GHz. \label{fig:sim-to-snr}}[0.25\textwidth]
  {\includegraphics[width=0.25\textwidth]{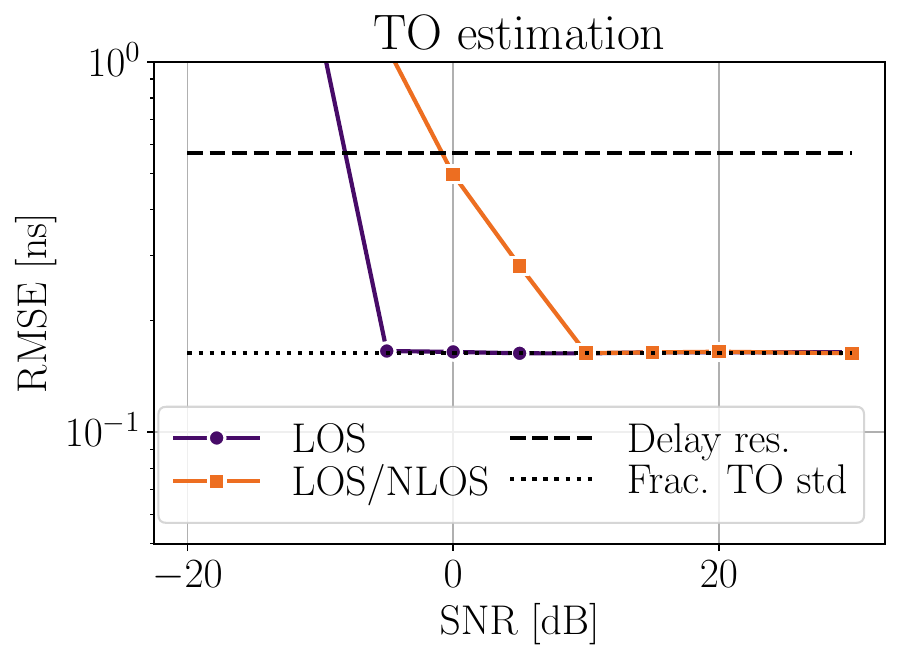}}
  \subcaptionbox{\ac{snr} = $5$~dB. \label{fig:sim-to-bw}}[0.25\textwidth]{{\includegraphics[width=0.25\textwidth]{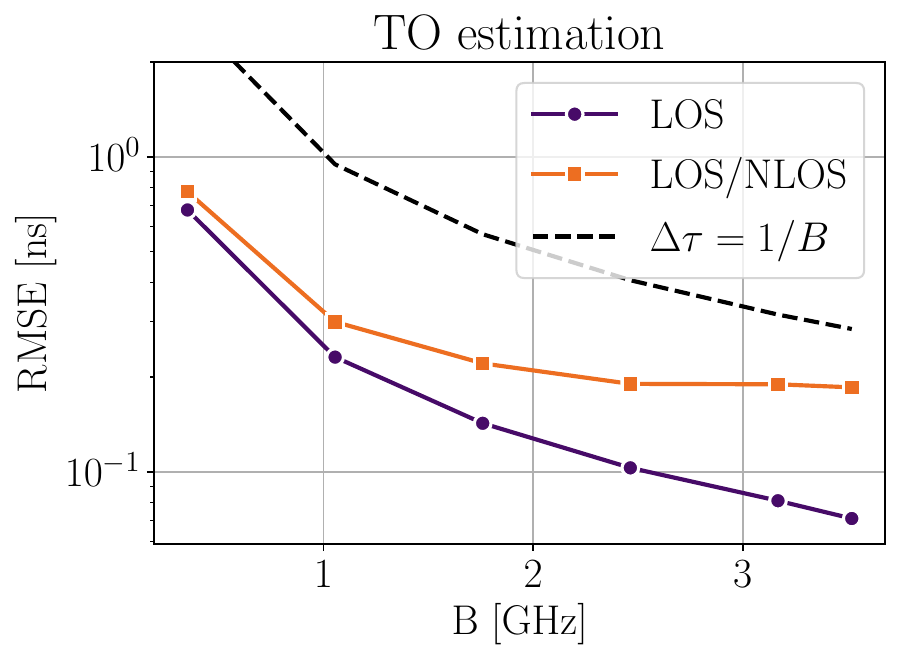}}}
		\subcaptionbox{$B=1.76$~GHz. \label{fig:sim-cfo-snr}}[0.237\textwidth]{\includegraphics[width=0.237\textwidth]{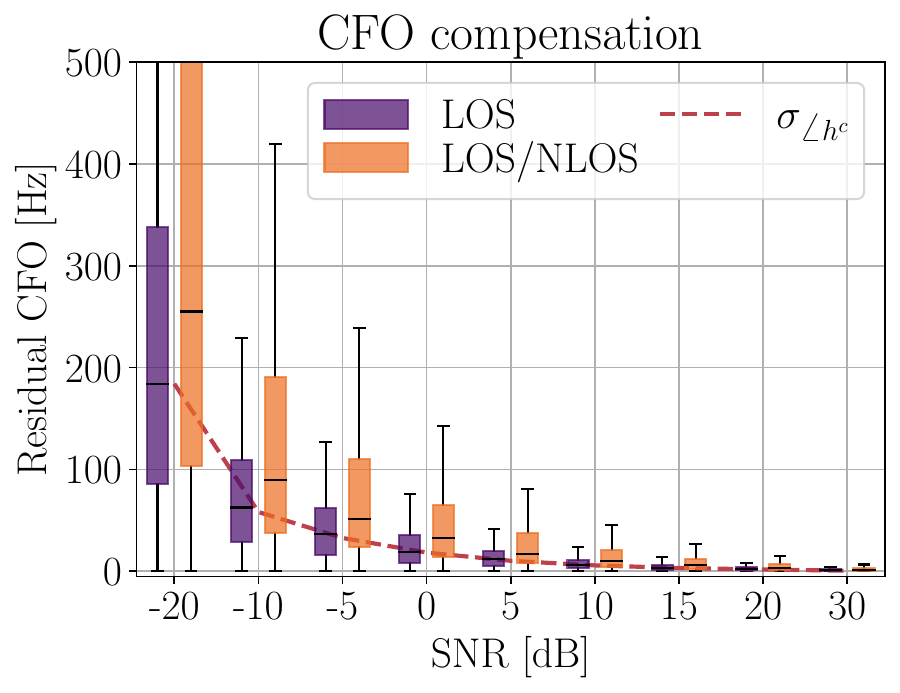}}
        \subcaptionbox{\ac{snr} = $5$~dB.\label{fig:sim-cfo-bw}}[0.237\textwidth]{\includegraphics[width=0.237\textwidth]{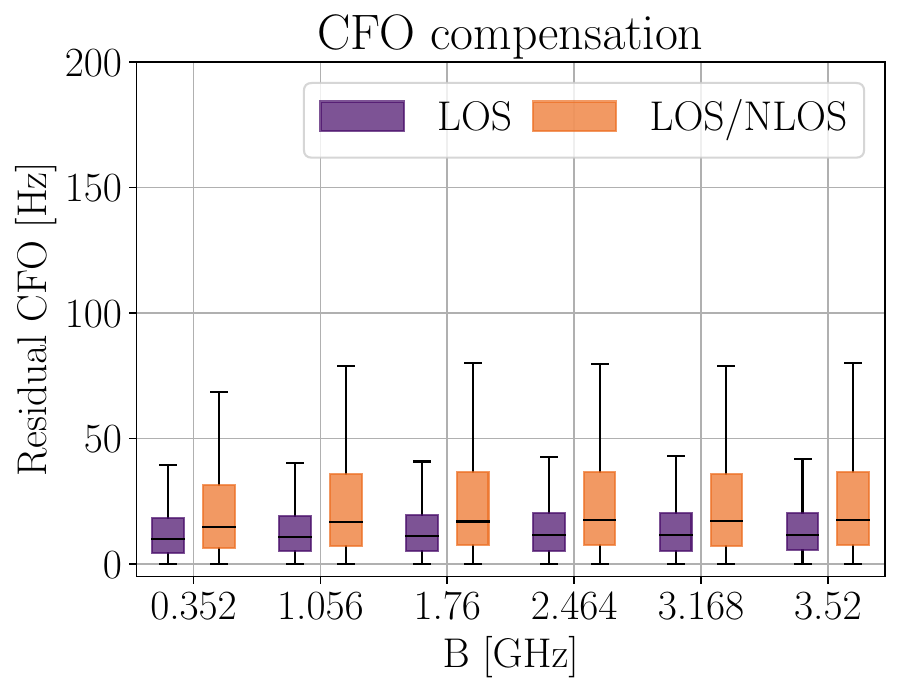}}
		\caption{\rev{Simulation results for \ac{to} estimation RMSE (a-b) and \ac{cfo} compensation error distribution (c-d), varying the \ac{snr} and the TX signal bandwidth . We report the results for \ac{los} and intermittent \ac{los}/\ac{nlos} conditions. }}
		\label{fig:simulation-res}
	\end{center}
 \vspace{-0.5cm}
\end{figure*}
\subsection{TO compensation error analysis}\label{sec:to-error}
\rev{We consider errors in the estimation of the relative \ac{to} from \eq{eq:rho_estimate}. This is a conservative approach, in that we do not account for the subsequent robustness improvement brought by the majority voting scheme in \eq{eq:maj-vote}. 
For simplicity, we neglect the time index $k$, following our assumption that the channel profile can be considered static in the short time between two subsequent packets.

The accuracy of the relative \ac{to} estimate in \eq{eq:rho_estimate} depends on (i)~the relative location of the \ac{cir} peaks, and (ii)~their sharpness. Indeed, cross terms due to the multiplication of \ac{cir} peaks caused by different propagation paths may result in strong, secondary cross-correlation peaks that cause ambiguity in the \ac{to} estimation. Moreover, despite the perfect autocorrelation properties of Golay sequences, in practice the \ac{cir} peaks are not exact Kronecker delta functions, but have a non-zero \textit{width}. This is caused by multiple factors, among which are the imperfect symbol synchronization (that causes a small error in the sampling point at the RX) and the fact that extended targets cause multiple reflections that overlap in the same or adjacent delay bins. However, because the \ac{cir} profiles depend on the underlying physical environment, which can be very different depending on the scenario, \ac{to} errors are difficult to model analytically without making restrictive assumptions.

For this reason, we take an alternative approach using the concept of \ac{af} in radar signal processing~\cite{richards2010principles}. The \ac{af} is defined as the cross-correlation of a waveform with its delayed and frequency shifted versions and it measures the sharpness of the main correlation peak and the secondary peaks level.
In our setting, we compute the \ac{af} of the \ac{cir} amplitude profiles used in \eq{eq:rho_estimate}, hence Doppler shift has no impact and the \ac{af} reduces to the autocorrelation of each \ac{cir} amplitude profile. The ideal \ac{af} in this situation is a Kronecker delta centered in zero. We obtain the \ac{af} of the \ac{cir} profiles in simulation and using \textit{real} measurements from our testbed implementation and setup described in \secref{sec:implementation} and \secref{sec:results}.  The resulting normalized \ac{af} for $1000$ \ac{cir} profiles is shown in \fig{fig:af-analysis} for LOS and intermittent LOS/NLOS scenarios. In both real and simulated \acp{cir}, the LOS \ac{af} exhibits a sharp peak at $0$~ns, which is well contained within the delay resolution of the system, shown by the black dashed lines. In LOS/NLOS, the \ac{af} shows a higher noise floor, which is due to the lower amplitude of the NLOS peaks. Nevertheless, the main peak is still clear and sharp around $0$~ns, showing the robustness of our approach. Note that the difference between simulated and real measurements is due to the much higher channel variability that can be obtained in simulation. The simulated \ac{af} is a more general assessment of JUMP's robustness for diverse channel realizations.  
\fig{fig:sim-to-snr} and \fig{fig:sim-to-bw} show the \ac{to} estimation \ac{rmse} varying the \ac{snr} and the system bandwidth, respectively. JUMP achieves an \ac{rmse} lower than the delay resolution $\Delta \tau$ (black dashed line) for an \ac{snr} of $-5$ dB in LOS and $0$ dB in LOS/NLOS. This is sufficient to correctly estimate the integer part of the \ac{to}. Note that this is exactly the purpose of JUMP's \ac{to} compensation method: we are not interested in estimating the \ac{to} itself, but only in computing the correct integer shift to be applied to consecutive \ac{cir} profiles to obtain consistent range measurements. At higher \ac{snr} values, the error converges to the standard deviation of the fractional part of the true \ac{to} used in the simulations (black dotted line). This is expected since the fractional part of the \ac{to} is neglected in~\eq{eq:rho_estimate}. 

\fig{fig:sim-to-bw} shows the \ac{to} compensation \ac{rmse} varying the TX bandwidth, $B$, for an \ac{snr} of $5$~dB. In LOS, the gain from using a larger $B$ scales similarly to the delay resolution. Conversely, in LOS/NLOS conditions JUMP cannot make full use of larger bands due to the worse actual \ac{snr}, and thus the error does not improve with the delay resolution beyond $2$~GHz.}

\rev{\subsection{CFO removal error analysis}\label{sec:cfo-error}
JUMP removes the \ac{cfo} using the phase of the static reference path. Such phase is affected by noise on the RX signal, which causes a residual phase error on the sensing path. We use expression $\angle{ h^{\rm c}[k, \hat{\ell}_{\rm s}]} = \mathrm{arctan}\left(h_Q^{\rm c}[k, \hat{\ell}_{\rm s}] /  h_I^{\rm c}[k, \hat{\ell}_{\rm s}]\right)$ introduced in \secref{sec:md-extr} to compute the residual phase error variance, $\sigma^2_{\angle{ h^{\rm c}}}$. The variance of $h_I^{\rm c}[k, \hat{\ell}_{\rm s}]$ and $h_Q^{\rm c}[k, \hat{\ell}_{\rm s}]$ is $\sigma_w^2 / (2G)$, and $\sigma^2_{\angle{ h^{\rm c}}}$ can be obtained by propagating the error induced by noise through the expression of $\angle{ h^{\rm c}[k, \hat{\ell}_{\rm s}]}$, as shown, e.g., in~\cite{ku1966notes}. For low $\sigma_w^2 / (2G)$, this gives
\begin{equation}\label{eq:res-phase-err}
    \sigma^2_{\angle{ h^{\rm c}}} =  \frac{\sigma_w^2}{2 G |h^{\rm c}[k, \hat{\ell}_{\rm s}]|^2} = \frac{1}{2 G  \Gamma} \frac{|\tilde{h}_1(kT)|^2}{ |h^{\rm c}[k, \hat{\ell}_{\rm s}]|^2},
\end{equation}
where we recall that $\tilde{h}_1(kT)$ is the gain of the \ac{los} path.
\eq{eq:res-phase-err} shows the dependency of the residual phase error on the power of the \ac{cir} tap corresponding to the reference path used for the \ac{cfo} removal. This is validated by the simulation result in \fig{fig:sim-cfo-snr}. We show the residual \ac{cfo} for different values of the \ac{snr} on the received signal, in both \ac{los} and intermittent \ac{los}/\ac{nlos} conditions. We also plot the theoretical error standard deviation from \eq{eq:res-phase-err} in  the \ac{los} case for \mbox{$G=L$} (rescaled by $2\pi T$) with a black dashed line, which matches well with the simulation result for higher values of the \ac{snr}. The best performance is obtained using the \ac{los} path as the reference since it is typically much stronger than any first order reflection. Nevertheless, JUMP shows excellent \ac{cfo} estimation even in intermittent \ac{los}/\ac{nlos}, where a static first order reflection is used as described in \secref{sec:static-selection}.
Finally, in \fig{fig:sim-cfo-bw} we report the residual \ac{cfo} distributions obtained fixing the \ac{snr} to $5$~dB and changing the TX bandwidth $B$ between $350$~MHz and $3.52$~GHz. The results demonstrate that the resulting error is independent of the bandwidth, as expressed by \eq{eq:res-phase-err}. This makes our \ac{cfo} removal approach also applicable to communication systems with lower bandwidth (e.g., 4G-LTE and 5G-NR), provided that a reference path can be identified.}

\begin{figure}[t!]
	\begin{center}   
		\centering
        \input{figures/overhead}
		\caption{\rev{Overhead introduced by JUMP on the IEEE~802.11ay PHY layer packet, as a function of the number of added TRN units, for different PSDU sizes. Solid lines correspond to MCS $8$, while dashed lines represent MCS $12$.}}
		\label{fig:oh-curves}
	\end{center}
 \vspace{-0.3cm}
\end{figure}
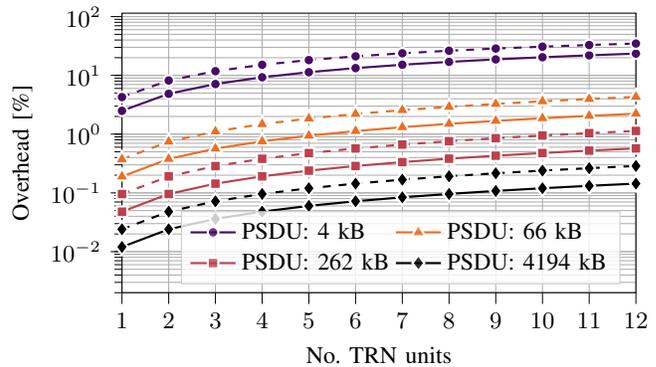

\rev{\subsection{Impact on communication: overhead}\label{sec:overhead}}
\rev{To perform sensing, JUMP requires obtaining: (i)~\ac{cir} estimates covering the full angular space in order to detect and track the targets, and (ii)~\ac{cir} estimates obtained with TX \acp{bp} pointing towards the targets of interest, with sufficient granularity to achieve a sufficient maximum Doppler frequency $f_{\rm D}^{\max} = 1/(2T)$. Condition~(i) is easily satisfied by beam training operations that are commonly performed in \ac{mmwave} systems to align the TX and RX beams for communication. As an example, in IEEE~802.11ay a full beam sweep is done periodically (e.g., every $100$~ms) by appending TRN units to the packets. Due to typical human movement speeds being in the order of $1-4$ meters per second, the beam training frequency is sufficient to fulfill JUMP's requirements without adding any overhead to the communication protocol. Condition (ii)~instead demands that additional channel estimation fields are appended to communication packets during normal traffic. Specifically, JUMP needs \textit{one} additional field \textit{per target}, and \textit{one} field for the reference static path, transmitted using a \ac{bp} that illuminates the corresponding direction. This adds overhead to the communication protocol, as shown in~\fig{fig:oh-curves}. We plot the overhead introduced by adding $1$ to $12$ TRN units, which corresponds to sensing up to $11$ targets concurrently, plus the reference path. Different colors correspond to different \ac{psdu} sizes used in the standard \cite{802.11ay}, while solid and dashed lines refer to \acp{mcs} $8$ and $12$, respectively. 
The overhead is computed as the ratio between the length of the added PHY layer symbols in the TRN units and the total number of symbols in the packet, considering the payload plus PHY and MAC headers~\cite{802.11ay}.
For \ac{psdu} sizes $66$~kB to $4194$~kB the overhead is less than $3$\%, with $11$ targets, which is a negligible impact on communication. Note that in practical scenarios, given the limited TX range at \ac{mmwave} frequencies, it is unlikely to sense such a high number of targets concurrently. For \ac{psdu} $4$~kB, the overhead can reach $10$-$30$\%. However, note that this \ac{psdu} size should be avoided anyways in IEEE~802.11ay deployments, as due to the high rates available in \ac{mmwave} communication it yields a very low MAC layer efficiency. Specifically, with a \ac{psdu} of $4$~kB, the MAC efficiency is $74$\% solely because of the MAC and PHY layers overheads.

Finally, we underline that in this paper we do not consider the irregularity of the inter-packet times in communication systems. We address this in a separate work, \cite{pegoraro2022sparcs}, proposing a method to reconstruct \ac{md} signatures from the irregular \ac{cir} estimates obtained from communication packets.}
\section{Implementation} \label{sec:implementation}

JUMP's prototype is based on the $60$~GHz \ac{sc} IEEE~802.11ay standard~\cite{802.11ay}, maintaining its packet structure. \ac{cir} estimation is performed by using trailing TRN fields.

\subsubsection{Testbed design} Our implementation is a customization of MIMORPH~\cite{Lacruz_MOBISYS2021}, an open-source project for \ac{mmwave} experimentation inclusing a Xilinx \ac{rfsoc} board and Sivers' \ac{mmwave} front-end~\cite{SIVERSIMA_BOTH}. \fig{fig:testbed} shows the main components of our testbed. The MIMORPH \ac{fpga} logic was modified to allow its operation as a TX, RX, or both functionalities simultaneously. 
In this way, we can emulate a monostatic \ac{jcs} system, which we use as a baseline for comparison. 
The \acp{adc} and \acp{dac} on the board are configured to operate at 3.52~GHz sampling frequency, fulfilling the requirements of \ac{mmwave} Wi-Fi standards (IEEE 802.11ad/ay), with 1.76~GHz of \ac{rf} bandwidth. 

The TX implements a loopback memory that feeds the \acp{dac} with the I/Q symbols to be transmitted, which are loaded from an external processor. To enable \ac{aod} estimation (\secref{sec:aod-est}), we implement a real-time antenna reconfiguration mechanism that allows to sweep through different \acp{bp} in the TRN fields of a single packet, as in~\cite{Lacruz_MOBISYS2021}. 

At the RX, after downconversion and sampling, packet detection is performed by searching for peaks in the autocorrelation of the received signal
\cite{Lacruz_MOBISYS2020,LiuTCASI2017}. 
In communication between asynchronous devices, in addition to \ac{to} and \ac{cfo}, the non-ideal sampling point at the RX causes a symbol timing offset. This introduces \ac{isi}. Although this is generally neglected in the \ac{jcs} literature~\cite{zhang2022integration,zhang2021overview}, in our experiments we found that it affects \ac{md} quality.
To avoid this, we implement a symbol synchronization block 
including a configurable fractional delay filter, based on a Farrow structure~\cite{Farrow1988Iscas}, followed by a fast Golay correlator~\cite{LiuTCASI2017}.
After proper sampling point selection and subsequent downsampling, the I/Q symbols are fed to the onboard RAM. 


\subsubsection{Different testbed configurations} 
\begin{figure}[t!]
     \centering
     \includegraphics[width=\columnwidth]{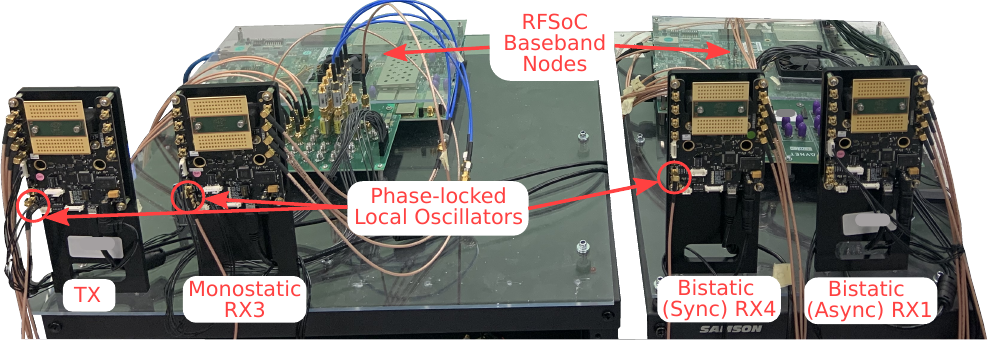}
     \caption{Testbed components.}
     \label{fig:testbed}
     \vspace{-3mm}
\end{figure}

Our implementation includes a TX, two JUMP receivers, RX1-2, and two receivers, RX3-4, which we use to assess and compare JUMP's performance against monostatic and phase-locked configurations.
JUMP's RXs are completely independent of the TX, as they do not share any clock source or oscillators. 
The monostatic RX, denoted by RX3, is located in the same location as the TX, as shown in \fig{fig:testbed}. For the monostatic case, the \ac{rfsoc} board concurrently operates as TX and RX (i.e., as a full-duplex system). To remove the \ac{cfo}, the local oscillator is fed from the TX antenna to the RX3 antenna (\fig{fig:testbed}).
The direct \ac{los} from TX to RX3 provides a reliable reference to align the \ac{cir} estimates, removing the \ac{to}.
%
With RX4, a phase-locked version of the bistatic \ac{jcs} system was implemented. This is similar to the monostatic RX3, but in this case, RX4 is co-located with JUMP's receiver, RX1. Similarly to the monostatic case, we phase-locked the \ac{mmwave} front-ends by sharing the TX local oscillator signal through a cable, thus eliminating the \ac{cfo}. RX4 is used as a reference to assess the quality of the JUMP reconstructed \ac{md} spectrum.

        
  
  
        
\subsubsection{System parameters}
We use the same values of $B$ and $T$ used in the simulation in~\secref{sec:analysis}. For \ac{md} extraction, we apply \ac{stft} with a Hanning window of size $W=256$. 
This yields a maximum resolvable Doppler frequency of $f_{\rm D}^{\max} = 1/(2T) = 1851.85$~Hz and a resolution of $\Delta f_D = 1/(WT) = 14.47$~Hz. The velocity corresponding to such bistatic Doppler frequency depends on $\gamma$ and $\beta$ as per \eq{eq:md-bistatic}.
For the \ac{aod} measurements, we set $\vartheta$ to $0.7$ multiplied by the maximum correlation value in \eq{eq:aoa-corr}. This allows capturing the \ac{aod} of different reflections in the same \ac{cir} tap while rejecting peaks due to sidelobes of the \acp{bp}.
$K=100$ frames are used for the selection of the static reference path, with a variance threshold \mbox{$\sigma_{\rm thr}^2 = 0.005$}~m$^2$ (see \secref{sec:static-selection}). \rev{ For the \ac{md} spectrum, we use $Q=2$, considering channel taps in a (bistatic) range of $\pm Q \Delta d = \pm 34$~cm from the target, which we empirically found to be suitable for human sensing, given typical human body sizes.}

\section{Experimental Results} \label{sec:results}

\begin{figure}[t!]
	\begin{center}   
    		\centering
    		\subcaptionbox{\texttt{Env1}.\label{fig:env1}}[3.5cm]{\includegraphics[width=3.17cm]{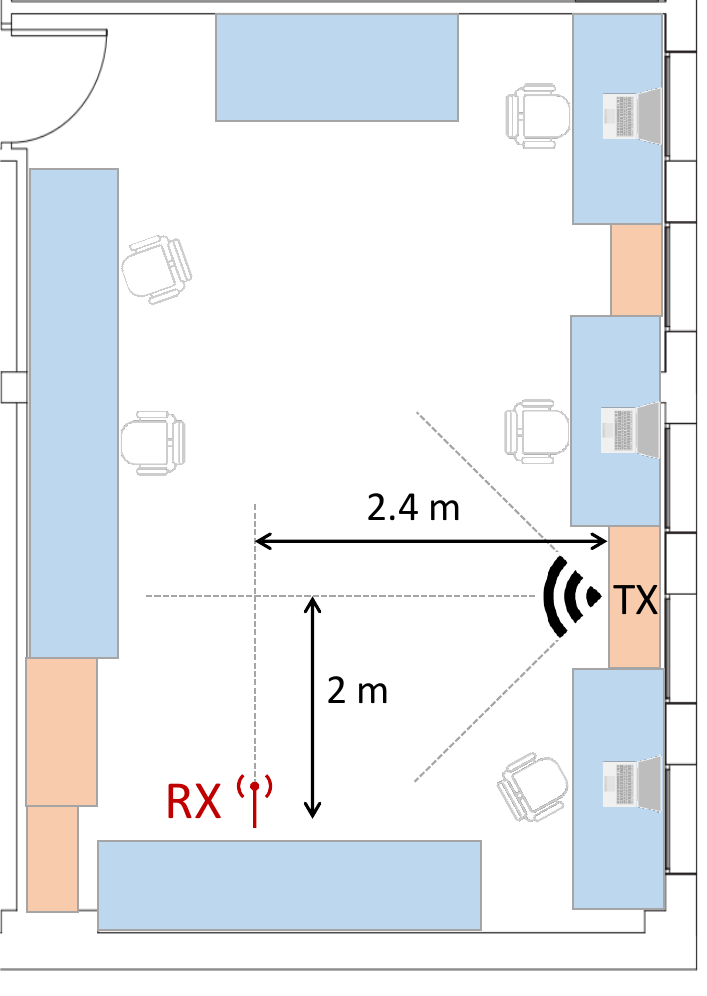}}
    		\subcaptionbox{\texttt{Env2}.\label{fig:env2}}[4.4cm]{\includegraphics[width=4.cm]{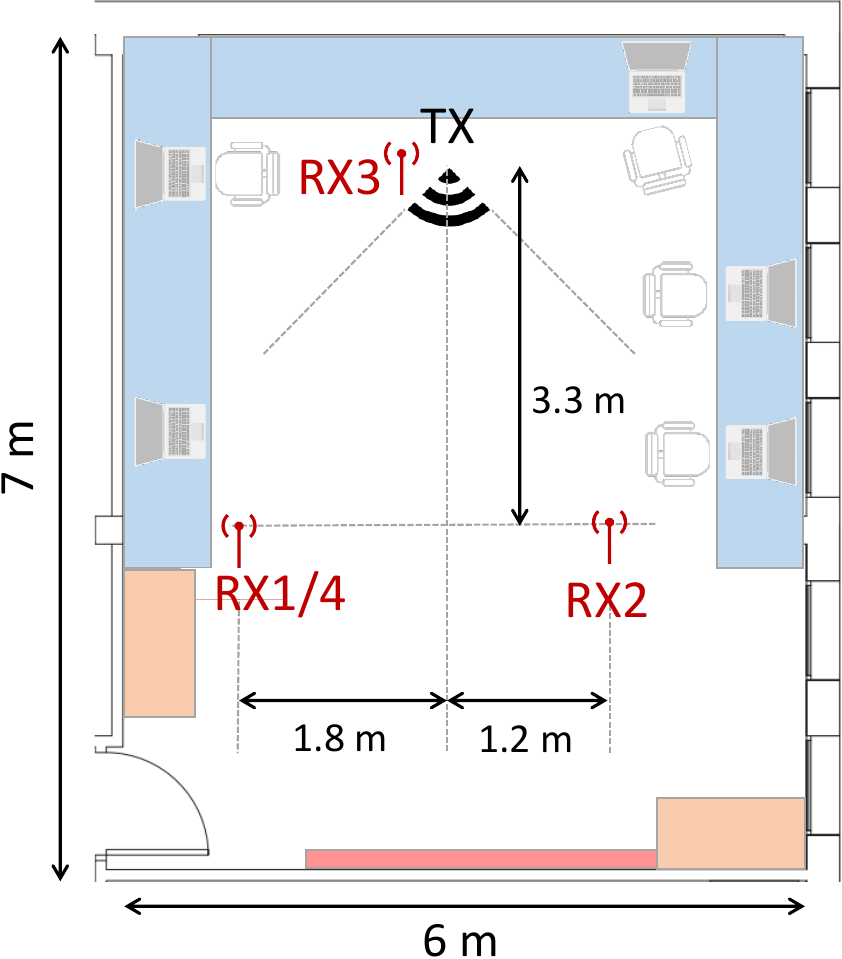}}
    		\caption{Schematic representation of the two environments.}
    		\label{fig:testbeds}
    	\end{center}
     \vspace{-3mm}
\end{figure}

JUMP was tested on the tasks of people tracking and \ac{md} extraction, in two different indoor environments to verify the system effectiveness under different conditions. The test areas are research laboratories, denoted by \texttt{Env1}, of dimensions $5 \times 8$~m, and \texttt{Env2}, of dimensions $6 \times 7$~m. \fig{fig:testbeds} shows them, along with the position of the TX and RX nodes.
The placement of the RXs with respect to the TX, in terms of relative distance and orientation, is shown in \fig{fig:testbeds} for \texttt{Env1} and \texttt{Env2}. We use \texttt{Env1} for baseline evaluations, using a single JUMP receiver, denoted by RX. In \texttt{Env2} instead, we perform comparisons with the monostatic and phase-locked systems, using RX1 and RX3-4, and we evaluate the performance of a \textit{multistatic} JUMP deployment using RX1-2 concurrently.
To reconstruct the ground truth movement trajectory of the subjects, we mark the floor at specific locations in the two environments and instruct them to move across the markers. To perform measurements under \ac{nlos} conditions, we use rectangular panels of absorbing material interposed between the TX and RX to block the \ac{los} signal. This is used to simulate common occlusions of the \ac{los} that can happen in dynamic environments, caused, e.g., by other subjects moving around.
We collect over $60$ \ac{cir} sequences in \texttt{Env1} and \texttt{Env2}, each with duration of about $12$ seconds ($40$k packets). 
\subsection{Bistatic tracking accuracy}\label{sec:tracking-acc}

\begin{figure}[t!]
	\begin{center}   
		\centering
		\subcaptionbox{Square.\label{fig:square}}[2.8cm]{\includegraphics[width=2.8cm]{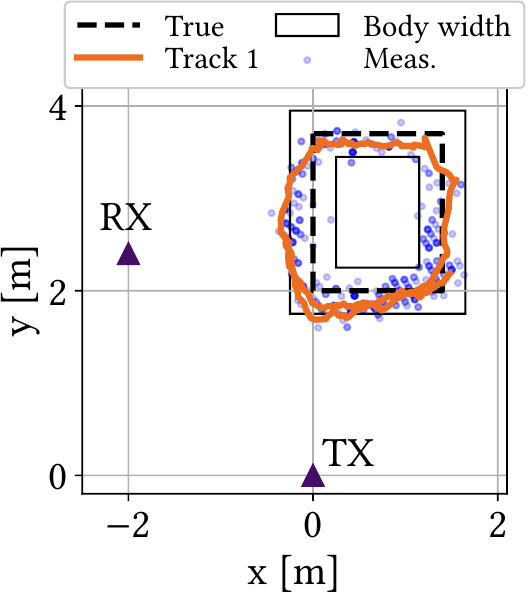}}
		\subcaptionbox{Line.\label{fig:line}}[2.8cm]{\includegraphics[width=2.8cm]{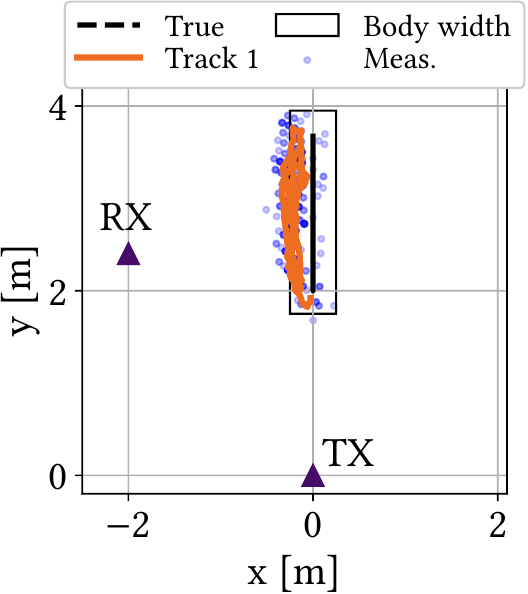}}
		\subcaptionbox{Triangle.\label{fig:triangle}}[2.8cm]{\includegraphics[width=2.8cm]{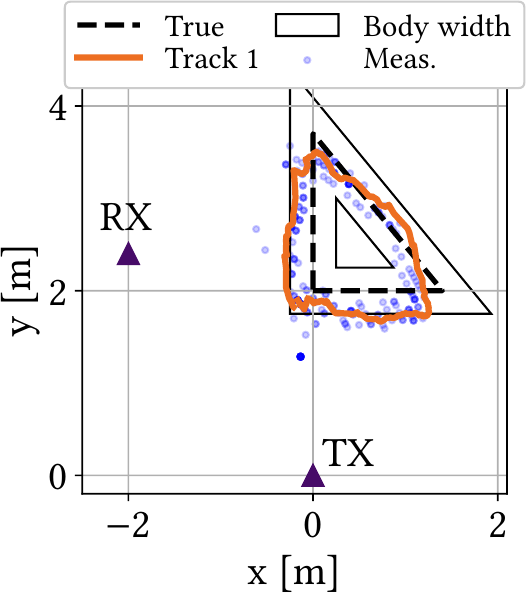}}
		\caption{Example estimated trajectories.}
		\label{fig:track-examples}
	\end{center}
 \vspace{-5mm}
\end{figure}

To evaluate JUMP's tracking accuracy, we compute the \ac{rmse} of the estimated movement trajectory of the subjects with respect to the ground truth. 
\fig{fig:track-examples} shows the tracks outputted by the \ac{ekf} in a \ac{los} setting in \texttt{Env1} for three different movement trajectories of a single person. Along with the tracks, we plot the ground truth with a solid black line and the measurement vectors, $\mathbf{z}$, with gray dots. Moreover, we also represent the body width of the subject during the movement with a black solid rectangle. The body width is non-negligible due to the high delay resolution of our system. Although it is not possible to know the exact reflection point of the signal on the subject's body, we expect the collected measurements to be biased towards the body side that is facing the TX/RX pair. This is clearly visible in \fig{fig:track-examples}. To account for the bias introduced by the body width, in the following analysis we compute the tracking error with respect to the trajectory of the body side facing the TX/RX, as this is more representative of the true tracking accuracy of the system. The body size was measured for each subject involved in the experiments.
\begin{figure}[t!]
	\begin{center}   
		\centering
		\subcaptionbox{LOS/NLOS tracking in the single and multi-target (MT) cases.\label{fig:box-multi}}[4.4cm]{\includegraphics[width=4.4cm]{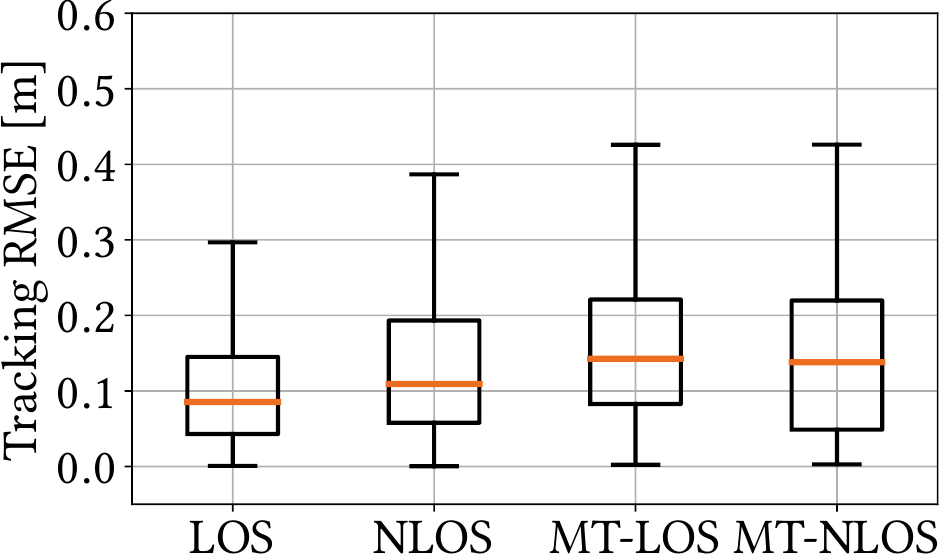}}
		\subcaptionbox{Tracking error depending on the target location.\label{fig:box-diff}}[3.5cm]{\includegraphics[width=3.5cm]{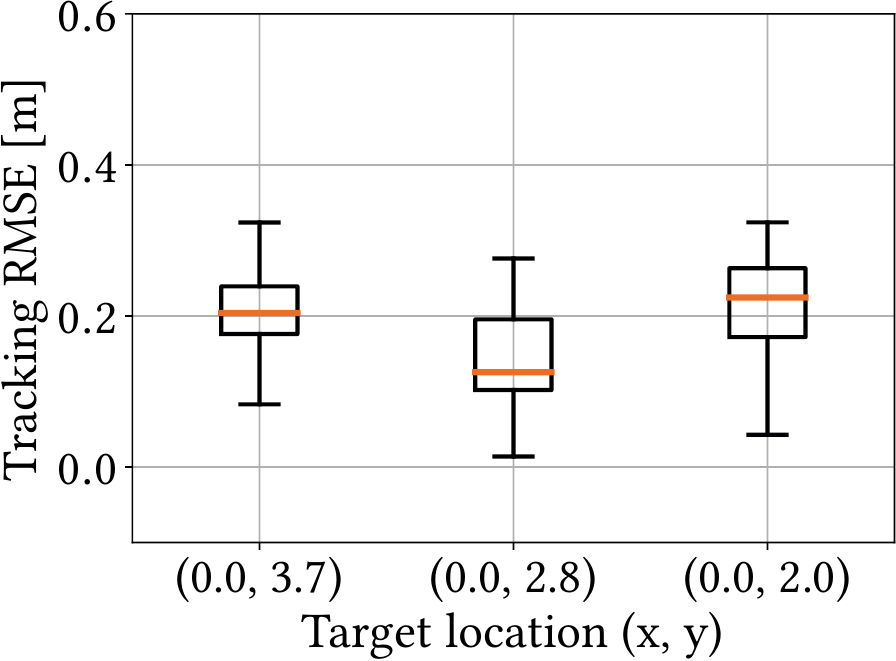}}	
		\caption{Boxplots of the normalized tracking RMSE.}
		\label{fig:los-nlos-comp}
	\end{center}
  \vspace{-5mm}
\end{figure}

\subsubsection{LOS tracking} As a baseline result, we compute the average \ac{rmse} for different measurement sequences obtained in \texttt{Env1}, obtaining a median \ac{rmse} of $8.5$~cm, as shown in the first boxplot of \fig{fig:box-multi}. This result provides a first assessment of the capability of our system to compensate for the \ac{to} due to clock asynchrony between TX and the RX. 
In \fig{fig:box-diff}, we show the \ac{rmse} obtained when tracking a subject sitting down and standing up at three different locations: $3.7$, $2.8$, and $2$~m in front of the TX. We observe that, on the one hand, distant targets cause reflections with high $d_{\rm tx}d_{\rm rx}$ product, which are harder to detect and track due to low \ac{snr}. On the other hand, targets too close to the \ac{los} link have the bistatic angle $\beta$ close to $\pi$, yielding a lower-ranging resolution. For these reasons, the median tracking errors in the first and third cases are higher than in the second one.

\subsubsection{Impact of NLOS} In the second boxplot in \fig{fig:box-multi} we show the \ac{rmse} distribution obtained in \textit{intermittent} \ac{los}/\ac{nlos} conditions. In collecting these \ac{cir} measurements, we block the \ac{los} link using the absorbing panel intermittently for time intervals of approximately $1-2$~s. This test is very challenging in terms of \ac{to} compensation, as the received packet can be detected from one of the reflections on surrounding objects, thus presenting a possibly large relative shift between subsequent packets. Still, our approach can successfully remove the \ac{to}, obtaining a median \ac{rmse} of $10.9$~cm, with the third quartile of the error distribution being less than $20$~cm.
\begin{figure}[t!]
	\begin{center}   
		\centering
		\subcaptionbox{RMSE CDF.\label{fig:rmse-cdf-mono}}[4.3cm]{\includegraphics[width=4.3cm]{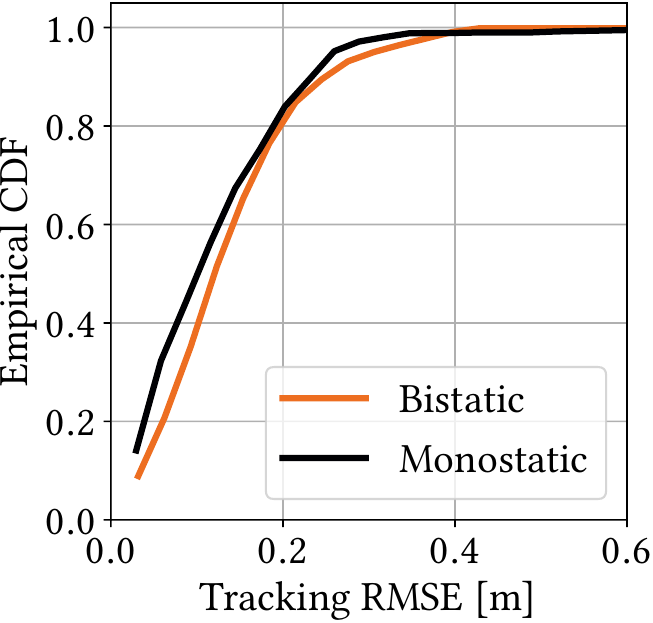}}
		\subcaptionbox{Example trajectory.\label{fig:traj-ex-mono}}[3.8cm]{\includegraphics[width=3.8cm]{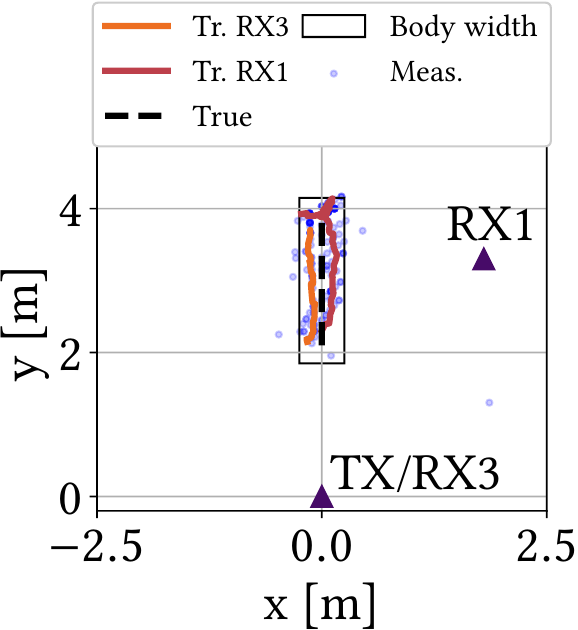}}
		\caption{Monostatic vs. bistatic tracking RMSE comparison.}
		\label{fig:track-bist}
	\end{center}
 \vspace{-5mm}
\end{figure}
\begin{figure}[t!]
	\begin{center}   
		\centering
		\includegraphics[width=4cm]{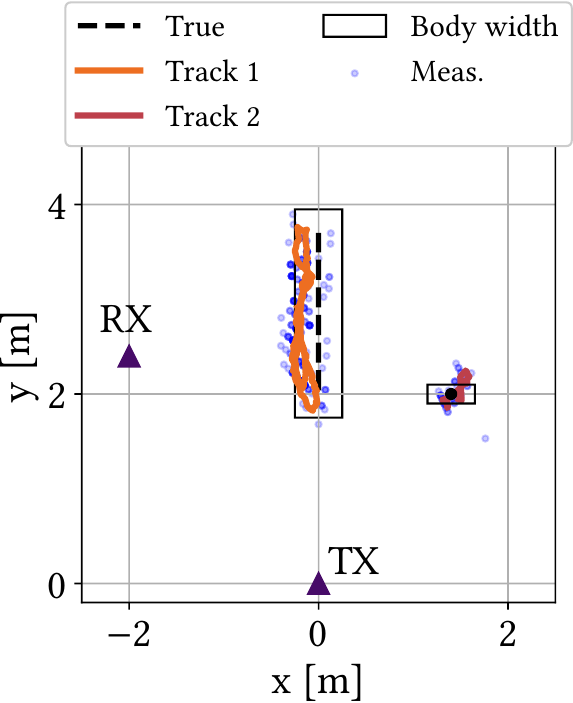}
		\includegraphics[width=4cm]{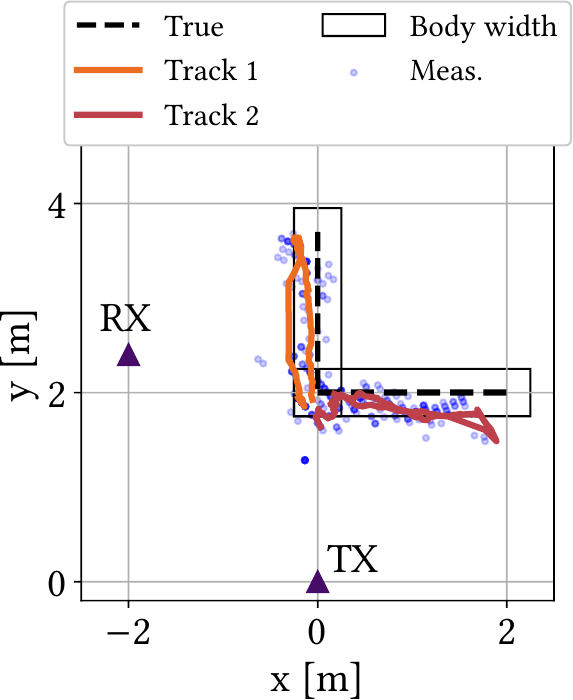}
		\caption{Multitarget tracking in \texttt{Env1}.}
		\label{fig:mt-scenario}
	\end{center}
  \vspace{-5mm}
\end{figure}

\subsubsection{Comparison with a monostatic system} In \fig{fig:track-bist} we compare our system with a \textit{monostatic} sensing configuration (see \fig{fig:traj-ex-mono}). This evaluation provides a comparison with existing approaches in the literature that adopt full-duplex monostatic sensing systems, e.g.,~\cite{barneto2021full, pegoraro2023rapid}. In addition, it allows evaluating the impact of the resolution as a function of the bistatic angle (see \secref{sec:bist-sens}). For this, we concurrently collect \ac{cir} sequences using RX3 and RX1, having the subject walk along different linear trajectories.
\fig{fig:rmse-cdf-mono} shows the empirical \ac{rmse} \ac{cdf} obtained with the monostatic and bistatic systems. Although less accurate, our system performs similarly to the monostatic configuration. This is because bistatic systems produce better sensing \ac{snr} in the region around RX1, which instead is a low \ac{snr} region for RX3. When multiple trajectories and locations are considered, our results show that this compensates for the degradation in the bistatic range resolution.

\subsubsection{Multitarget scenario} Finally, we evaluate the tracking accuracy when multiple subjects concurrently move in the environment. \fig{fig:mt-scenario} shows two example results obtained with two subjects when one is walking and the other is sitting down/standing up repeatedly (left), and when both are walking on different trajectories (right). 
The system can accurately track multiple targets obtaining the \ac{rmse} distributions shown in the last two boxplots of \fig{fig:box-multi}. We report the results for both the \ac{los} and the intermittent \ac{los}-\ac{nlos} cases. The median \ac{rmse} slightly degrades compared to the single subject case ($14$~cm), due to the more frequent outliers caused by the more challenging multi-target tracking task.

\subsection{micro-Doppler quality}\label{sec:md-quality}
\begin{figure}[t!]
	\begin{center}   
		\centering
		\subcaptionbox{Tracking.\label{fig:traj-ex-lin}}[3.cm]{\includegraphics[width=2.7cm]{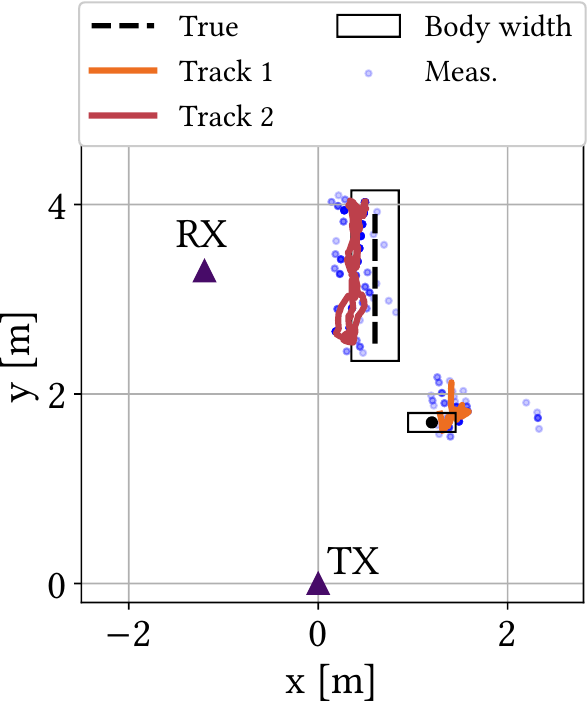}}
		\subcaptionbox{Extracted \ac{md}.\label{fig:traj-ex-lin-md}}[5.3cm]{\includegraphics[width=5.3cm]{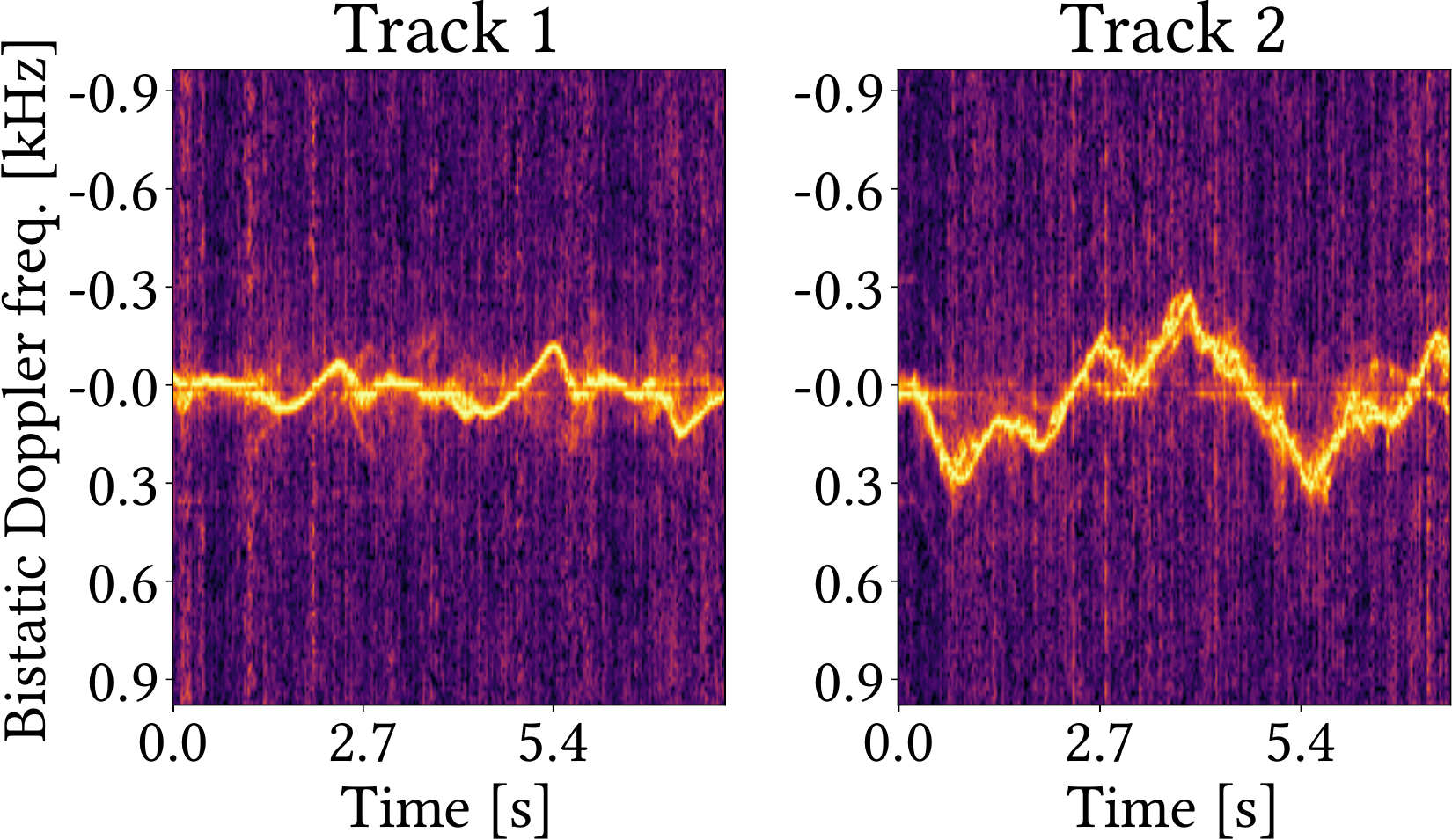}}
		\caption{\ac{md} extracted in a multitarget scenario.}
	\label{fig:multitarget}
  \vspace{-0.5cm}
	\end{center}
\end{figure}

\begin{figure*}[t!]
	\begin{center}   
		\centering
		\subcaptionbox{Walking.\label{fig:md-comp-walk}}[5.5cm]{\includegraphics[width=5.3cm]{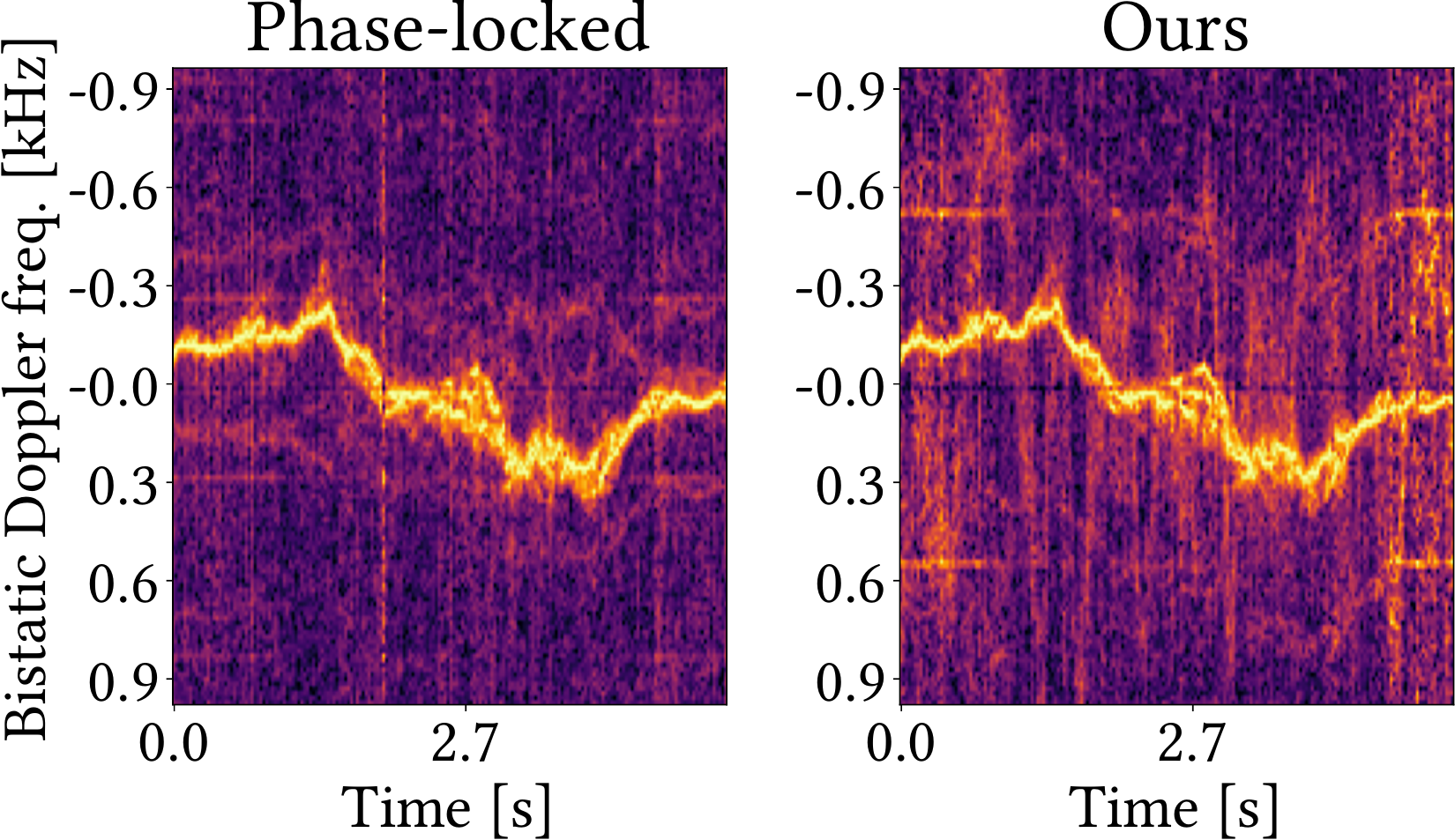}}
		\subcaptionbox{Sitting.\label{fig:md-comp-sit}}[5.5cm]{\includegraphics[width=5.3cm]{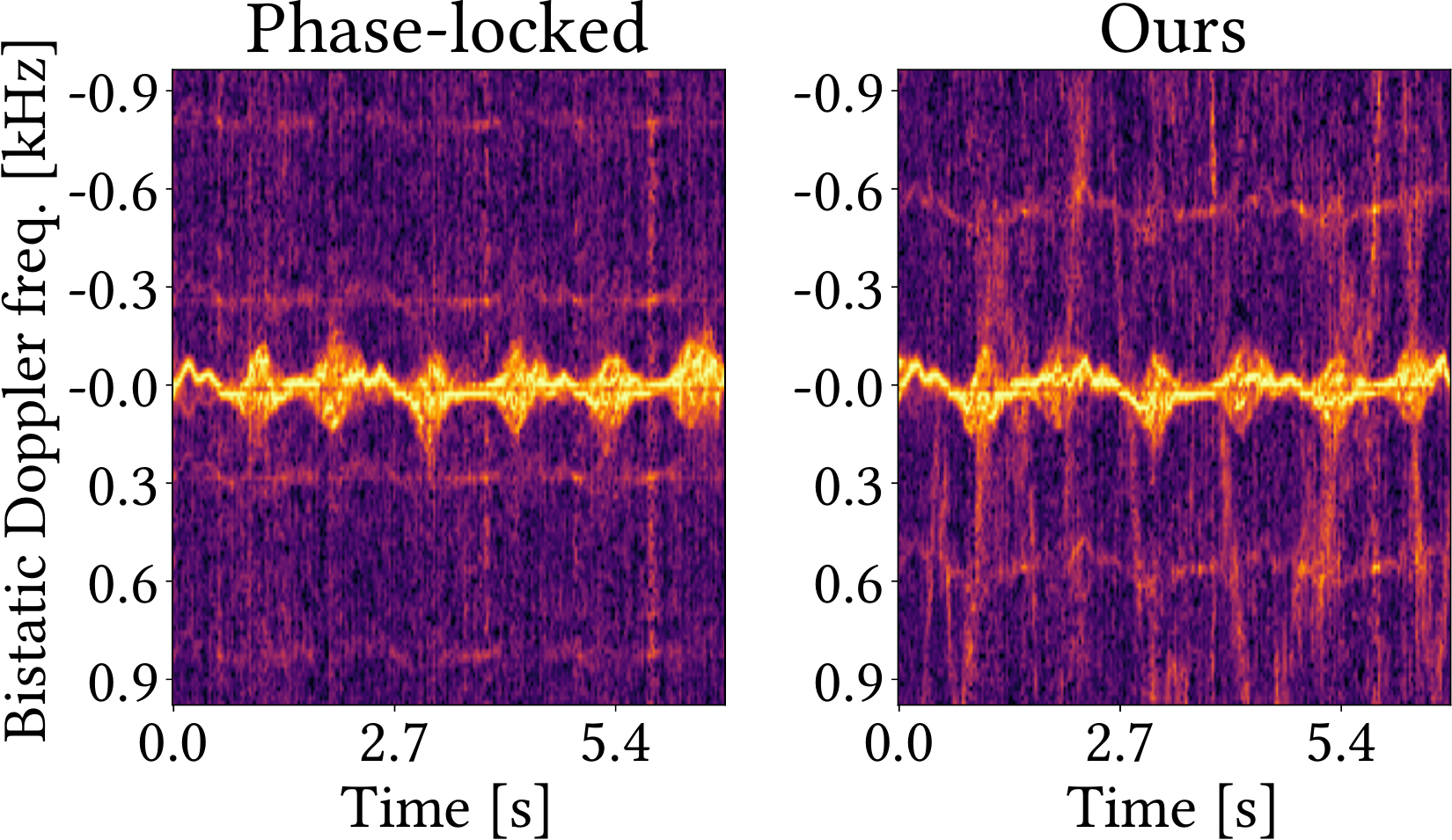}}
		\subcaptionbox{Hand gestures.\label{fig:md-comp-wave}}[5.5cm]{\includegraphics[width=5.3cm]{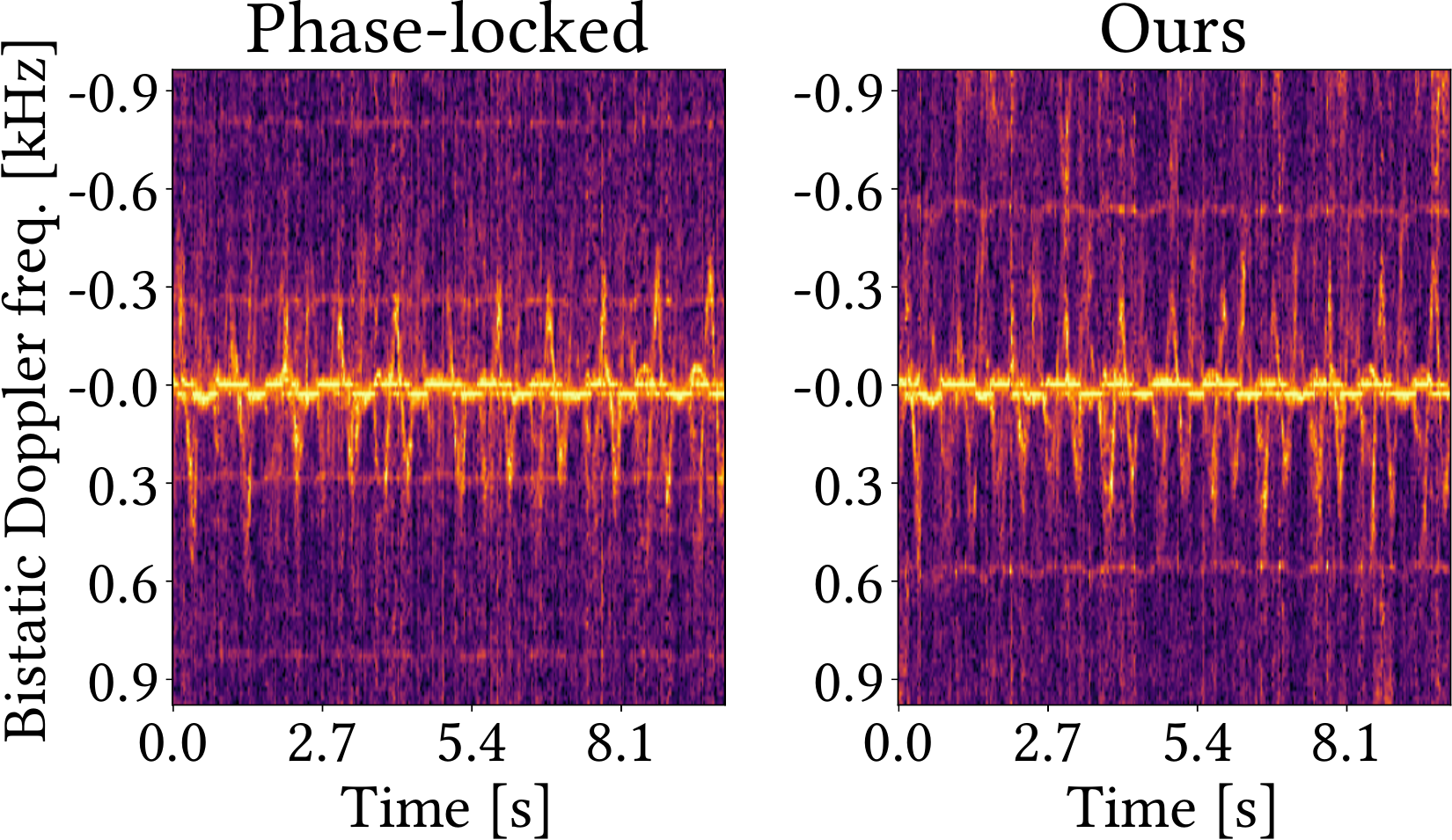}}
		\caption{\ac{md} signatures for three different activities obtained with a phase-locked system and with JUMP.}
		\label{fig:qualitative-md-locked}
	\end{center}
 \vspace{-0.3cm}
\end{figure*}

\begin{figure*}[t!]
     \centering
     \subcaptionbox{\texttt{Env2} with reflector.\label{fig:setup-refl}}[6.8cm]{\includegraphics[width=6.5cm]{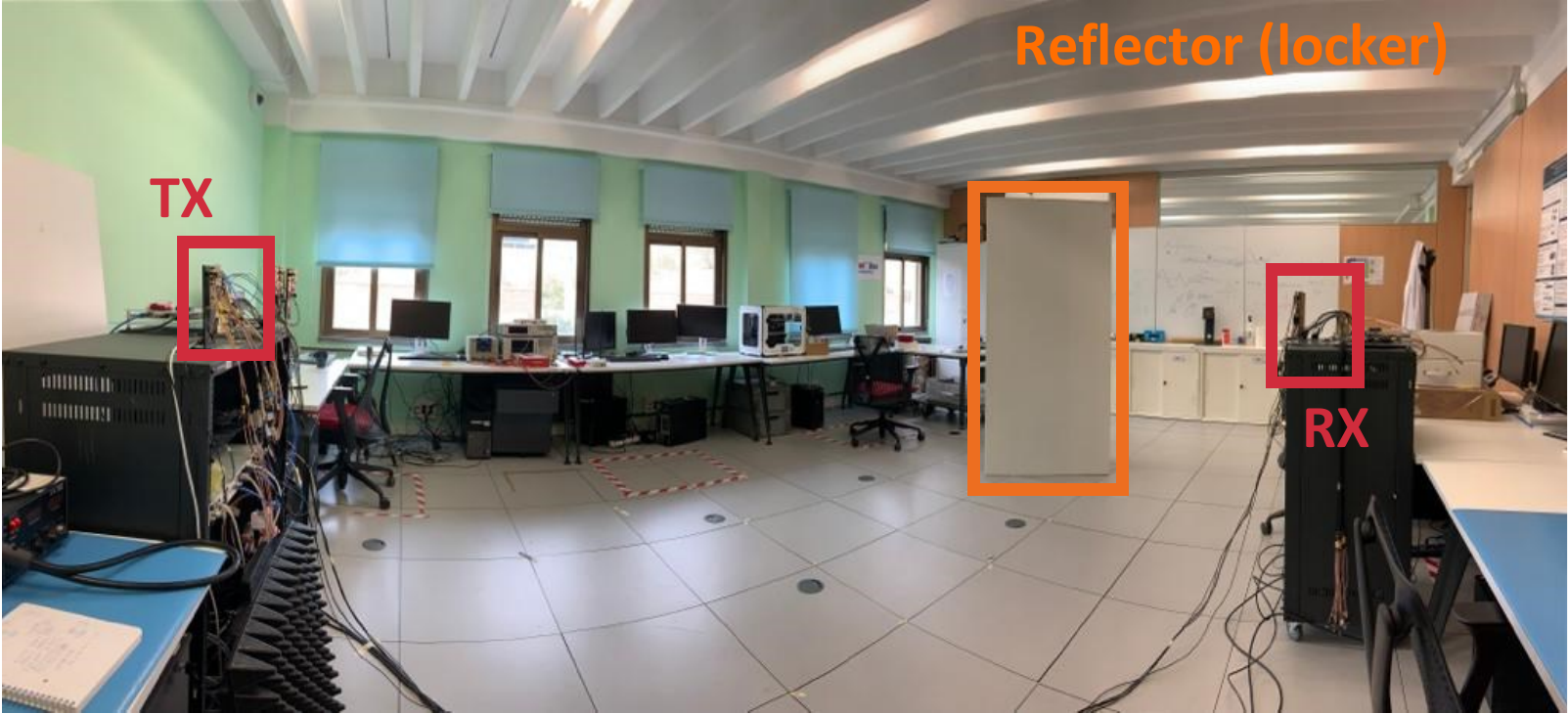}}
     \subcaptionbox{W/o. reflector.\label{fig:md-norefl}}[5cm]{\includegraphics[width=5cm]{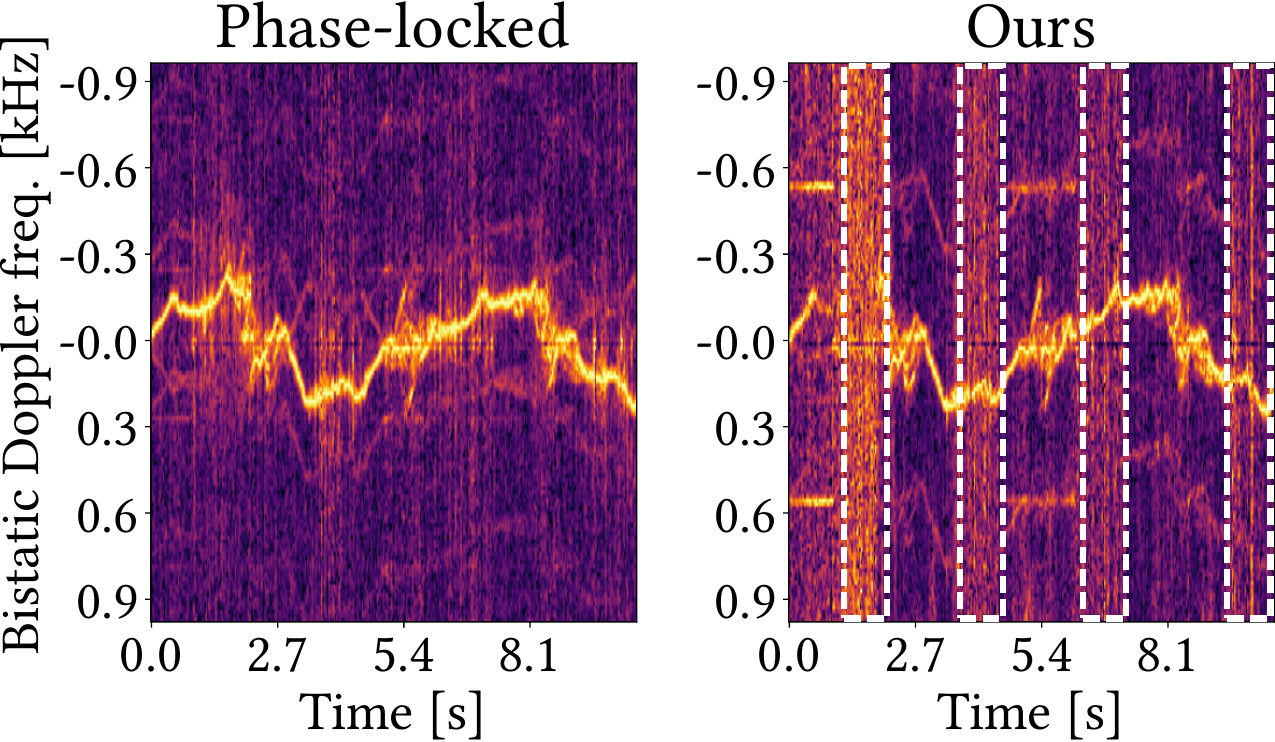}}
     \subcaptionbox{W. reflector (4.8 m).\label{fig:md-refl}}[5cm]{\includegraphics[width=5cm]{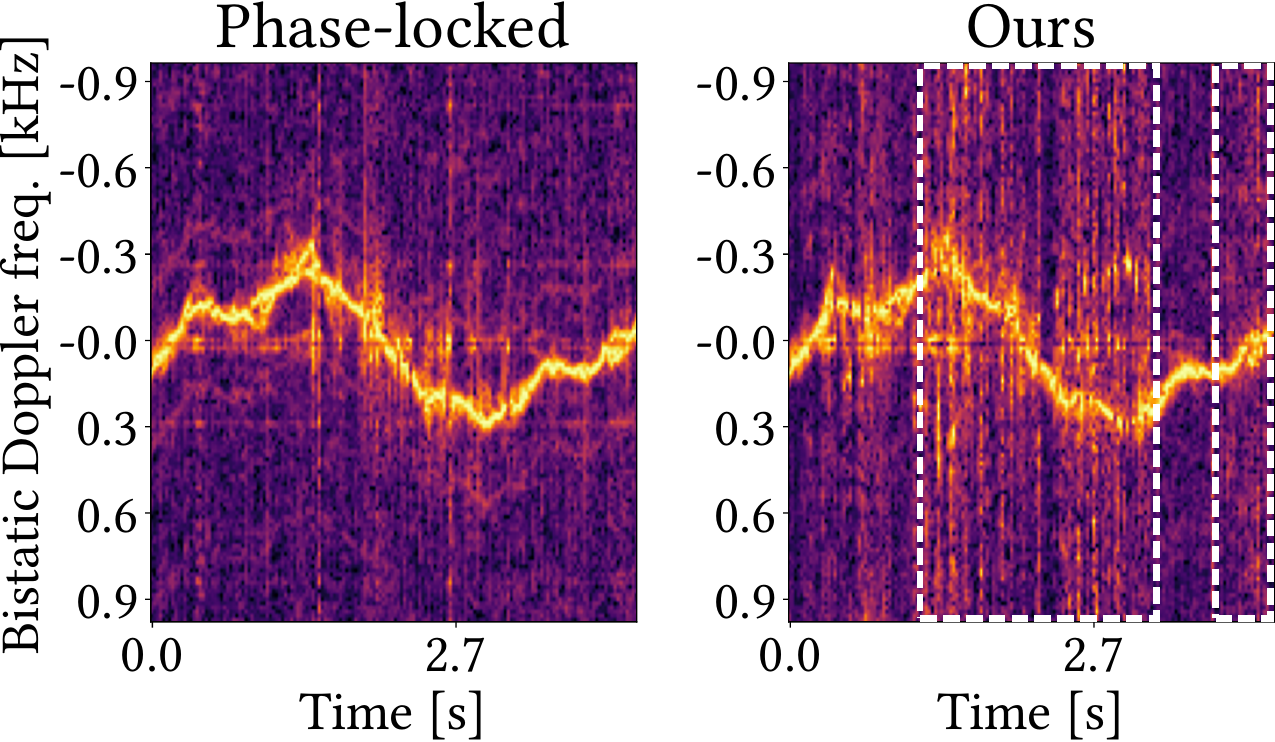}}
     \caption{\ac{md} reconstruction with intermittent LOS/NLOS. NLOS regions are enclosed in white dashed rectangles}
     \label{fig:los-nlos-mdcomp}
     \vspace{-0.3cm}
\end{figure*}
\begin{figure}
     \centering
     \includegraphics[width=4.8cm]{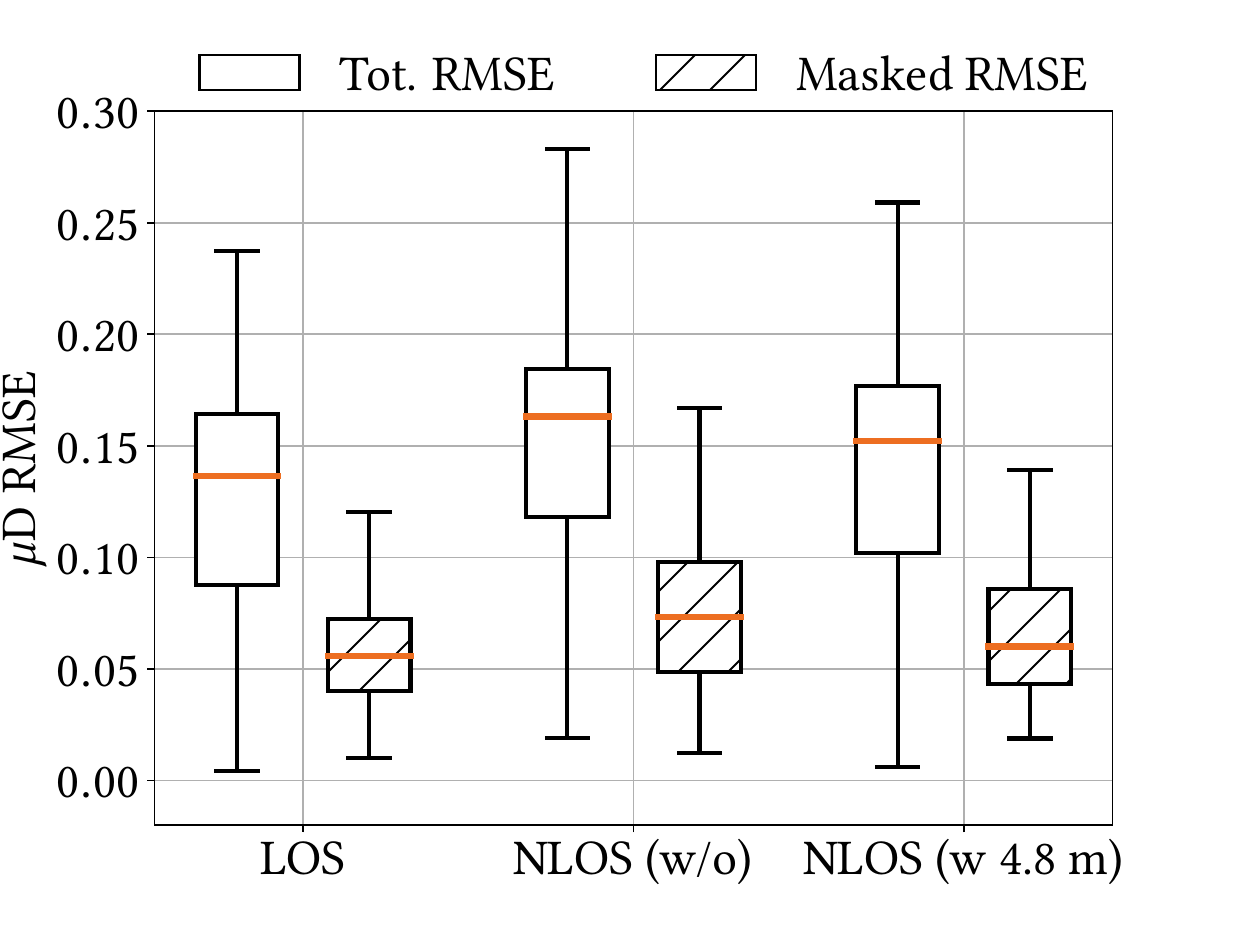}
     \includegraphics[width=2.6cm]{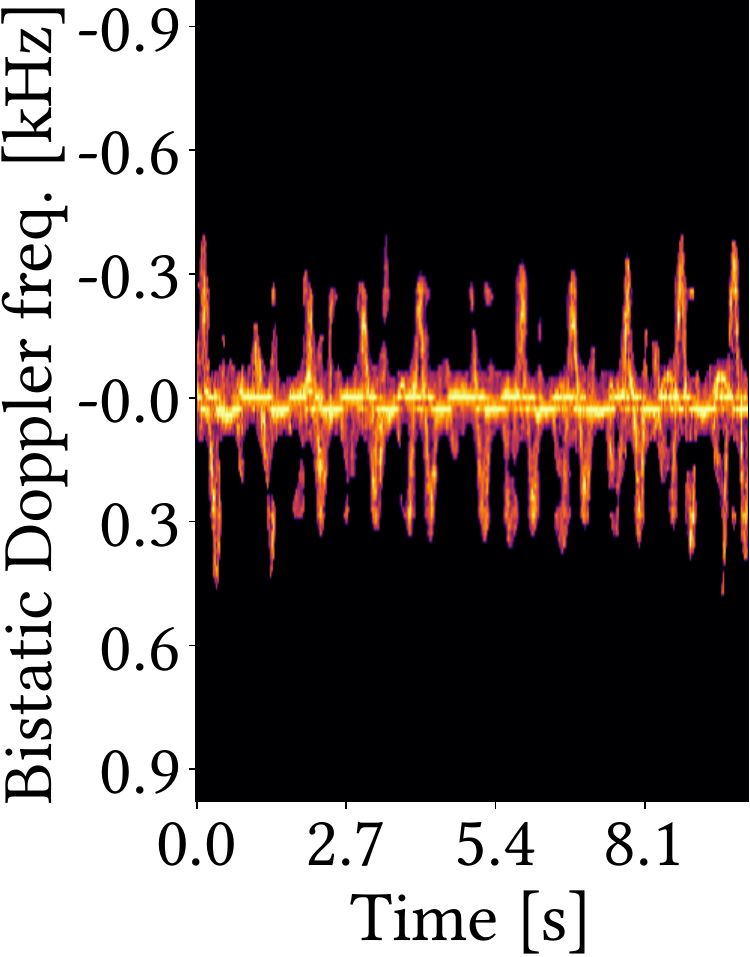}
     \caption{(Left) Normalized  \ac{rmse} distributions on the \ac{md} spectrograms, using the phase-locked ones as reference, in: LOS, NLOS with a strong reflector (metal locker) placed 4.8~m from the TX, and in NLOS with no strong reflectors. (Right) Example masked locked \ac{md} obtained from \fig{fig:md-comp-wave}.}
     \label{fig:md-quality}
     \vspace{-0.3cm}
\end{figure}
Next, we evaluate the quality of the \ac{md} signatures obtained with our \ac{cfo} removal technique in \texttt{Env2}. 
The \acp{md} shown here and in the rest of this section are obtained as $\log \mu\mathrm{D}[w,q]$, and normalized frame-wise in the interval $[0, 1]$. Darker colors correspond to lower energy levels, while lighter ones represent high energy.
We show an example result in \fig{fig:multitarget}, with two subjects, one sitting down and standing up (Track 1) and one walking (Track 2). 
Notice that the \ac{md} signatures after the removal of the \ac{cfo} contain a strong center frequency due to the torso's movement and sidebands due to the limbs. 
To provide a ground-truth reference for the \ac{md} spectrograms, we co-locate JUMP's receiver RX1 and the phase-locked RX4 at coordinates $(1.8, 3.3)$~m. 

\subsubsection{Comparison with a phase-locked system} In \fig{fig:qualitative-md-locked}, the \ac{md} of our system is compared with the phase-locked system for three different activities: (a)~walking, (b)~sitting/standing, and (c)~hand gestures. These tests are performed in a \ac{los} setting. The cleaned \ac{md} reflects the phase-locked one very accurately, even for fine-grained Doppler shifts involved in hand gestures.
To quantitatively evaluate the difference between the phase-locked spectra and JUMP's ones, we compute the frame-wise, \textit{masked} \ac{rmse} of the spectrograms, after temporally aligning them. A direct computation of the \ac{rmse} would not give a clear indication of the quality of the \ac{md}, as it would equally weigh differences in the background noise floor and the body \ac{md} contribution. For this reason, we apply a 2D Gaussian filter, with parameter $\sigma=2$, to the phase-locked \ac{md} to filter out background noise, as shown in \fig{fig:md-quality} (right). Then we restrict the \ac{rmse} computation to the spectrogram elements in which the filtered spectrogram lies over a threshold of $0.45$.
The \ac{rmse} distribution over all the collected frames is reported in \fig{fig:md-quality} (\ac{los}). For completeness, we also report the standard \ac{rmse}, calculated over the entire spectrogram.

\begin{figure*}[t!]
\centering
     \subcaptionbox{Walking.\label{fig:test2-multist}}[3cm]{\includegraphics[width=2.8cm]{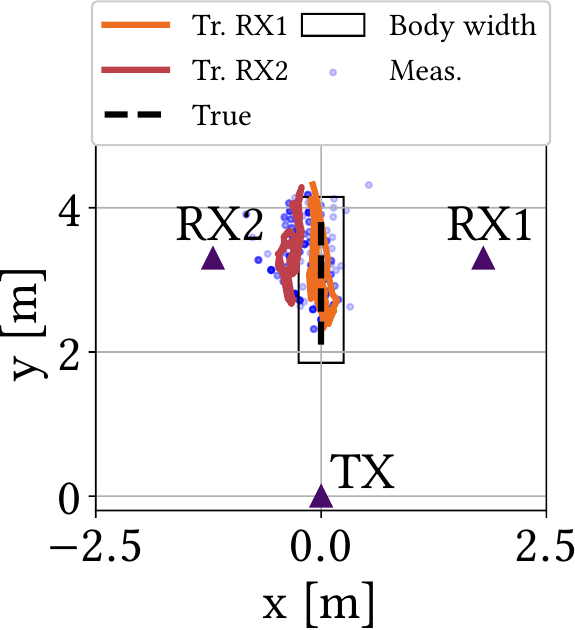}}
     \subcaptionbox{Walking \ac{md}.\label{fig:md-multist-1}}[5.3cm]{\includegraphics[width=5.3cm]{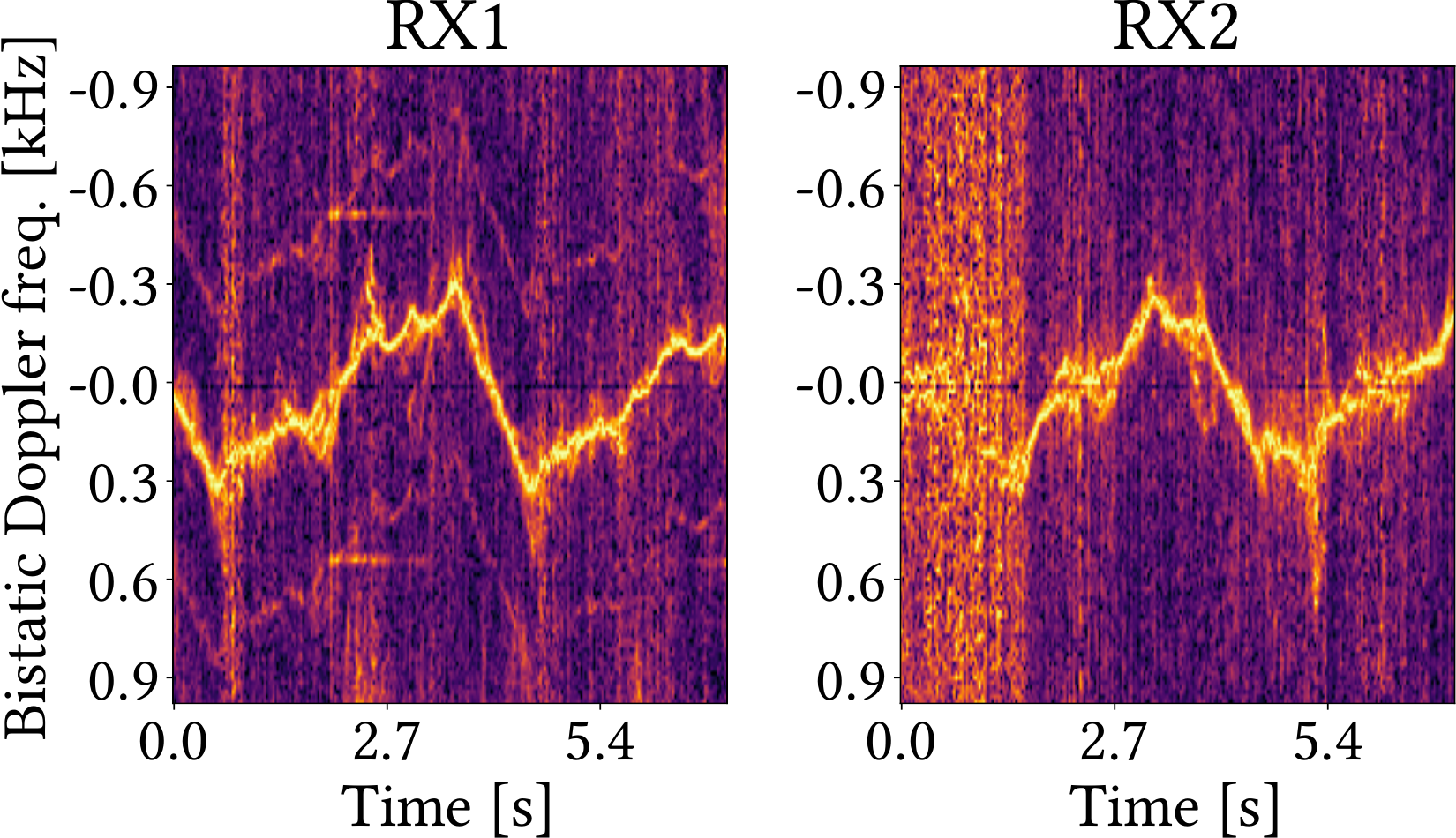}}
     \subcaptionbox{Sit/stand.\label{fig:multist-track-sit-stand}}[3cm]{\includegraphics[width=2.8cm]{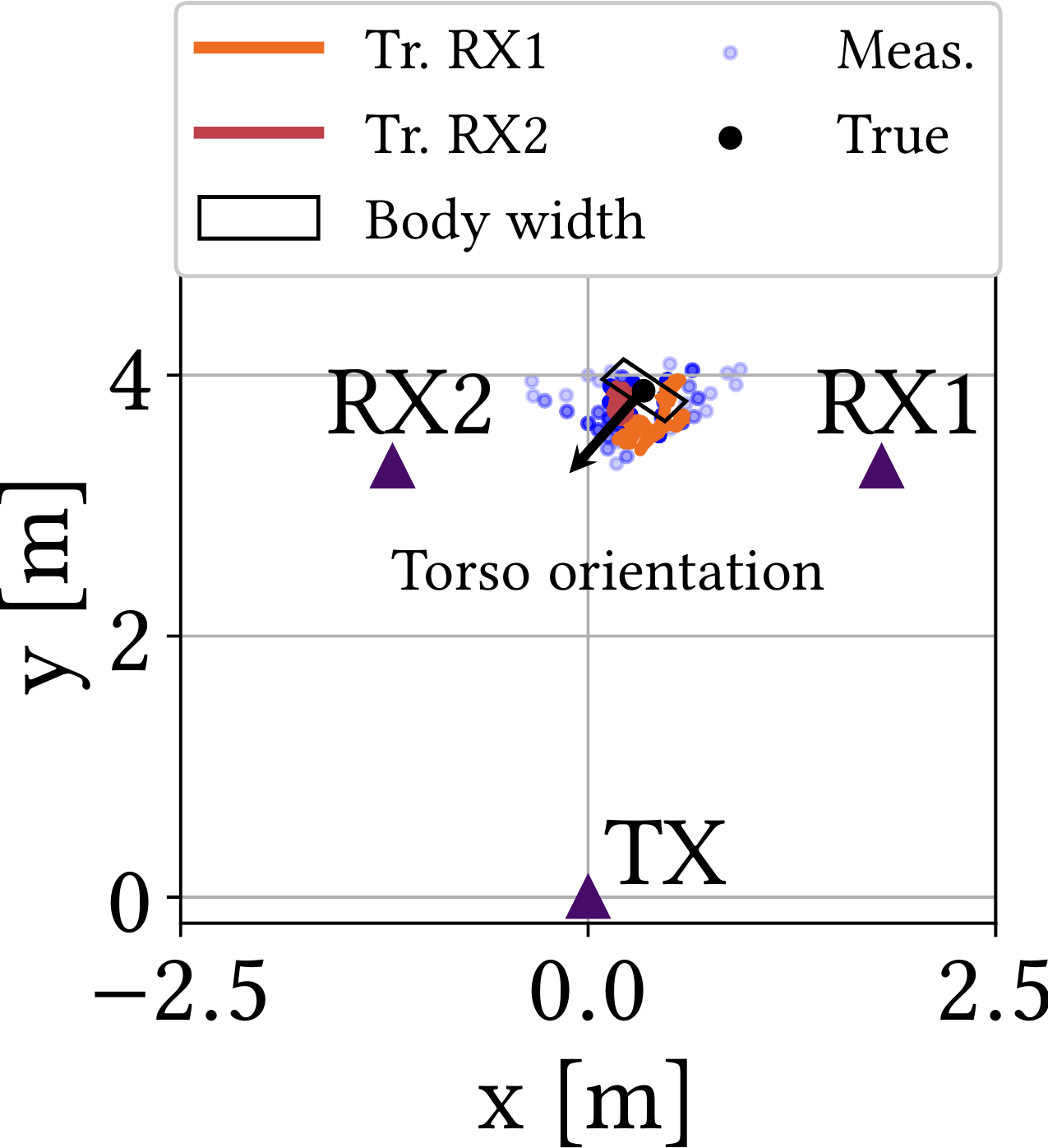}}
     \subcaptionbox{Sit/stand \ac{md}.\label{fig:multist-md-sit-stand}}[5.3cm]{\includegraphics[width=5.3cm]{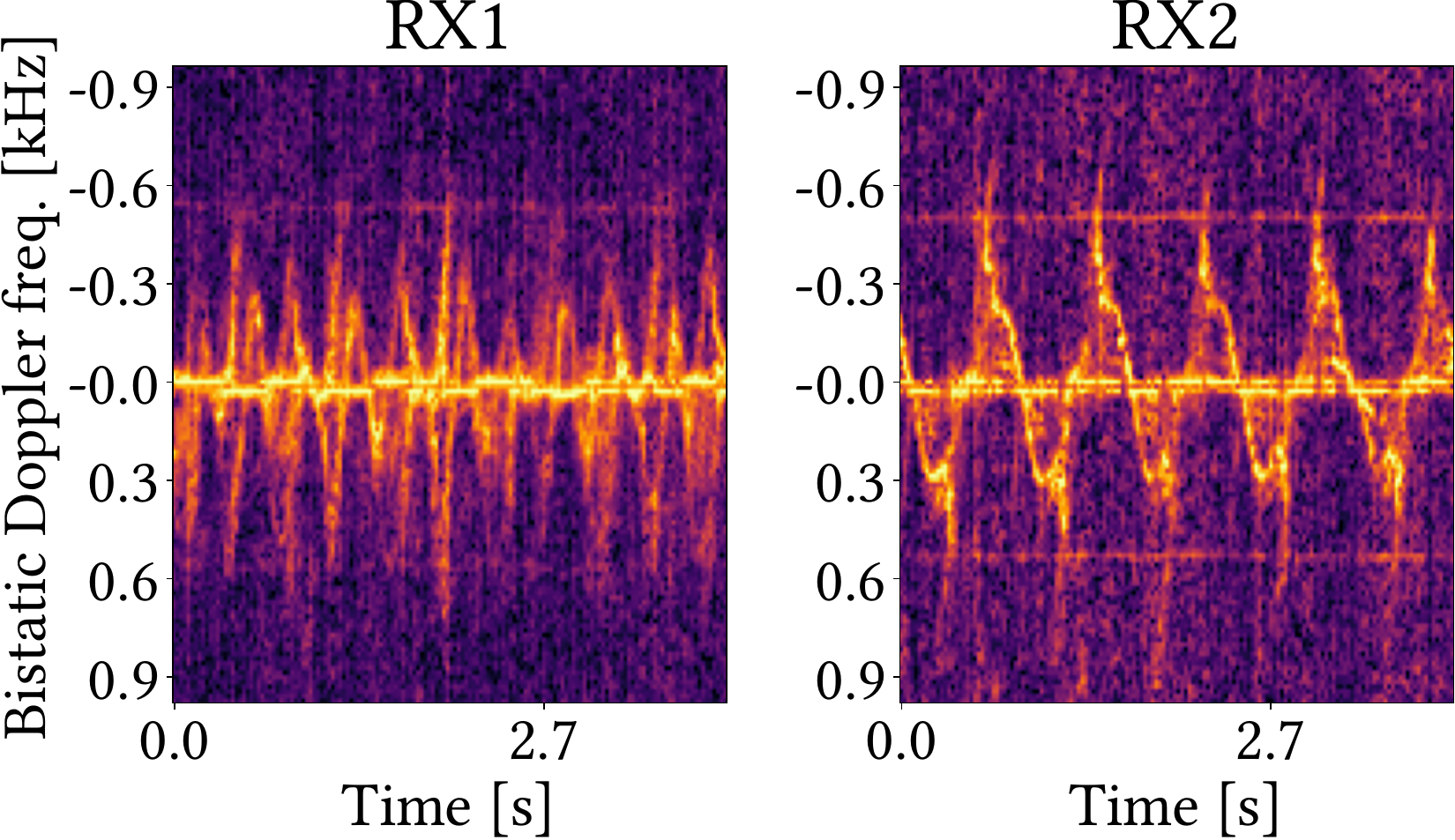}}
     \caption{Independent multistatic tracking and \ac{md} extraction using multiple RXs.}
     \label{fig:multist-md}
     \vspace{-0.5cm}
\end{figure*}

\subsubsection{Impact of \ac{nlos}} In \ac{nlos} situations, our algorithm removes the \ac{cfo} using the strongest static multipath reflection as a reference. Depending on the corresponding signal's strength and reflector's location, the resulting \ac{md} spectrum can be of higher or lower quality. In \fig{fig:los-nlos-mdcomp}, we show two \ac{md} spectrogram examples obtained in \texttt{Env2} using the same setup described above. We intermittently block this evaluation's \ac{los} link using the absorbing panel. The \ac{nlos} regions in the \ac{md} are highlighted using white dashed rectangles. In \fig{fig:md-norefl}, we used the standard \texttt{Env2} setup, without reflectors close to the testbed devices, while in \fig{fig:md-refl}, we placed a metal locker at $4.8$~m in front of the TX. In the first case, the \ac{md} reconstruction is particularly challenging, as the algorithm's reference multipath reflections are weak due to the large distance of the reflectors. Despite this, our method successfully recovers the \ac{md} in 3 out of 4 \ac{los} occlusion events, while the spectrum appears corrupted during the first one. We verified, as exemplified in \fig{fig:md-refl}, that when the metal locker is present on the scene, the \ac{md} corruption no longer occurs, as the algorithm can always find a reliable and strong reflection for removing the \ac{cfo}. 
Quantitative results for this scenario are reported in \fig{fig:md-quality},  without (w/o) and with (w) the reflector. In the worst case of \ac{nlos} without reflectors, JUMP obtains a normalized \ac{rmse} of $0.07$.

\subsection{Multistatic scenario}\label{sec:multistatic}

The proposed method easily scales to multiple RXs, allowing thorough and robust sensing of the environment. In \fig{fig:multist-md}, we show two measurement sequences obtained by concurrently operating RX1 and RX2 in \texttt{Env2}, with a subject walking (a-b) and performing hand gestures (c-d).
In the first case, the subject starts moving close to the TX-RX2 \ac{los}, thus yielding a noisy \ac{md}. However, the spectrum obtained from RX1 does not show noisy regions thanks to the better position of RX1 to capture the movement. 
In addition, the shape of the \ac{md} is slightly different due to the different viewpoints for RX1-2. This effect is captured even more clearly in \fig{fig:multist-md-sit-stand}, where the subject is waving his arms in front of RX1 and RX2, with the torso oriented midway between TX and RX2. 
In this position, the full movement of the arms is not clearly visible to RX1, as it is blocked by the subject's back. 
This results in very different \ac{md} patterns that provide richer information about the movement.


\section{Discussion and future work}\label{sec:discussion}
\subsubsection{Multistatic data fusion} The results provided in \secref{sec:multistatic} pave the way for multistatic data fusion across multiple receivers. As shown in \fig{fig:multist-md}, each JUMP receiver returns an independent view of the physical environment to be monitored. Leveraging multiple receivers would then provide a means to improve tracking accuracy, e.g., by combining their estimated tracks to solve possible occlusion events. Moreover, multiple receivers usually have different observation angles for the same targets. The fusion of \ac{md} spectrum features coming from such diverse viewpoints allows for enhanced diversity and, overall, for an improved understanding of the underlying movement, especially for extended targets with many moving parts. Target identification or motion recognition algorithms can exploit such enhanced feature representations to boost their performance.
 \subsubsection{\ac{nlos} operation} Our target tracking accuracy results in the presence of \ac{nlos} events show that, for indoor human sensing, our assumption that most background reflections remain constant in between subsequent tracking frames holds. Moreover, when blocked, the \ac{los} path gradually disappears across subsequent \ac{cir} estimations, yielding correct and clear correlation peaks. 
In general, however, the tradeoff between inter-packet time and the level of dynamicity in the environment plays a key role. As shown in~\cite{pegoraro2023rapid}, a minimum packet rate has to be ensured to accurately capture the Doppler spectrum of the physical targets, to avoid aliasing and low resolution.
In the case of very challenging dynamic environments, with few static reflections, the JUMP correlation approach could be replaced by optimization-based association mechanisms between multipath reflections appearing in different \ac{cir} estimates. This aspect, and the possible improvements that it may bring, are left to future investigations. 
\rev{\subsubsection{Mobility} The analysis in \secref{sec:rx-mob} shows that JUMP's \ac{cfo} removal step can be extended to also compensate for RX movement. However, this requires estimating the RX velocity vector using external sensors (e.g., an onboard accelerometer). The resulting compensation is highly sensitive to errors in such estimation, which may degrade performance in practical settings. Therefore, we identify mobility as a key challenge for future \ac{jcs} research that needs further attention. Interesting solutions may involve enhancing the accelerometer estimate using information from the wireless channel.}

\section{Concluding remarks} \label{sec:conclusion}
 The problem of integrating sensing functionalities in bistatic asynchronous communication systems is addressed in this paper. For that, we designed and prototyped JUMP, a \ac{jcs} scheme that overcomes timing and frequency offsets due to clock asynchrony following two main pathways. First, it leverages the correlation in the channel propagation paths across different packets, second, it identifies a static reference path whose phase is used to remove the frequency offset from the reflections on sensing targets. An extensive \ac{cir} measurement campaign has been carried out targeting indoor human movement sensing, using a 60~GHz IEEE 802.11ay-based implementation of the system. Results show that JUMP is competitive with synchronous full-duplex and phase-locked solutions. It achieves a median worst-case multitarget tracking error of $14$~cm and a human \ac{md} spectrum normalized reconstruction error of $0.07$ in intermittent \ac{nlos} conditions.

\bibliographystyle{IEEEtran}
\bibliography{references.bib}

\begin{IEEEbiography}[{\includegraphics[width=1in,height=1.25in,clip,keepaspectratio]{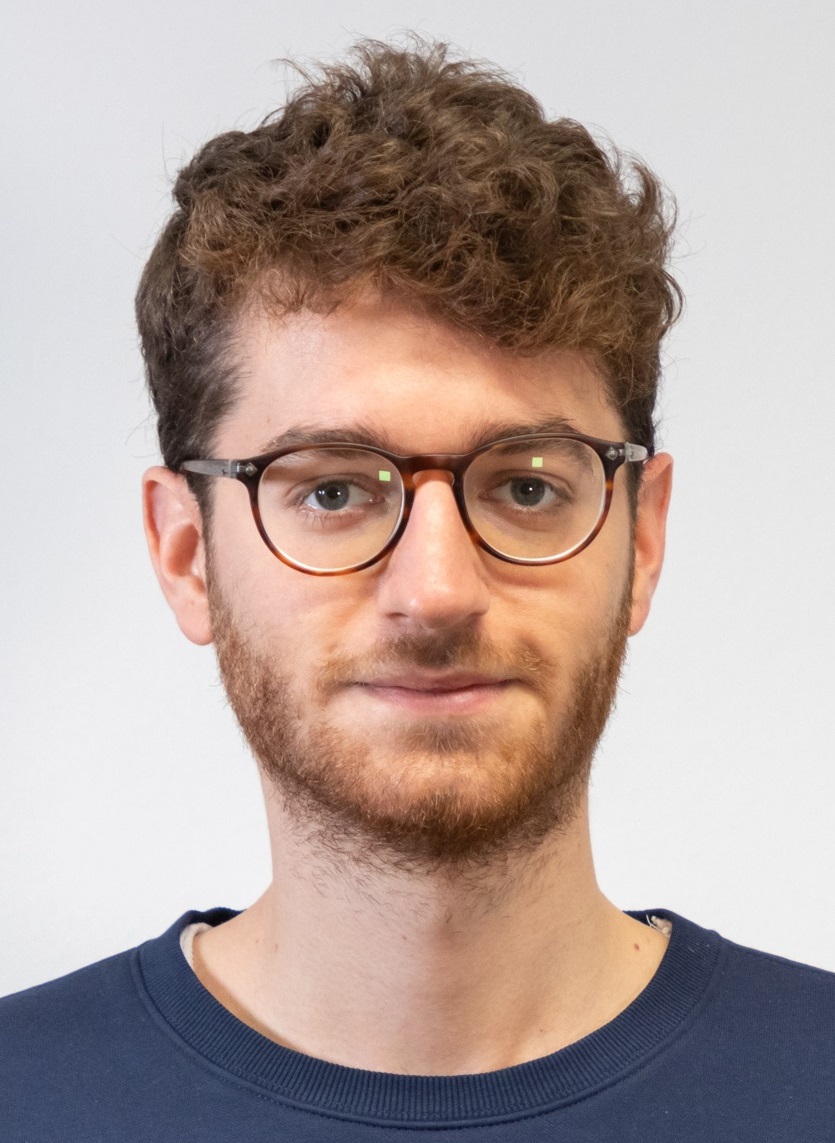}}]
{Jacopo Pegoraro} (M'23) received his Ph.D. in Information Engineering from the University of Padova, Padua, Italy, in 2023. He is currently working as a postdoctoral researcher with the SIGNET Research Group, Department of Information Engineering, in the same University. He was a visiting research scholar at the New York University, Tandon school of Engineering in 2022, working on joint communication and sensing. His research interests include signal processing, sensor fusion and machine learning with applications to \ac{mmwave} sensing.
\end{IEEEbiography}

\begin{IEEEbiography}[{\includegraphics[width=1in,height=1.25in,clip,keepaspectratio]{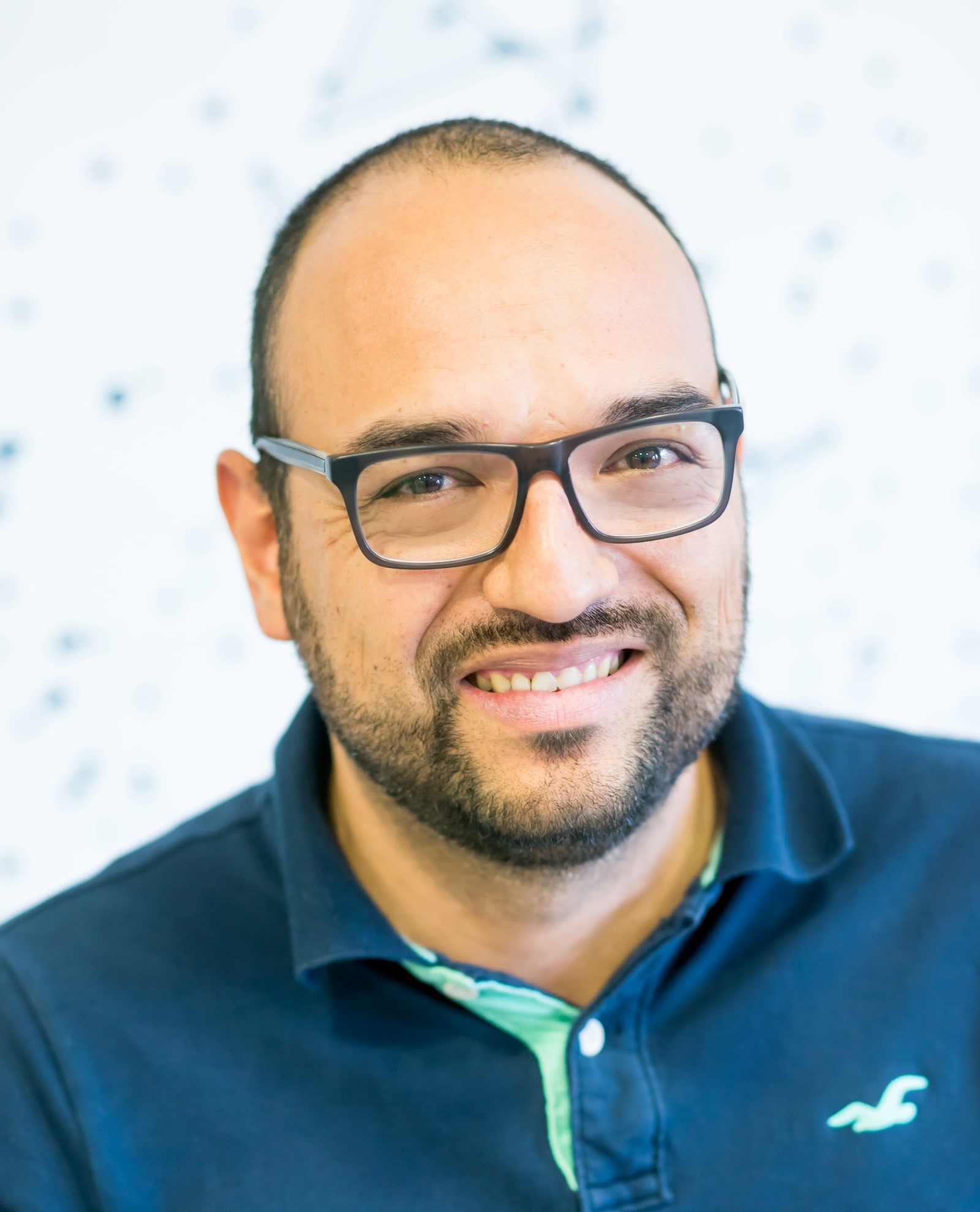}}]{Jesus O. Lacruz }
is a Research Engineer at IMDEA Networks, Spain since 2017. He received his Bachelor degree in Electrical Engineering from Universidad de Los Andes, Venezuela in 2009 and the PhD degree in Electronic Engineering from Universidad Politecnica de Valencia, Spain in 2016. His research interests lie in the design and implementation of fast signal processing algorithms for digital communication systems in FPGA devices.
\end{IEEEbiography}

\begin{IEEEbiography}[{\includegraphics[width=1in,height=1.25in, clip,keepaspectratio]{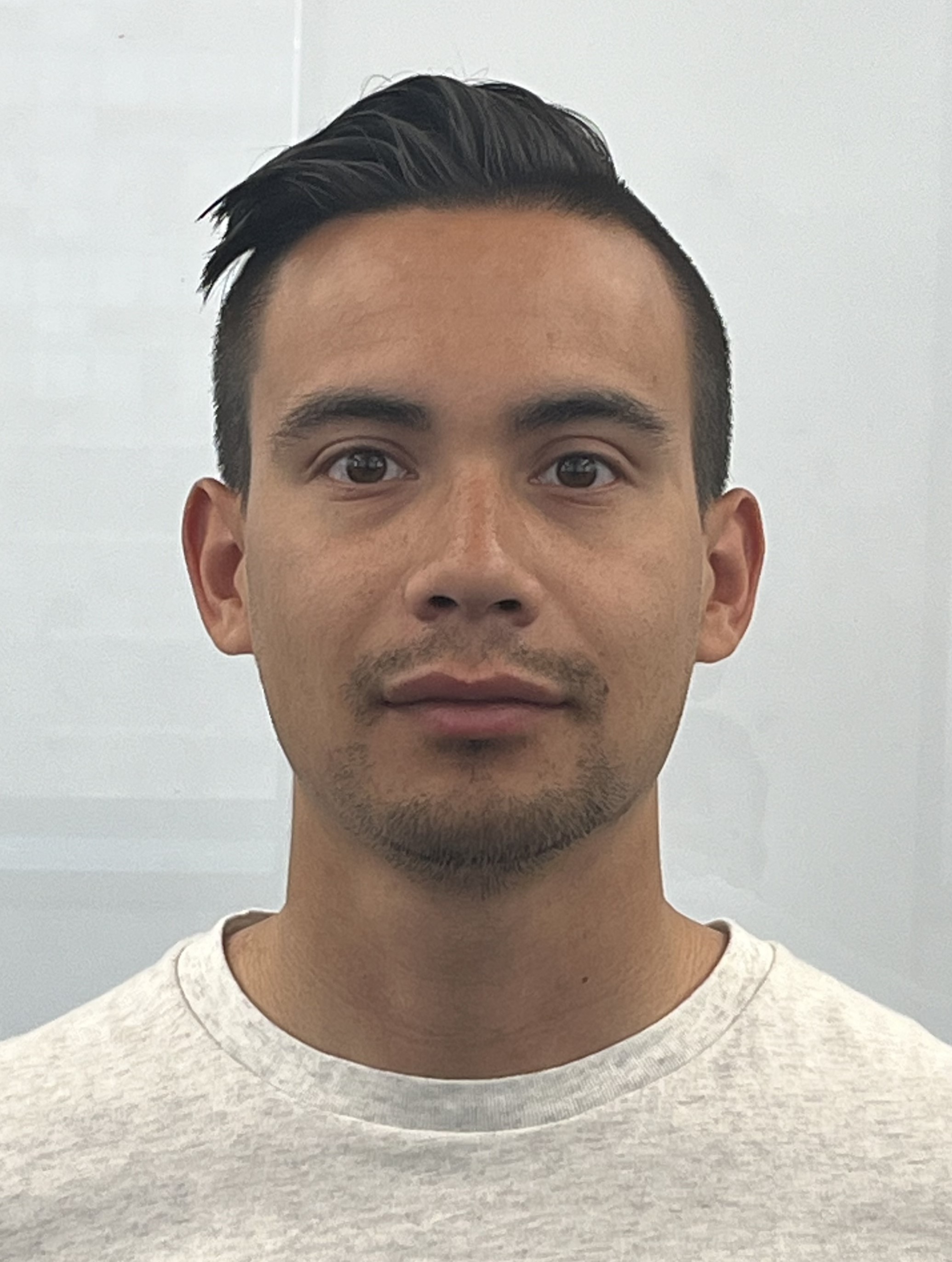}}]{Tommy Azzino}  received the B.Sc. degree in information engineering and the M.Sc. degree in telecommunications engineering from the University of Padova, Italy, in 2016 and 2019, respectively.
He is currently pursuing the Ph.D. degree in electrical and computer engineering with the NewYork University Tandon School of Engineering,Brooklyn, NY, USA, under the supervision of Prof.S. Rangan.
He held visiting research positions with the Department of Commerce,National Institute of Standards and Technology (NIST), Gaithersburg, MD,USA, in 2018, Nokia Bell Labs France, Paris, France, in 2019. 
He took part in the 2018 Edition of the Seeds for the Future Project promoted by Huawei Technologies, Shenzhen, China, in 2018. 
His research interests include 5G mobile cloud/edge computing, 5G network simulation and prototyping, MAC and network-layer resource allocation, wireless communications, and machine learning for wireless communications.

\end{IEEEbiography}

\begin{IEEEbiography}[{\includegraphics[width=1in,height=1.25in,clip,keepaspectratio]{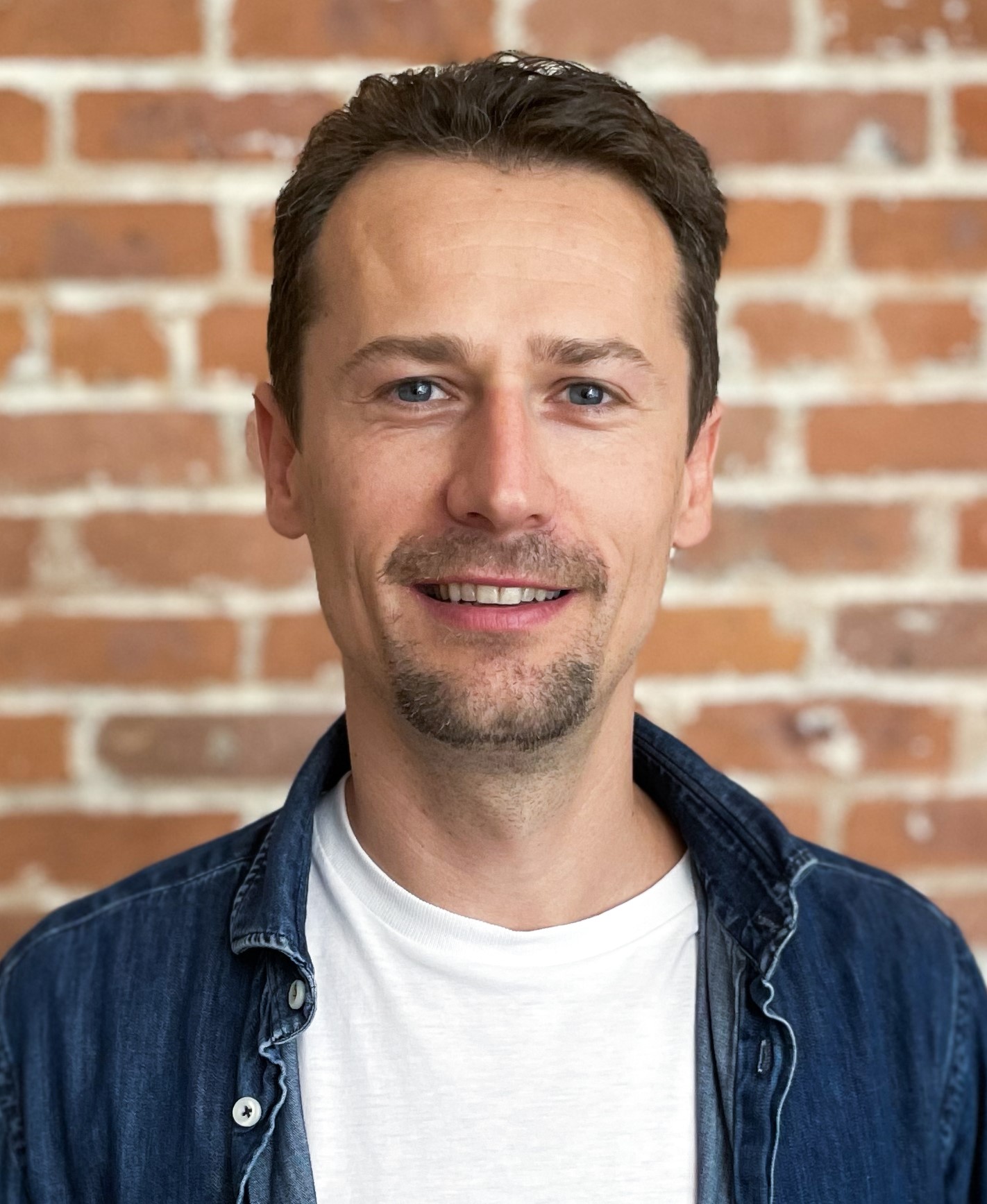}}]{Marco Mezzavilla } is Research Faculty at the NYU Tandon School of Engineering and co-founder of Pi-Radio, a spin-off company of NYU that develops mmWave/sub-THz software defined radios (SDRs). He received the B.A.Sc., the M.Sc., and the Ph.D. at the University of Padua, Italy, all in Electrical Engineering. He held visiting research positions at the NEC Network Laboratories in Heidelberg (Germany, 2009), at the Centre Tecnològic Telecomunicacions Catalunya (CTTC) in Barcelona (Spain, 2010), and at Qualcomm Research in San Diego (USA, 2012). He joined New York University (NYU) Tandon School of Engineering in 2014, where he currently leads several research projects that focus on mmWave and sub-THz radio access technologies for 5G and beyond. He served as co-chair and reviewer for several IEEE and ACM conferences, and as guest editor of various journals and magazines. His current research interests lie at the intersection between communications, cybersecurity, and robotics. He is a senior member of IEEE.
\end{IEEEbiography}

\begin{IEEEbiography}[{\includegraphics[width=1in,height=1.25in,clip,keepaspectratio]{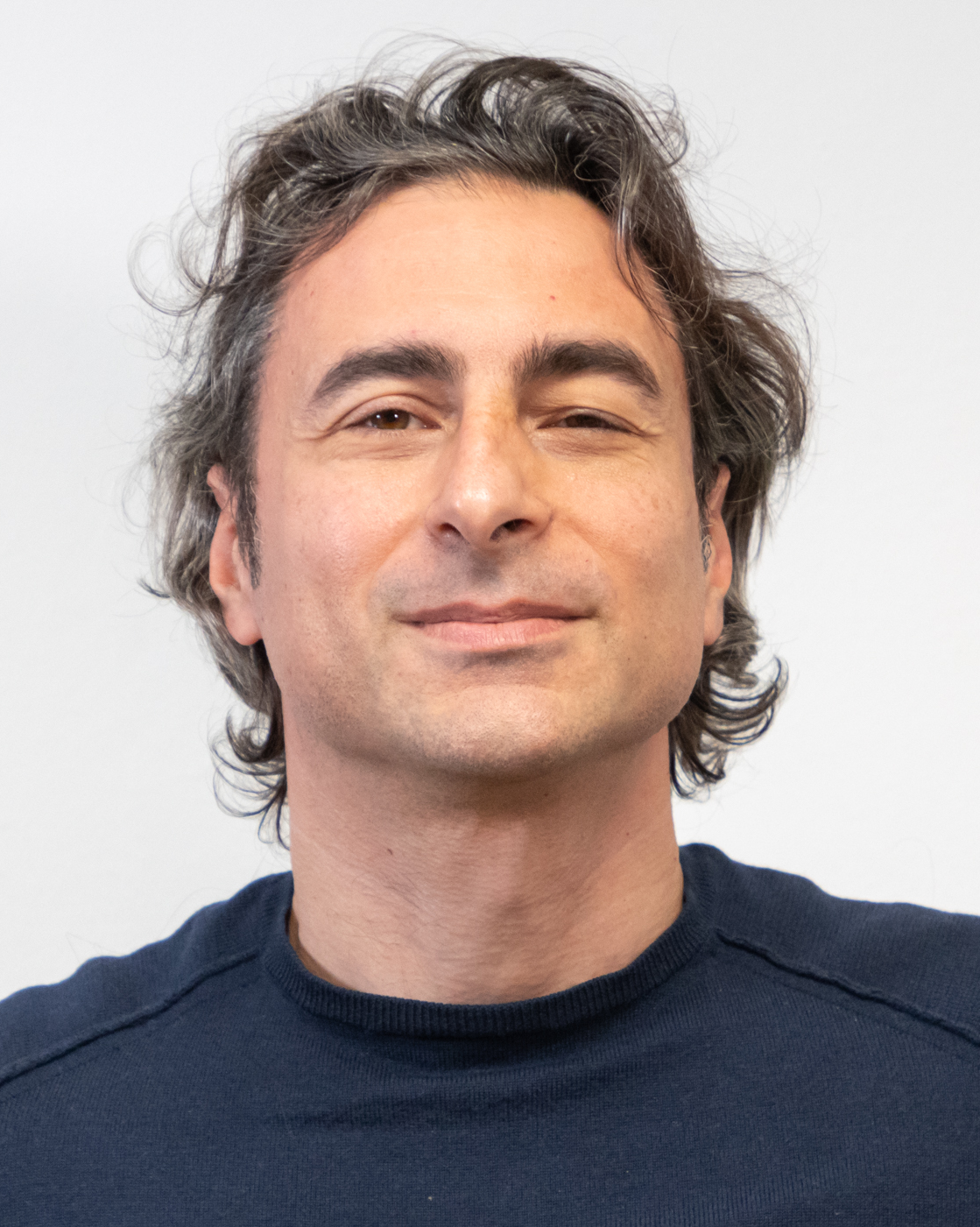}}]%
{Michele Rossi} (SM'13) is a Professor of Wireless Networks in the Department of Information Engineering (DEI) at the University of Padova (UNIPD), Italy, where is the head of the Master's Degree in ICT for internet and Multimedia (\url{http://mime.dei.unipd.it/}). He also teaches Human Data Analysis at the Data Science Master's degree at the Department of Mathematics (DM) at UNIPD (\url{https://datascience.math.unipd.it/}). Since 2017, he has been the Director of the DEI/IEEE Summer School of Information Engineering (\url{http://ssie.dei.unipd.it/}). His research interests lie broadly in wireless sensing systems, green mobile networks, edge and wearable computing. Over the years, he has been involved in several EU projects on wireless sensing and IoT and has collaborated with major companies such as Ericsson, DOCOMO, Samsung and INTEL. His research is currently supported by the European Commission through the H2020 projects MINTS (no. 861222) on ``mmWave networking and sensing'' and GREENEDGE (no. 953775) on ``green edge computing for mobile networks'' (project coordinator). Dr. Rossi has been the recipient of seven best paper awards from the IEEE and currently serves on the Editorial Boards of the IEEE Transactions on Mobile Computing, and of the Open Journal of the Communications Society.
\end{IEEEbiography}

\begin{IEEEbiography}[{\includegraphics[width=1in,height=1.25in,clip,keepaspectratio]{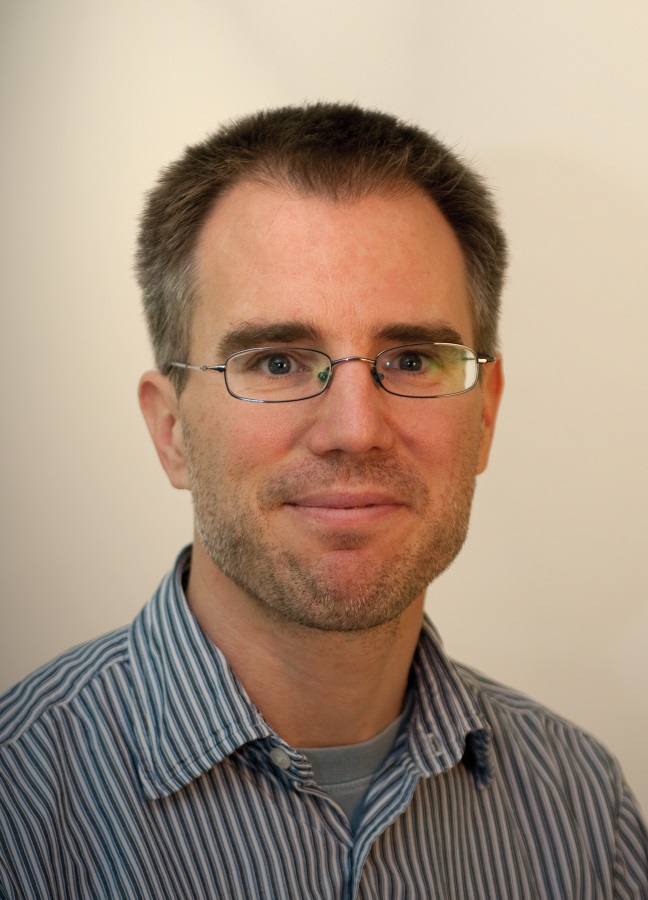}}]{Joerg Widmer}
(M'06-SM'10-F'20) 
is Research Professor and Research Director of IMDEA Networks in Madrid, Spain. Before, he held positions at DOCOMO Euro-Labs in Munich, Germany and EPFL, Switzerland. He was a visiting researcher at the International Computer Science Institute in Berkeley, USA, University College London, UK, and TU Darmstadt, Germany. His research focuses on wireless networks, ranging from extremely high frequency millimeter-wave communication and MAC layer design to mobile network architectures. Joerg Widmer authored more than 150 conference and journal papers and three IETF RFCs, and holds 13 patents. He was awarded an ERC consolidator grant, the Friedrich Wilhelm Bessel Research Award of the Alexander von Humboldt Foundation, a Mercator Fellowship of the German Research Foundation, a Spanish Ramon y Cajal grant, as well as eight best paper awards. He is an IEEE Fellow and Distinguished Member of the ACM.
\end{IEEEbiography}

\begin{IEEEbiography}[{\includegraphics[width=1in,height=1.25in,clip,keepaspectratio]{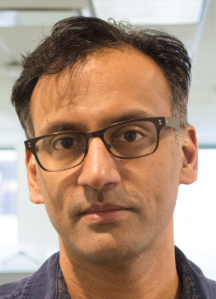}}]{Sundeep Rangan} received the B.A.Sc. at the University of Waterloo, Canada and the M.Sc. and Ph.D. at the University of California, Berkeley, all in Electrical Engineering. He has held postdoctoral appointments at the University of Michigan, Ann Arbor and Bell Labs. In 2000, he co-founded (with four others) Flarion Technologies, a spin-off of Bell Labs, that developed Flash OFDM, the first cellular OFDM data system and pre-cursor to 4G cellular systems including LTE and WiMAX. In 2006, Flarion was acquired by Qualcomm Technologies. Dr. Rangan was a Senior Director of Engineering at Qualcomm involved in OFDM infrastructure products. He joined NYU Tandon (formerly NYU Polytechnic) in 2010 where he is currently a Professor of Electrical and Computer Engineering. He is a Fellow of the IEEE and the Associate Director of NYU WIRELESS, an industry-academic research center.
\end{IEEEbiography}

\vfill

\end{document}

%% file: ACRONYMS.tex
\acrodef{prop}[\textit{MIMORPH}]{MIMO Radio Platform for Heterogeneous wireless systems}

\acrodef{3gpp}[3GPP]{3rd Generation Partnership Project}
\acrodef{abft}[A-BFT]{Association Beamforming Training}
\acrodef{ack}[ACK]{Acknowledge}
\acrodef{adc}[ADC]{Analog-to-Digital Converter}
\acrodef{af}[AF]{Ambiguity Function}
\acrodef{dac}[DAC]{Digital-to-Analog Converter}
\acrodef{aoa}[AoA]{Angle of Arrival}
\acrodef{aod}[AoD]{Angle of Departure}
\acrodef{ap}[AP]{Access Point}
\acrodef{amc}[AMC]{Advanced Mezzanine Card}
\acrodef{awv}[AWV]{Antenna Wave Vector}
\acrodef{axi}[AXI]{Advanced eXtensible Interface}
\acrodef{ber}[BER]{Bit Error Rate}
\acrodef{bft}[BFT]{Beamforming Training}
\acrodef{bp}[BP]{Beam Pattern}
\acrodef{bpsk}[BPSK]{Binary Phase-Shift Keying}
\acrodef{brp}[BRP]{Beam Refinement Phase}
\acrodef{cs}[CS]{Compressed Sensing}
\acrodef{cdf}[CDF]{Cumulative Distribution Function}
\acrodef{cef}[CEF]{Channel Estimation Field}
\acrodef{cacfar}[CA-CFAR]{Cell-Averaging Constant False-Alarm Rate}
\acrodef{cacc}[CACC]{Cross-Antenna Cross-Correlation}
\acrodef{casr}[CASR]{Cross-Antenna Signal Ratio}
\acrodef{cfo}[CFO]{Carrier Frequency Offset}
\acrodef{cir}[CIR]{Channel Impulse Response}
\acrodef{csi}[CSI]{Channel State Information}
\acrodef{csirs}[CSI-RS]{CSI-Reference Signal}
\acrodef{cv}[CV]{Constant Velocity}
\acrodef{cnn}[CNN]{Convolutional Neural Network}
\acrodef{cots}[COTS]{Commercial-Off-The-Shelf}
\acrodef{dl}[DL]{Deep Learning}
\acrodef{dma}[DMA]{Direct Memory Access}
\acrodef{dmg}[DMG]{Directional Multi Gigabit}
\acrodef{dti}[DTI]{Data Transfer Interval}
\acrodef{dft}[DFT]{Discrete Fourier Transform}
\acrodef{dtft}[DTFT]{Discrete-Time Fourier Transform}
\acrodef{edmg}[EDMG]{Enhanced Directional Multi Gigabit}
\acrodef{ekf}[EKF]{Extended Kalman Filter}
\acrodef{elu}[ELU]{Exponential-Linear Unit}
\acrodef{fmcw}[FMCW]{Frequency-Modulated Continuous-Wave}
\acrodef{fov}[FOV]{Field-of-View}
\acrodef{ft}[FT]{Fourier Transform}
\acrodef{fo}[FO]{Frequency Offset}
\acrodef{fpga}[FPGA]{Field Programmable Gate Array}
\acrodef{gpio}[GPIO]{General Purpose Input/Output}
\acrodef{gsps}[GSPS]{Giga-Samples per Second}
\acrodef{gps}[GPS]{Global Positioning Systems}
\acrodef{har}[HAR]{Human Activity Recognition}
\acrodef{ht}[HT]{High Throughput}
\acrodef{if}[IF]{Intermediate Frequency}
\acrodef{ifs}[IFS]{Inter-Frame Spacing}
\acrodef{iht}[IHT]{Iterative Hard Thresholding}
\acrodef{imu}[IMU]{Inertial Measurement Unit}
\acrodef{isac}[ISAC]{Integrated Sensing And Communication}
\acrodef{isi}[ISI]{Inter-Symbol Interference}
\acrodef{jcs}[JCS]{Joint Communication \& Sensing}
\acrodef{nnjpdaf}[NN-JPDAF]{Nearest-Neighbors Joint Probabilistic Data Association Filter}
\acrodef{los}[LOS]{Line-of-Sight}
\acrodef{lbm}[LBM]{Loop-Back Memory}
\acrodef{mae}[MAE]{Mean Absolute Error}
\acrodef{mcs}[MCS]{Modulation and Coding Scheme}
\acrodef{md}[$\mu$D]{micro-Doppler}
\acrodef{mimo}[MIMO]{Multiple Input Multiple Output}
\acrodef{mmwave}[mmWave]{Millimeter-Wave}
\acrodef{msps}[MSPS]{Mega-Samples per Second}
\acrodef{mu}[MU]{Multiple User}
\acrodef{music}[MUSIC]{MUltiple SIgnal Classification}
\acrodef{nac}[NAC]{Normalized Auto Correlation}
\acrodef{nco}[NCO]{Numerical Controlled Oscillator}
\acrodef{nlos}[NLOS]{Non-Line-of-Sight}
\acrodef{ofdm}[OFDM]{Orthogonal Frequency Division Multiplexing}
\acrodef{per}[PER]{Packet Error Rate}
\acrodef{phy}[PHY]{Physical Layer}
\acrodef{pl}[PL]{Programmable Logic}
\acrodef{pov}[POV]{Point-of-View}
\acrodef{ps}[PS]{Processing System}
\acrodef{psdu}[PSDU]{Physical layer Service Data Unit}
\acrodef{rcs}[RCS]{Radar Cross Section}
\acrodef{rf}[RF]{Radio Frequency}
\acrodef{rfsoc}[RFSoC]{Radio Frequency System on a Chip}
\acrodef{rmse}[RMSE]{Root Mean Squared Error}
\acrodef{rss}[RSS]{Received Signal Strength}
\acrodef{rom}[ROM]{Read Only Memories}
\acrodef{sc}[SC]{Single Carrier}
\acrodef{sdr}[SDR]{Software Defined Radio}
\acrodef{siso}[SISO]{Single Input Single Output}
\acrodef{sls}[SLS]{Sector Level Sweep}
\acrodef{snr}[SNR]{Signal-to-Noise Ratio}
\acrodef{soc}[SoC]{System on a Chip}
\acrodef{spb}[SPB]{Signal Processing Blocks}
\acrodef{srrc}[SRRC]{Square-Root-Raised-Cosine}
\acrodef{ssb}[SSB]{Synchronization Signal Block}
\acrodef{ssr}[SSR]{Super Sample Rate}
\acrodef{sta}[STA]{Station}
\acrodef{stf}[STF]{Short Training Field}
\acrodef{stft}[STFT]{Short Time Fourier Transform}
\acrodef{su}[SU]{Single User}
\acrodef{tf}[TF]{Time-Frequency}
\acrodef{toa}[ToA]{Time of Arrival}
\acrodef{to}[TO]{Timing Offset}
\acrodef{usrp}[USRP]{Universal Software Radio Peripheral}
\acrodef{vht}[VHT]{Very High Throughput}
\acrodef{wlan}[WLAN]{Wireless Local Area Network}
\acrodef{wvd}[WVD]{Wigner-Ville Distribution}

%% file: figures/overhead.tex
\begin{tikzpicture}
\tikzstyle{every node}=[font=\small]
\definecolor{black003}{RGB}{0,0,3}
\definecolor{chocolate23711033}{RGB}{237,110,33}
\definecolor{darkgray176}{RGB}{176,176,176}
\definecolor{indianred1896475}{RGB}{189,64,75}
\definecolor{indigo7111104}{RGB}{71,11,104}
\definecolor{lightgray204}{RGB}{204,204,204}

\begin{axis}[
width=0.95\columnwidth,
height=5.3cm,
legend cell align={left},
legend columns=2,
legend style={
  fill opacity=0.8,
  draw opacity=1,
  text opacity=1,
  at={(0.97,0.03)},
  anchor=south east,
  draw=lightgray204
},
log basis y={10},
tick align=outside,
tick pos=left,
x grid style={darkgray176},
xlabel={No. TRN units},
xmajorgrids,
xmin=1, xmax=12,
xminorgrids,
xtick style={color=black},
xtick={1,2,3,4,5,6,7,8,9,10,11,12},
xticklabels={
  \(\displaystyle {1}\),
  \(\displaystyle {2}\),
  \(\displaystyle {3}\),
  \(\displaystyle {4}\),
  \(\displaystyle {5}\),
  \(\displaystyle {6}\),
  \(\displaystyle {7}\),
  \(\displaystyle {8}\),
  \(\displaystyle {9}\),
  \(\displaystyle {10}\),
  \(\displaystyle {11}\),
  \(\displaystyle {12}\)
},
y grid style={darkgray176},
ylabel={Overhead [\%]},
ymajorgrids,
ymin=0.002, ymax=114.575646433038,
yminorgrids,
ymode=log,
ytick style={color=black}
]
\addplot [thick, indigo7111104, mark=*, mark size=2, mark options={solid,draw=white}]
table {%
1 2.49480271339417
2 4.86815452575684
3 7.12871313095093
4 9.28433227539062
5 11.342155456543
6 13.3086881637573
7 15.1898746490479
8 16.9911518096924
9 18.7175025939941
10 20.373514175415
11 21.9633960723877
12 23.4910297393799
};
\addlegendentry{PSDU: 4 kB}
\addplot [thick, chocolate23711033, mark=triangle*, mark size=3, mark options={solid,draw=white}]
table {%
1 0.189603388309479
2 0.378489196300507
3 0.566661477088928
4 0.754124104976654
5 0.940881252288818
6 1.12693679332733
7 1.31229507923126
8 1.49695932865143
9 1.68093395233154
10 1.86422276496887
11 2.04682874679565
12 2.22875714302063
};
\addlegendentry{PSDU: 66 kB}
\addplot [thick, indianred1896475, mark=square*, mark size=2, mark options={solid,draw=white}]
table {%
1 0.0478908121585846
2 0.0957357659935951
3 0.143534943461418
4 0.19128842651844
5 0.238996222615242
6 0.286658406257629
7 0.334275156259537
8 0.381846368312836
9 0.429372221231461
10 0.476852774620056
11 0.5242879986763
12 0.571678161621094
};
\addlegendentry{PSDU: 262 kB}
\addplot [thick, black003, mark=diamond*, mark size=3, mark options={solid,draw=white}]
table {%
1 0.0120056420564651
2 0.0240084007382393
3 0.0360082872211933
4 0.0480052754282951
5 0.0599993951618671
6 0.0719906315207481
7 0.0839790031313896
8 0.0959645062685013
9 0.10794710367918
10 0.11992684751749
11 0.13190370798111
12 0.143877729773521
};
\addlegendentry{PSDU: 4194 kB}
\addplot [thick, indigo7111104, dashed, mark=*, mark size=2, mark options={solid,draw=white}, forget plot]
table {%
1 4.27046251296997
2 8.19112682342529
3 11.8032779693604
4 15.1419563293457
5 18.23708152771
6 21.114372253418
7 23.7960319519043
8 26.3013706207275
9 28.6472129821777
10 30.8483333587646
11 32.9177093505859
12 34.8668251037598
};
\addplot [thick, chocolate23711033, dashed, mark=triangle*, mark size=3, mark options={solid,draw=white}, forget plot]
table {%
1 0.373948335647583
2 0.745110213756561
3 1.11351680755615
4 1.47919881343842
5 1.84218621253967
6 2.20250844955444
7 2.56019473075867
8 2.91527462005615
9 3.26777625083923
10 3.61772704124451
11 3.96515488624573
12 4.31008625030518
};
\addplot [thick, indianred1896475, dashed, mark=square*, mark size=2, mark options={solid,draw=white}, forget plot]
table {%
1 0.0954729989171028
2 0.190763860940933
3 0.285873115062714
4 0.380801260471344
5 0.475548923015594
6 0.570116460323334
7 0.66450434923172
8 0.758713364601135
9 0.852743804454803
10 0.946596324443817
11 1.04027092456818
12 1.13376891613007
};
\addplot [thick, black003, dashed, mark=diamond*, mark size=3, mark options={solid,draw=white}, forget plot]
table {%
1 0.0239918455481529
2 0.047972172498703
3 0.071941003203392
4 0.0958983302116394
5 0.119844220578671
6 0.143778547644615
7 0.167701482772827
8 0.191612973809242
9 0.215512990951538
10 0.239401459693909
11 0.26327857375145
12 0.287144303321838
};
\end{axis}

\end{tikzpicture}